\documentclass[12pt,english,floatfix,nofootinbib,superscriptaddress,aps,prd,preprint]{revtex4}
\usepackage[utf8]{inputenc}       
\usepackage[english]{babel}       
\usepackage{amsmath,amsthm,amssymb,amsfonts,mathrsfs,amsbsy} 
\usepackage{tensor}               
\usepackage{slashed}              
\usepackage{esint}                
\usepackage{cancel}               
\usepackage{graphicx}             
\usepackage{natbib}
\usepackage{float}                
\usepackage{subfig}               
\usepackage[font=small,labelfont=bf]{caption} 
\usepackage{multirow}             
\usepackage{array}                
\usepackage{tikz}                 
\usetikzlibrary{quotes,angles,arrows,decorations.markings} 
\usepackage{makecell} 

\usepackage{lipsum}

\usepackage{xcolor}               
\usepackage{color}                
\usepackage{textcomp}             
\usepackage{units}                

\newcommand{\be}{\begin{equation}}
\newcommand{\ee}{\end{equation}}
\newcommand{\Be}{\begin{eqnarray}}
\newcommand{\Ee}{\end{eqnarray}}

\newcommand{\mincir}{\raise
-3.truept\hbox{\rlap{\hbox{$\sim$}}\raise4.truept\hbox{$<$}\ }}
\newcommand{\magcir}{\raise
-3.truept\hbox{\rlap{\hbox{$\sim$}}\raise4.truept\hbox{$>$}\ }}

\newcolumntype{Y}{>{\centering\arraybackslash}X}
\providecommand{\U}[1]

\usepackage[dvips]{epsfig}        
\usepackage[dvips]{graphicx}      
\usepackage{hyperref}             
\hypersetup{
    colorlinks=true,              
    breaklinks=true,              
    citecolor=blue,               
    linkcolor=[rgb]{0,0.5,0.9},   
    urlcolor=red,                 
    filecolor=green               
}

\newcommand{\ie}{\begin{equation}}
\newcommand{\fe}{\end{equation}}
\newcommand{\se}{\begin{eqnarray}}
\newcommand{\ff}{\end{eqnarray}}

\newcommand{\nl}{\text{NLED}}
\newcommand{\eff}{\text{eff}}

\begin{document}

\title{Gravitational signatures of a nonlinear electrodynamics in $f(R,T)$ gravity}


\author{A. A. Ara\'{u}jo Filho}
\email{dilto@fisica.ufc.br}
\affiliation{Departamento de Física, Universidade Federal da Paraíba, Caixa Postal 5008, 58051--970, João Pessoa, Paraíba,  Brazil.}
\affiliation{Departamento de Física, Universidade Federal de Campina Grande Caixa Postal 10071, 58429-900 Campina Grande, Paraíba, Brazil.}


\author{N. Heidari}
\email{heidari.n@gmail.com}

\affiliation{Center for Theoretical Physics, Khazar University, 41 Mehseti Street, Baku, AZ-1096, Azerbaijan.}
\affiliation{School of Physics, Damghan University, Damghan, 3671641167, Iran.}

\author{I. P. Lobo}
\email{lobofisica@gmail.com}

\affiliation{Department of Chemistry and Physics, Federal University of Para\'iba, Rodovia BR 079 - km 12, 58397-000 Areia-PB,  Brazil}
\affiliation{Departamento de Física, Universidade Federal de Campina Grande Caixa Postal 10071, 58429-900 Campina Grande, Paraíba, Brazil.}


\author{V. B. Bezerra}
\email{valdir@fisica.ufpb.br}
\affiliation{Departamento de Física, Universidade Federal da Paraíba, Caixa Postal 5008, 58051--970, João Pessoa, Paraíba,  Brazil.}


\date{\today}

\begin{abstract}

In this work, we investigate a nonlinear electrodynamics model in the context of $f(R,T)$ gravity. We begin by outlining the general features of the theory and analyzing the event horizon under conditions ensuring its real and positive definiteness. We then examine light trajectories, focusing on critical orbits, shadow radii, and geodesics of massless particles. The parameters $\alpha$ and $\beta$, associated with the nonlinear extension of the Reissner--Nordström spacetime, are constrained using observational data from the Event Horizon Telescope (EHT). Subsequently, we analyze the thermal aspects of the system, including Hawking temperature, entropy, and heat capacity. Quasinormal modes are computed for scalar, vector, tensor, and spinorial perturbations, with the corresponding time--domain profiles explored as well. Gravitational lensing is then studied in both weak and strong deflection limits, along with the stability of photon spheres. Finally, we examine additional topological aspects, including topological thermodynamics and  the topological photon sphere.

\end{abstract}


\maketitle

\tableofcontents


\section{Introduction}

Extensions of General Relativity (GR) have garnered considerable attention in recent years, primarily due to their potential to resolve key observational inconsistencies within the $\Lambda$CDM framework. These include the persistent $H_0$ tension \cite{Hu:2023jqc}, the $S_8$ discrepancy \cite{Joseph:2022jsf}, and emerging evidence from DESI indicating a possible variation in the cosmological constant \cite{DESI:2025zgx}. In addition to observational motivations, several theoretical challenges—such as the occurrence of singularities \cite{Senovilla:1998oua}, the unresolved nature of dark energy \cite{Frusciante:2019xia}, and the underlying mechanism driving cosmic inflation \cite{Vazquez:2018qdg}—have further spurred interest in modifications to GR. Beyond modifications arising from additional terms in the Einstein--Hilbert action—motivated by various theoretical arguments—recent studies have explored the possibility of relaxing the coupling between gravity and matter as a means to introduce novel effects that may help resolve the aforementioned tensions in GR \cite{BarrosoVarela:2024htf}. This approach, known as \emph{non--minimal coupling}, can be implemented in various forms (for a comprehensive review, see \cite{Velten:2021xxw}).

One of the most extensively investigated realizations of such effects involves incorporating the stress--energy tensor of matter—or specific matter fields—into the gravitational sector of the Lagrangian via a general function $f(R,T)$, where $R$ denotes the Ricci scalar and $T$ the trace of the stress--energy tensor \cite{Harko:2011kv}. In these models, dynamical violations of the equivalence principle may emerge, opening up important observational possibilities, especially in the strong--gravity regime.

Extreme environments offer a valuable testing ground for examining the boundaries of fundamental physical theories. In particular, nonlinear extensions of Maxwell electrodynamics have garnered significant interest for their potential to resolve singularities in both cosmological and astrophysical settings~\cite{PhysRevD.65.063501, Sorokin:2021tge}. In fact, some physical processes indicate that we should modify Maxwell's electromagnetism. For instance, vacuum polarization induces small deviations from QED, such as birefringence \cite{Battesti:2012hf}, which could be phenomenologically described as effects of nonlinear electrodynamics \cite{Sorokin:2021tge,Boillat:1970gw,Bialynicka-Birula:1970nlh}. From a theoretical perspective, nonlinear modifications of Maxwell electrodynamics arise in quantum gravity theories, such as ModMax theory, string theory and M--theory \cite{Gibbons:2001gy,Bandos:2020hgy,heidari2025gravitational,2024arXiv241102907E,panah2025some}, and have been investigated in a myriad of scenarios, from cosmology and astrophysics to condensed matter physics \cite{Ayon-Beato:2000mjt,Pereira:2018mnn,Gaete:2014nda,Novello:2006ng,Pan:2011vi,Neves:2023uxi}. Experimental efforts have been undertaken to search for traces of these effects, including Pb--Pb collisions at the LHC \cite{ATLAS:2017fur} and investigations of light--by--light scattering and photon splitting at the ROKK-1M facility \cite{Akhmadaliev:2001ik}.

Among various astrophysical phenomena, black hole observables have become especially promising tools for probing fundamental physics—an interest amplified by the landmark detections from the LIGO and VIRGO collaborations~\cite{LIGOScientific:2016aoc}, as well as high--resolution imaging provided by the Event Horizon Telescope (EHT) collaboration \cite{EventHorizonTelescope:2019dse,EventHorizonTelescope:2022wkp}.

Gravitational wave astronomy has emerged as a powerful tool for probing a wide range of phenomena, including gravitational lensing effects~\cite{heidari2023gravitational,020,araujo2024gravitational,019,aa2024implications,araujo2023analysis}. Traditionally, research in this area has focused on light deflection in weak gravitational fields, particularly within the context of the Schwarzschild metric and more general static, spherically symmetric spacetimes \cite{021}. However, the strong--field regime near compact objects such as black holes is anticipated to exhibit pronounced deviations from classical predictions, potentially enabling critical tests of both general relativity and nonlinear electrodynamics~\cite{022,mohan2025strong}.

Black hole shadows serve as powerful observational features for investigating the behavior of gravity in the strong-field regime. These dark regions, set against the backdrop of luminous emission from accretion of matter, result from extreme light deflection near the event horizon and carry information about the geometry of the surrounding spacetime. Theoretical studies of this phenomenon date back to Bardeen in the 1970s~\cite{Cunningham}, with further developments by Falcke, Melia, and Agol~\cite{Falcke:1999pj}, who proposed that the shadow of $Sgr A^{*}$ could be detectable via very long baseline interferometry operating at submillimeter wavelengths. This prediction was confirmed in 2019, when the Event Horizon Telescope (EHT) collaboration released the first image of a black hole in the center of the galaxy $M87$, followed by the imaging of $Sgr A^{*}$. These results marked a turning point in gravitational physics, enabling detailed comparisons between general relativity and alternative theories in highly curved spacetimes~\cite{Vagnozzi:2019apd, Bambi:2019tjh, Allahyari:2019jqz, Kumar:2020hgm, Afrin:2021imp, Khodadi:2021gbc, Afrin:2021wlj, Khodadi:2022pqh, Fu:2021fxn, Afrin:2022ztr, Afrin:2023uzo, Ghosh:2022kit, Afrin:2024khy, Khodadi:2024ubi,Liu:2024lve,Nojiri:2024qgx,Nojiri:2024nlx,Nojiri:2024txy,karmakar2023thermodynamics}.

The investigation of light deflection near compact astrophysical objects has evolved considerably, especially with the development of lensing formalisms tailored to spacetimes dominated by strong gravitational fields. An important step in this direction was taken when Virbhadra and Ellis formulated a version of the lens equation suitable for black holes embedded in asymptotically flat geometries~\cite{031,virbhadra2000schwarzschild}. Their results revealed that intense curvature around such objects can give rise to multiple highly deflected images, symmetrically distributed with respect to the optical axis—a phenomenon absent in weak--field regimes. This approach soon became a basis for subsequent refinements~\cite{032,033,034}, enabling a more precise description of light propagation in non--perturbative gravitational domains.

Over time, this analytical framework has been extended to diverse gravitational settings, taking into account a wide variety of spacetime geometries~\cite{Pantig:2022ely,grespan2023strong,Kuang:2022xjp,Ovgun:2018fnk,metcalf2019strong,Ovgun:2018tua,virbhadra2002gravitational,aa2024antisymmetric,Okyay:2021nnh,bisnovatyi2017gravitational,ezquiaga2021phase,virbhadra1998role,Li:2020dln,Pantig:2022gih,cunha2018shadows,oguri2019strong,Virbhadra:2022ybp}, including metrics derived from extensions of general relativity~\cite{heidari2023gravitational,40,nascimento2024gravitational,chakraborty2017strong} and solutions featuring nontrivial topological structures such as wormholes~\cite{38.3,ovgun2019exact,38.1,38.5,38.4,38.2,Lobo:2020jfl}. Moreover, charged and rotating configurations, like the Reissner--Nordström and Kerr--like spacetimes, have also been investigated in this context~\cite{37.5,hsieh2021gravitational,036.2,jusufi2018gravitational,37.4,hsieh2021strong,37.1,37.2,37.6,036.1,37.3,036}. More recent efforts have focused on the role of gravitational-induced image deformation and its measurable optical features, broadening the theoretical landscape of lensing phenomena~\cite{virbhadra2024conservation,virbhadra2022distortions}.

Moreover, small disturbances in the region surrounding a black hole lead to characteristic oscillatory responses, which dominate the gravitational wave signal during the phase known as the ringdown \cite{Konoplya:2019hlu,Konoplya:2013rxa,Kokkotas:2010zd,karmakar2024quasinormal,gogoi2023quasinormal,karmakar2022quasinormal,Konoplya:2007zx,gogoi2024quasinormal,Konoplya:2011qq}. These oscillations occur at complex frequencies referred to as quasinormal modes, which depend solely on the black hole’s intrinsic parameters. The real part of each mode determines the oscillation frequency, while the imaginary component controls the rate at which the amplitude decays over time. Because quasinormal spectra are directly linked to the black hole’s mass, charge, and angular momentum, they offer a precise characterization of the object and also establish a connection with its shadow properties~\cite{Jusufi:2020dhz} and the greybody factors \cite{Konoplya:2024vuj,Konoplya:2024lir}. Recently, it has been claimed that quasinormal modes have been detected, but this is still a matter of debate where the control of uncertainties play a significant role \cite{Franchini:2023eda}. We expect that future runs of the LIGO/Virgo/KAGRA collaboration can help to clarify this issue.

In this paper, it is explored the implications of a nonlinear electrodynamics scenario embedded in the context of $f(R,T)$ gravity. The analysis begins with a revision of the theoretical structure, followed by a discussion on the criteria ensuring a physically meaningful event horizon. Attention is then directed toward the propagation of light, emphasizing the behavior of massless particles along null geodesics, the characterization of shadow radii, and the determination of critical circular orbits. The thermodynamic profile of the system is subsequently detailed through the evaluation of the Hawking temperature, entropy, and heat capacity. A comprehensive treatment of quasinormal mode spectra is performed for scalar, vector, tensor, and spinorial field perturbations, including their evolution in the time domain. The investigation proceeds with the study of gravitational lensing in both weak and strong regimes, along with an assessment of photon sphere stability. The work finishes by addressing topological characteristics, with emphasis on the formulation of topological thermodynamics and the identification of a topological photon sphere.


\section{The black hole solution and the general features}

A static and spherically symmetric black hole configuration emerges within the framework of $f(R,T)$ gravity, where the curvature scalar $R$ and the trace $T$ of the stress--energy tensor enter a modified gravitational action. In this context, the matter content arises from a nonlinear extension of classical electrodynamics. The specific model investigated in Ref. \cite{Rois:2024iiu} adopts the function $f(R,T) = R + \beta T$, where the constant $\beta$ introduces a non--minimal coupling between the geometry and the matter sector.

The electromagnetic contribution stems from a generalized Lagrangian density of the form $\mathcal{L}_{\text{nl}}(F) = f_0 + F + \alpha F^p$, where $F = \frac{1}{4}F_{\mu\nu}F^{\mu\nu}$ represents the standard Maxwell invariant and $\alpha$, $p$, and $f_0$ are constants characterizing the deviation from linear electrodynamics. Notably, $f_0$ plays the role of an effective cosmological constant term, while the parameters $\alpha$ and $p$ encode the strength and structure of the nonlinear corrections.

The field strength tensor $F_{\mu\nu}$, defined through the antisymmetric derivative of the gauge field $A_\mu$, takes the usual form $F_{\mu\nu} = \partial_\mu A_\nu - \partial_\nu A_\mu$. For a purely magnetic configuration, the only nonvanishing component is $F_{23} = Q \sin\theta$, corresponding to a magnetic monopole with charge $Q$. Under this configuration, the Maxwell invariant becomes $F = Q^2/(2r^4)$, satisfying the field equations obtained by varying the full action with respect to the vector potential $A_\gamma$
\ie\label{eq:action}
    S[g_{\mu\nu},A_{\gamma}]=\int \sqrt{-g} \, \mathrm{d}^4x \left[f(R,T)+2\kappa^2{\cal L}_{\nl}(F)\right]\, .
\fe
Here, the parameter $\kappa$ denotes the coupling strength associated with the matter Lagrangian and is determined by requiring consistency with the Newtonian limit of the gravitational theory.

A spherically symmetric spacetime arising from the field equations associated with the action \eqref{eq:action}, expressed in the coordinate system $(t, r, \theta, \phi)$, admits the following metric form
$\mathrm{d} s^2 = f(r) \mathrm{d}t^2 - f^{-1}(r)\mathrm{d}r^2 - r^2(\mathrm{d}\theta^2+\sin^2 \theta \mathrm{d}\phi^2)$ reads
\ie
\label{generalfr}
f(r) = 1 - \frac{2M}{r} + \frac{Q^{2}}{r^{2}} - \frac{\Lambda_{\eff}}{3}r^2+\frac{2^{1-p}}{3-4p}\alpha[2 \beta(p-1)-1]Q^{2p}r^{2-4p} .
\fe

An effective cosmological term arises in the model as $\Lambda_{\eff} = 2(2\beta + 1)f_0$, with $M$ denoting the mass parameter associated with the black hole. The appearance of the term proportional to $r^{2 - 4p}$ originates from the nonlinear structure of the electromagnetic sector and is governed by the coefficient $\alpha$. The parameter $\beta$, introduced through the non--minimal matter--geometry coupling, alters both the magnitude of the nonlinear electromagnetic corrections and the effective cosmological term. In addition, as we shall show, these latter parameters will be constrained by astrophysical observations based on data from the EHT.

In parallel, the exponent $p$ not only influences the degree of nonlinearity but also modifies the radial scaling behavior of the magnetic charge contribution. Notably, for $p = 1$, the standard Reissner--Nordström form is recovered, allowing the nonlinear effects to be absorbed into a redefined effective charge $Q_{\eff}^2 \equiv Q^2(1 + \alpha)$. In contrast, when $p > 1$, the corrections decay more rapidly with distance, as one should naturally expect.

\begin{figure}
    \centering
     \includegraphics[scale=0.51]{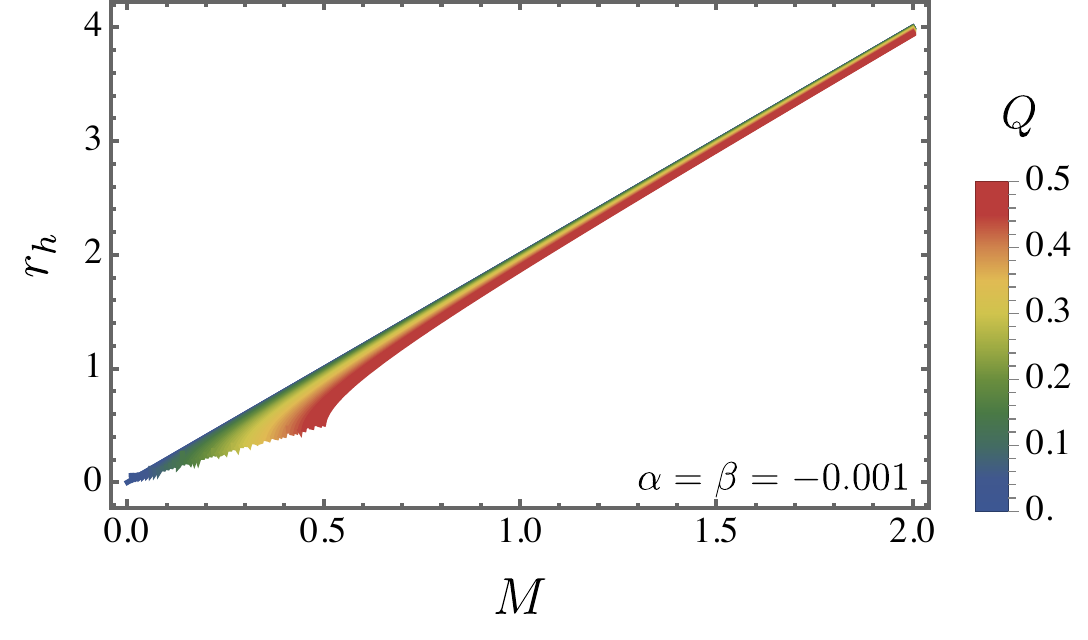}
    \caption{The behavior of the event horizon radius $r_h$ as a function of the black hole mass $M$ is depicted for different values of the magnetic charge $Q$, with the parameters fixed at $\alpha = \beta = -0.001$.}
    \label{horizon}
\end{figure}

Taking $p = 2$ in Eq. (\ref{generalfr}) and neglecting the cosmological constant term, the resulting expression simplifies to the following form, as presented in Ref. \cite{Rois:2024iiu}:
\ie
\label{metricfuntion}
f(r) = 1 - \frac{2M}{r} + \frac{Q^{2}}{r^{2}} - \frac{\alpha(2 \beta -1)Q^{4}}{10 r^{6}}.
\fe
The above function, has six distinct roots. Nevertheless, only three of them turn out to be physical (real and positive defined quantities). In other words, we have only one event horizon $r_{h}$, which is written as
\ie
\label{eventhhh}
r_{h} = \left(M + \sqrt{M^2-Q^2}\right) + \frac{\alpha  \left(2 \beta  Q^4-Q^4\right)}{20 \left(M + \sqrt{M^2-Q^2}\right)^3 \left(M^2-Q^2 + M \sqrt{M^2-Q^2}\right)},
\fe
where we have naturally considered $\alpha$ and $\beta$ small. It is important to mention that when $\alpha \to 0$ and $\beta \to 0$, we recover the event horizon of the Reisser--Nordstrom case. In fact, when $\alpha\to 0$, we have the standard Maxwell Lagrangian, whose stress-energy tensor has null trace, which is the reason why the non-minimal coupling effect (that depends on $\beta$) also vanishes. Notice that similar to this latter black hole case, here, a relation must be imposed for the parameters $M$, $Q$, $\alpha$, $\beta$ in order to $r_{h} > 0$, which are
\ie
2 \alpha \beta  Q^4-Q^4 >0, \quad \quad \text{and} \quad \quad M > Q.
\fe
Notice that these conditions must be fulfilled simultaneously. Fig. \ref{horizon} provides a qualitative illustration, while Table \ref{eventhorizon} presents the corresponding quantitative values of the event horizon $r_h$. Both indicate that increasing the magnetic charge $Q$—with fixed parameters $\alpha = \beta = -0.01$—leads to a reduction in the event horizon radius. Additionally, for fixed $Q$, a decrease in the coupling parameters $\alpha = \beta$ results in a smaller value of $r_h$.

\begin{table}[!h]
\begin{center}
\begin{tabular}{c c c|| c c c } 
 \hline\hline\hline
 $Q$ & $\alpha=\beta$ &  $r_{h}$ & $Q$ & $\alpha=\beta$ &  $r_{h}$  \\ [0.2ex] 
 \hline
 0.6  & -0.01 & 1.80001 & 0.6  & -0.01 & 1.800010  \\ 

 0.7  & -0.01 & 1.71416 & 0.6  & -0.02 & 1.800020   \\
 
 0.8  & -0.01 & 1.60005 & 0.6  & -0.03 & 1.800035  \\
 
 0.9  & -0.01 & 1.43607  & 0.6  & -0.04 & 1.800030  \\
 
 0.99  & -0.01 & 1.14312 & 0.6  & -0.05 & 1.800040   \\ 
 [0.2ex] 
 \hline \hline \hline
\end{tabular}
\caption{\label{eventhorizon} The left side of the table presents the numerical values of the event horizon $r_h$ for various values of the magnetic charge $Q$, with the parameters fixed at $\alpha = \beta = -0.01$. On the right side, the corresponding horizon radii are listed for different values of $\alpha = \beta$, while keeping the magnetic charge $Q$ constant.}
\end{center}
\end{table}


\section{The journey of light}

In spacetimes influenced by nonlinear electromagnetic corrections, the parameters $\alpha$ and $\beta$ significantly affect the geometry and its observational consequences. These couplings impact the field dynamics, resulting in notable deviations in photon trajectories. Of particular importance is the photon sphere, which determines the conditions for circular light paths and strongly affects the gravitational lensing profile. The black hole shadow, shaped by the bending and confinement of light near this region, stands out as a measurable feature through which such theoretical frameworks can be constrained. Furthermore, a detailed examination of null geodesics within this context provides valuable information on how light behaves under these nonlinear modifications. The following subsections address each of these elements in more detailed manner.


\subsection{Geodesics}

This section is devoted to the investigation of geodesic motion. To this end, we begin by formulating
\ie
\frac{\mathrm{d}^{2}x^{\mu}}{\mathrm{d}\tau^{2}} + \Gamma\indices{^\mu_\alpha_\beta}\frac{\mathrm{d}x^{\alpha}}{\mathrm{d}\tau}\frac{\mathrm{d}x^{\beta}}{\mathrm{d}\tau} = 0. \label{geogeo}
\fe

Here, $\tau$ denotes a generic affine parameter that parametrizes the trajectory. From this setup, one obtains a system of four coupled differential equations, each governing the evolution along one of the spacetime coordinates. The full set of equations describing the geodesic motion is presented as follows:
\ie
\frac{\mathrm{d} t^{\prime}}{\mathrm{d} \tau} = -\frac{2 r' t' \left(10 M r^5+3 \alpha  (2 \beta -1) Q^4-10 Q^2 r^4\right)}{10 r^6 (r-2 M)+\alpha  (1-2 \beta ) Q^4 r+10 Q^2 r^5} ,
\fe
\ie
\begin{split}
& \frac{\mathrm{d} r^{\prime}}{\mathrm{d} \tau} = \frac{\left(t'\right)^2 \left(10 r^5 (2 M-r)+\alpha  (2 \beta -1) Q^4-10 Q^2 r^4\right) \left(10 M r^5+3 \alpha  (2 \beta -1) Q^4-10 Q^2 r^4\right)}{100 r^{13}} \\
& -\frac{\left(r'\right)^2 \left(-10 M r^5+\alpha  (3-6 \beta ) Q^4+10 Q^2 r^4\right)}{10 r^6 (r-2 M)+\alpha  (1-2 \beta ) Q^4 r+10 Q^2 r^5}  - \left(\theta '\right)^2 \left(2 M+\frac{\alpha  (2 \beta -1) Q^4}{10 r^5}-\frac{Q^2}{r}-r\right)\\
& + r \sin ^2(\theta ) \left(\varphi '\right)^2 \left(-\frac{2 M}{r}+\frac{Q^4 (\alpha -2 \alpha  \beta )}{10 r^6}+\frac{Q^2}{r^2}+1\right).
\end{split}
\fe
\ie
\frac{\mathrm{d} \theta^{\prime}}{\mathrm{d} \tau} = \sin (\theta ) \cos (\theta ) \left(\varphi '\right)^2-\frac{2 \theta ' r'}{r},
\fe
and, finally,  
\ie
\frac{\mathrm{d} \varphi^{\prime}}{\mathrm{d} \tau} = -\frac{2 \varphi ' \left(r'+r \theta ' \cot (\theta )\right)}{r}.
\fe

\begin{figure}
    \centering
     \includegraphics[scale=0.64]{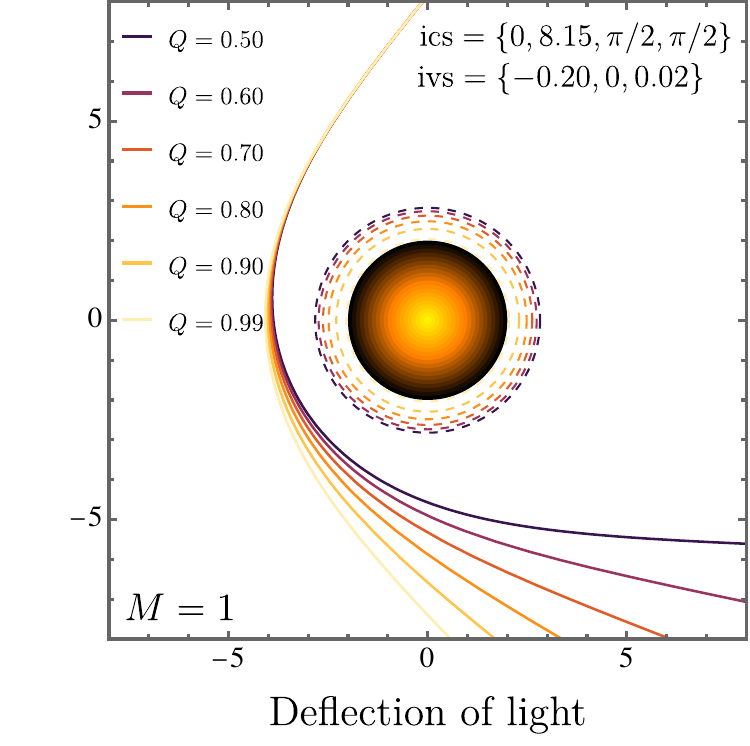}
     \includegraphics[scale=0.64]{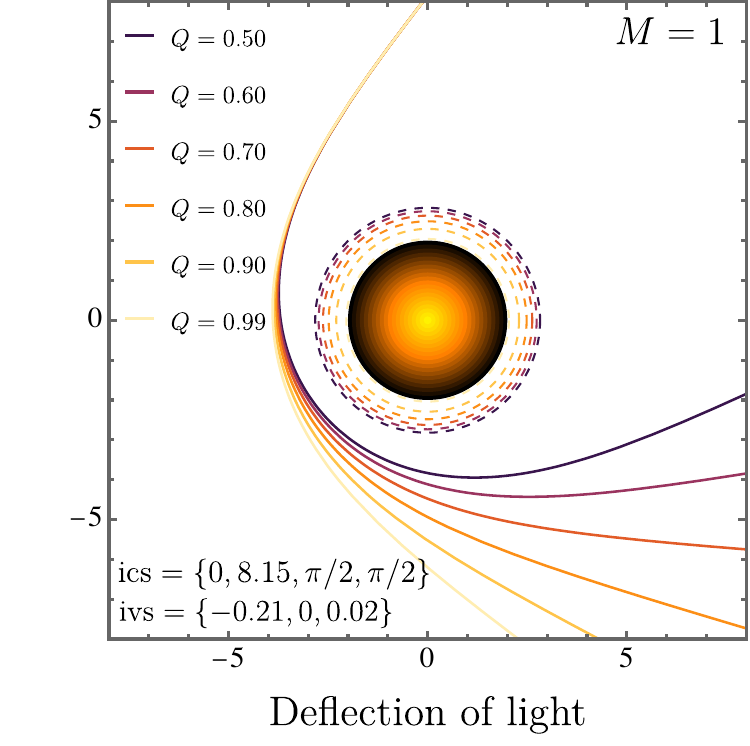}
     \includegraphics[scale=0.64]{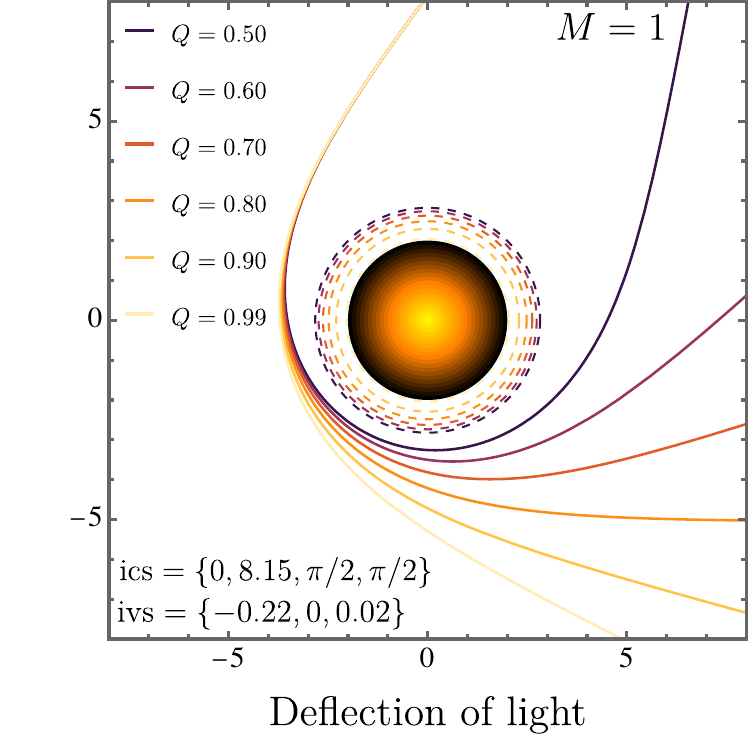}
     \includegraphics[scale=0.64]{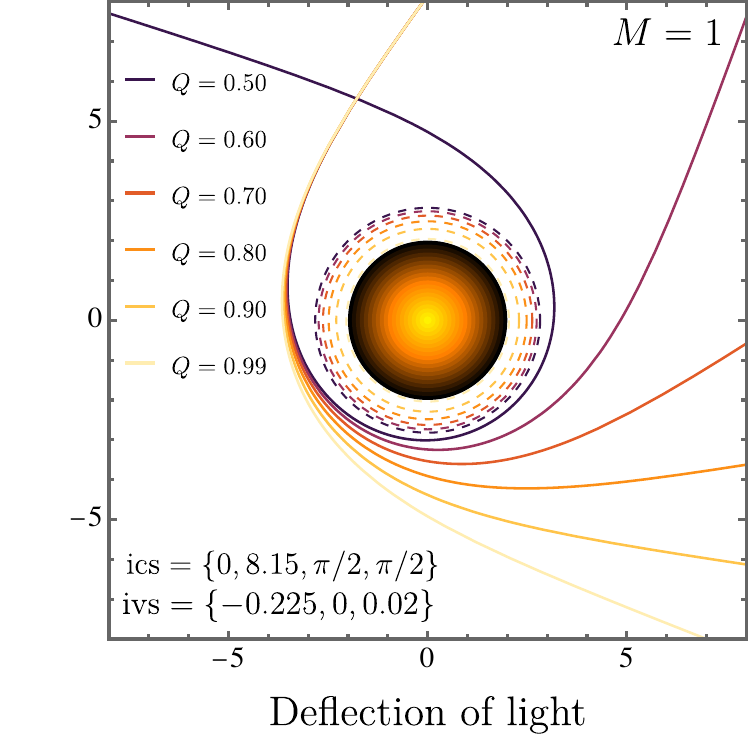}
    \caption{The geodesics, obtained numerically, are presented for $Q = 0.5$, $0.6$, $0.7$, $0.8$, $0.9$, and $0.99$, with fixed values $\alpha = \beta = -0.1$. The large black disk denotes the event horizon, while the dot--dashed lines indicate the photon sphere.}
    \label{geodesicslight}
\end{figure}

Fig. \ref{geodesicslight} illustrates the trajectories of massless particles, obtained numerically for varying values of the magnetic charge: $Q = 0.5$, $0.6$, $0.7$, $0.8$, $0.9$, and $0.99$, while keeping $\alpha = \beta = -0.1$ fixed. In the plot, the solid black disk marks the location of the event horizon, and the dot--dashed curves represent the photon sphere radius, i.e., both of them are shown just for a pictorial purpose only. One can observe that as $Q$ increases, the bending of light becomes progressively weaker, indicating a reduction in the deflection angle. 

\begin{figure}
    \centering
     \includegraphics[scale=0.64]{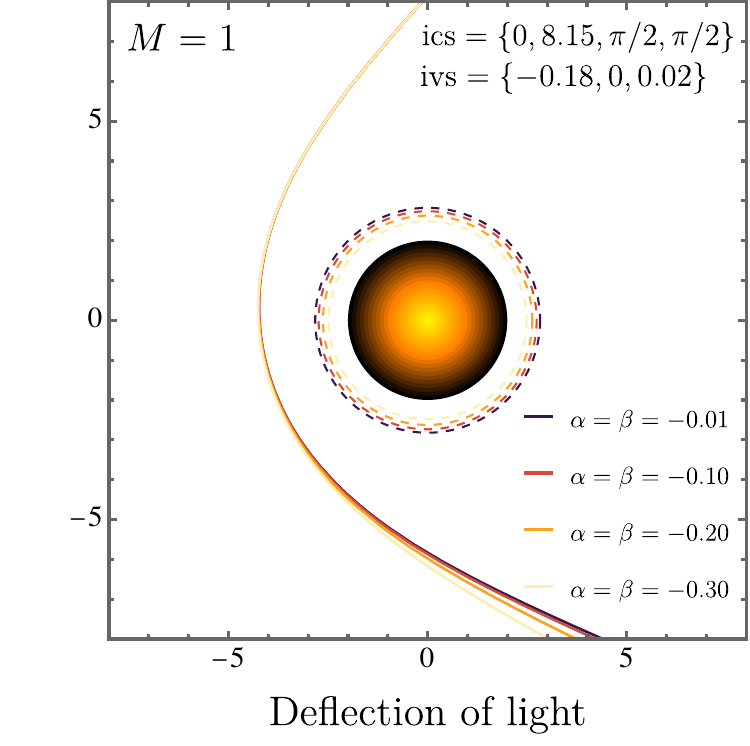}
     \includegraphics[scale=0.64]{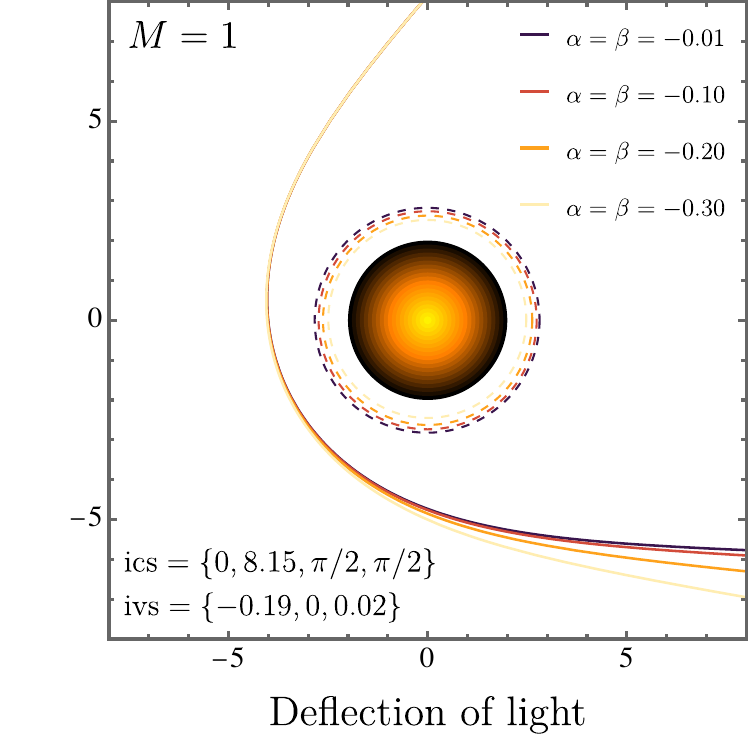}
     \includegraphics[scale=0.64]{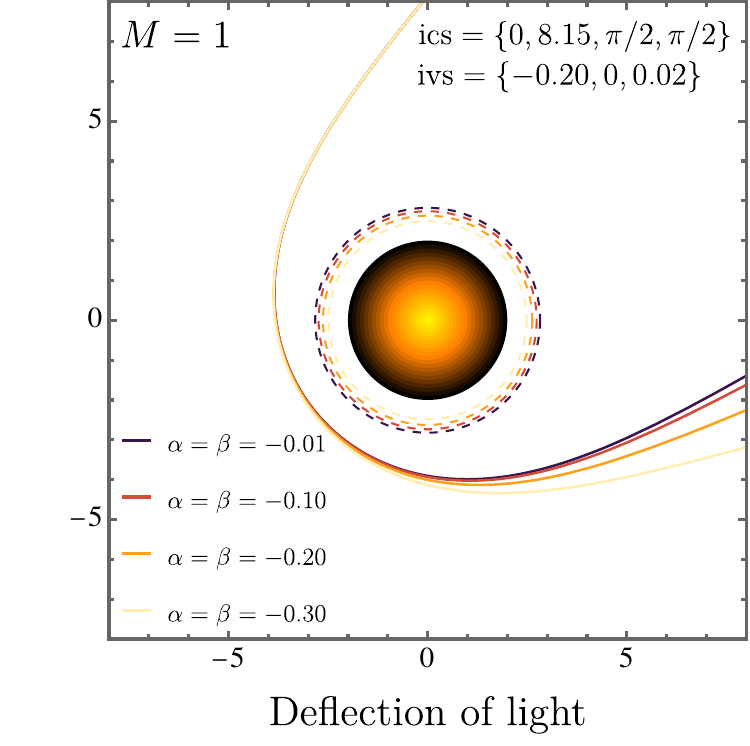}
     \includegraphics[scale=0.64]{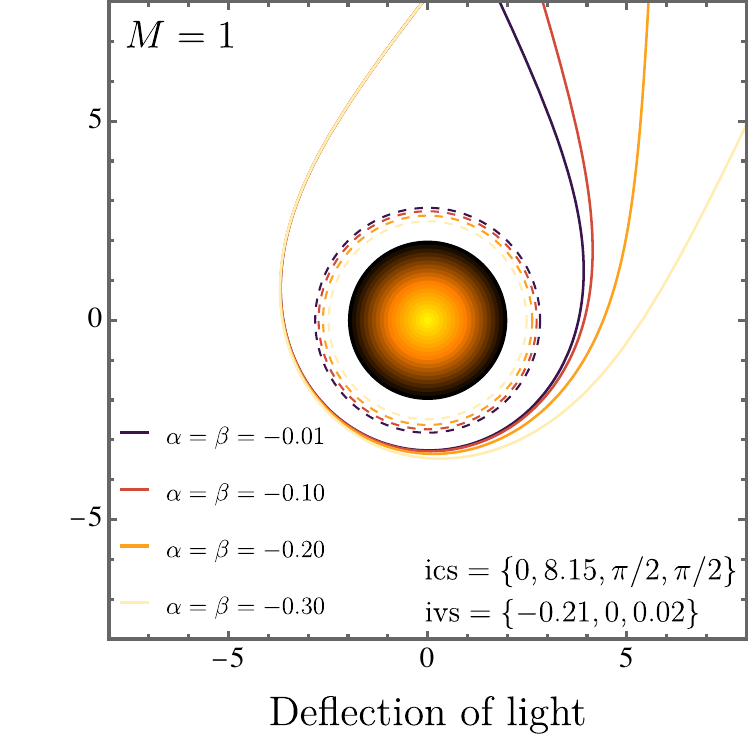}
    \caption{The numerically obtained geodesics are shown for $\alpha = \beta = -0.01$, $-0.1$, $-0.2$, and $-0.3$, with the charge fixed at $Q = 0.5$. The large colored disk represents the event horizon, and the dot--dashed lines mark the photon sphere corresponding to the chosen parameter values.}
    \label{geodesicslight2}
\end{figure}

On the other hand, Fig. \ref{geodesicslight2} illustrates how the parameters $\alpha$ and $\beta$ affect the trajectory of light. Overall, for $Q = 0.5$, we observe that decreasing these parameters leads to a weaker deflection, according to the numerical conditions adopted here.

In the following analysis, we constrain the parameters $\alpha$ and $\beta$ using observational data from the Event Horizon Telescope (EHT) related to $SgrA^{*}$ \cite{vagnozzi2022horizon,akiyama2022first}.

\subsection{Photon spheres and shadows}

As a starting point for the analysis, a general expression for the metric tensor $\mathrm{g}_{\mu\nu}$ is introduced
\ie
\mathrm{d}s^{2} = \mathrm{g}_{\mu \nu}\mathrm{d}x^\mu \mathrm{d}x^\nu  =  - \mathrm{A}(r)\mathrm{d}t^2 + \mathrm{B}(r)\mathrm{d}{r^2} + \mathrm{C}(r)\mathrm{d}\theta ^2 + \mathrm{D}(r)\,{{\mathop{\rm \sin}\nolimits} ^2}\theta \mathrm{d}\varphi ^2.
\fe

In this context, the functions $\mathrm{A}(r)$, $\mathrm{B}(r)$, $\mathrm{C}(r)$, and $\mathrm{D}(r)$ correspond to the radial profiles of the metric components. To proceed with the analysis, the Lagrangian formalism is employed, allowing for a systematic derivation of the equations of motion associated with photon trajectories
\ie
\mathcal{L} = \frac{1}{2}{\mathrm{g}_{\mu \nu }}{{\dot x}^\mu }{{\dot x}^\nu },
\fe
so that
\ie
\mathcal{L} = \frac{1}{2}\Big[ - \mathrm{A}(r){{\dot t}^2} + \mathrm{B}(r){{\dot r}^2} + \mathrm{C}(r){{\dot \theta }^2} + \mathrm{D}(r){{\mathop{\rm \sin}\nolimits} ^2}\, \theta {{\dot \varphi }^2}\Big].
\fe

Restricting the analysis to the equatorial plane ($\theta = \frac{\pi}{2}$) and utilizing the Euler--Lagrange formalism, one obtains two conserved quantities arising from the spacetime symmetries: the energy $\mathrm{E}$ associated with time translation invariance, and the angular momentum $\mathrm{L}$ linked to rotational symmetry. These constants are expressed as:
\ie
\label{sdcvfdsd}
\mathrm{E} = \mathrm{A}(r)\dot t \quad\mathrm{and}\quad \mathrm{L} = \mathrm{D}(r)\dot \varphi,
\fe
and by incorporating the condition for light--like particles, characterized by a null spacetime interval, the resulting expression for their motion simplifies to:
\ie
\label{lihhhfj}
- \mathrm{A}(r){{\dot t}^2} + \mathrm{B}(r){{\dot r}^2} + \mathrm{D}(r){{\dot \varphi }^2} = 0.
\fe

Accordingly, after performing the necessary algebraic steps to replace Eq. (\ref{sdcvfdsd}) into Eq. (\ref{lihhhfj}), the expression takes the following form:
\ie
\frac{{{{\dot r}^2}}}{{{{\dot \varphi }^2}}} = {\left(\frac{{\mathrm{d}r}}{{\mathrm{d}\varphi }}\right)^2} = \frac{{\mathrm{D}(r)}}{{\mathrm{B}(r)}}\left(\frac{{\mathrm{D}(r)}}{{\mathrm{A}(r)}}\frac{{{\mathrm{E}^2}}}{{{\mathrm{L}^2}}} - 1\right).
\fe

Also, notice that
\ie
\frac{\mathrm{d}r}{\mathrm{d}\lambda} = \frac{\mathrm{d}r}{\mathrm{d}\varphi} \frac{\mathrm{d}\varphi}{\mathrm{d}\lambda}  = \frac{\mathrm{d}r}{\mathrm{d}\varphi}\frac{\mathrm{L}}{\mathrm{D}(r)}, 
\fe
with
\ie
\Dot{r}^{2} = \left( \frac{\mathrm{d}r}{\mathrm{d}\lambda} \right)^{2} =\left( \frac{\mathrm{d}r}{\mathrm{d}\varphi} \right)^{2} \frac{\mathrm{L}^{2}}{\mathrm{D}(r)^{2}}.
\fe

Thus far, a broad framework has been established to identify the conditions for critical photon orbits—commonly referred to as the photon sphere—in a general static, spherically symmetric geometry. The next step involves tailoring this approach to the particular spacetime under consideration, leading to the following result:
$\mathrm{A}(r) = 1 - \frac{2M}{r} + \frac{Q^{2}}{r^{2}} - \frac{\alpha(2 \beta -1)Q^{4}}{10 r^{6}}$, $\mathrm{B}(r) =  \left(1 - \frac{2M}{r} + \frac{Q^{2}}{r^{2}} - \frac{\alpha(2 \beta -1)Q^{4}}{10 r^{6}}\right)^{-1}$, $\mathrm{C}(r) = r^{2}$ and $\mathrm{D}(r) = r^{2}\sin^{2}\theta$. Thereby,
\ie
\Dot{r}^{2} = \mathrm{E}^{2} + \mathcal{V}(r,\alpha,\beta,Q),
\fe
with $\mathcal{V}(r,\alpha,\beta,Q)$ being
\ie
\mathcal{V}(r,\alpha,\beta,Q) = \frac{L^2 \left(1 - \frac{2M}{r} + \frac{Q^{2}}{r^{2}} - \frac{\alpha(2 \beta -1)Q^{4}}{10 r^{6}}\right)}{r^2}.
\fe

To locate the position of the photon sphere (critical orbits), we must solve the equation $\mathrm{d}\mathcal{V}/\mathrm{d}r = 0$. Notably, this equation produces three distinct roots; nevertheless, only one of them represent a physical solution (being a real and positive defined quantity)
\ie
\begin{split}
\label{photonsphererph}
 r_{ph} =  & \, \,\frac{1}{2} \left(\sqrt{9 M^2-8 Q^2}+3 M\right) \\
& +\frac{32 \alpha  (2 \beta -1) Q^4}{5 \left(\sqrt{9 M^2-8 Q^2}+3 M\right)^3 \left(3 M \left(\sqrt{9 M^2-8 Q^2}+3 M\right)-8 Q^2\right)}.
\end{split}
\fe
It is worth noting that, similar to the procedure used for deriving the expression for the event horizon, the expression for $r_{ph}$ was also obtained under the assumption that both $\alpha$ and $\beta$ are small. Moreover, as it is straightforward to verify from the above expression, if we take into account the limit where $\alpha \to 0$ and $\beta \to 0$, we recover the photon sphere solution to the Reisser--Nordström case. In addition, we present the quantitative values of $r_{ph}$ in Tab. \ref{crsitsicsaldordbidtsd} by considering different values for $Q$, $\alpha$, and $\beta$.
In a general panorama, we observe that, for fixed values of $\alpha = \beta = -0.01$, increasing the magnetic charge $Q$ leads to a reduction in the photon sphere radius. Furthermore, although the variations are relatively small, a decrease in the coupling parameters $\alpha = \beta$ also results in a slight increase in the photon sphere size.

\begin{table}[!h]
\begin{center}
\begin{tabular}{c c c|| c c c } 
 \hline\hline\hline
 $Q$ & $\alpha=\beta$ &  $r_{ph}$ & $Q$ & $\alpha=\beta$ &  $r_{ph}$  \\ [0.2ex] 
 \hline
 0.60  & -0.01 & 2.73694 & 0.99  & -0.01 & 2.03854  \\ 

 0.70  & -0.01 & 2.62695 & 0.99  & -0.02 & 2.03876   \\
 
 0.80  & -0.01 & 2.48491 & 0.99  & -0.03 & 2.03899  \\
 
 0.90  & -0.01 & 2.29379  & 0.99  & -0.04 & 2.03922  \\
 
 0.99  & -0.01 & 2.03854 & 0.99  & -0.05 & 2.03947   \\ 
 [0.2ex] 
 \hline \hline \hline
\end{tabular}
\caption{\label{crsitsicsaldordbidtsd} The numerical values corresponding to the critical photon orbits are presented. On the left panel, the magnetic charge $Q$ is varied with fixed parameters $\alpha = \beta$, while on the right panel, $Q$ remains constant and the values of $\alpha = \beta$ are modified.
}
\end{center}
\end{table}

An important question naturally emerges here: is the critical orbit $r_{ph}$ stable or unstable? This issue will be properly examined in the context of gravitational lensing under the weak deflection limit, where the stability of photon trajectories will be investigated through the application of the Gauss--Bonnet theorem to the optical geometry \cite{Gibbons:2008rj}.

With these elements established, the shadow radius can now be precisely determined and is given by the following expression:
\ie
\begin{split}
& \mathcal{R} = \sqrt{\frac{\mathrm{D}(r)}{\mathrm{A}(r)}}\bigg|_{r = r_{ph}}
 \\
& = \sqrt{\frac{\left(\gamma +\frac{1}{2} \left(\sqrt{9 M^2-8 Q^2}+3 M\right)\right)^2}{\frac{Q^2}{\left(\gamma +\frac{1}{2} \left(\sqrt{9 M^2-8 Q^2}+3 M\right)\right)^2}-\frac{2 M}{\gamma +\frac{1}{2} \left(\sqrt{9 M^2-8 Q^2}+3 M\right)}-\frac{\alpha  (2 \beta -1) Q^4}{10 \left(\gamma +\frac{1}{2} \left(\sqrt{9 M^2-8 Q^2}+3 M\right)\right)^6}+1}},
\end{split}
\fe
where $\gamma$ is defined below
\ie
\gamma \equiv \frac{32 \alpha  (2 \beta -1) Q^4}{5 \left(\sqrt{9 M^2-8 Q^2}+3 M\right)^3 \left(3 M \left(\sqrt{9 M^2-8 Q^2}+3 M\right)-8 Q^2\right)}. 
\fe

\begin{figure}
    \centering
     \includegraphics[scale=0.51]{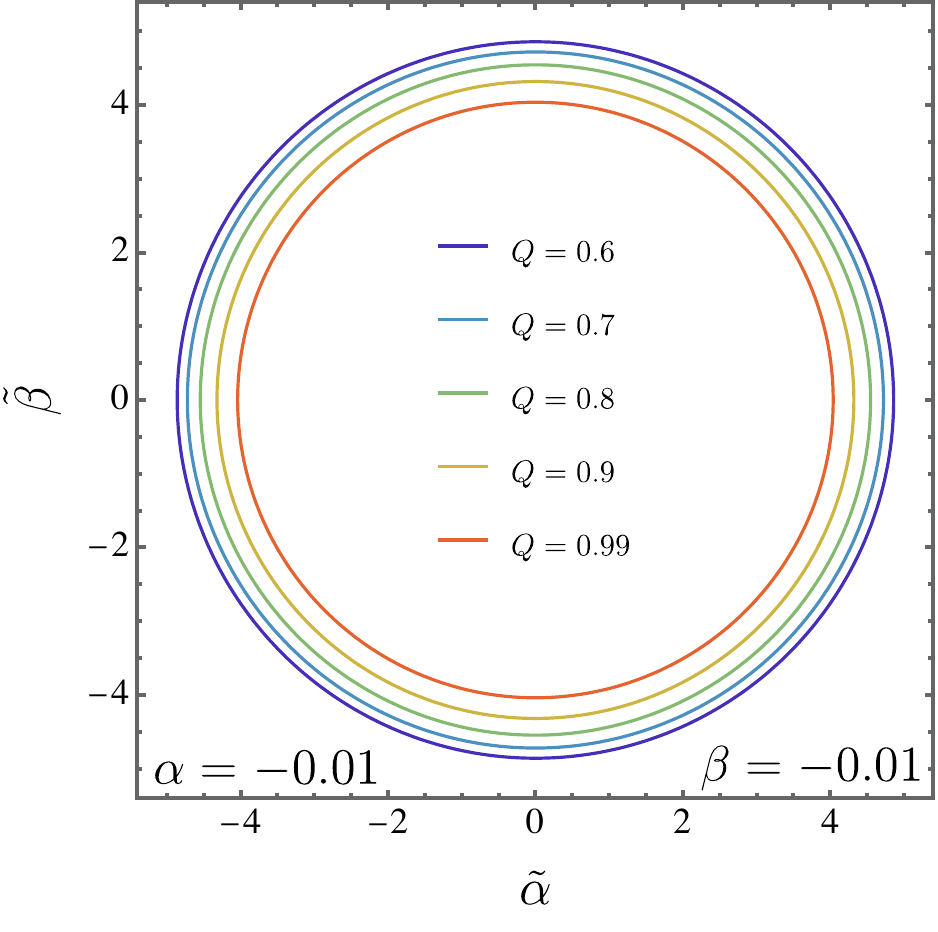}
     \includegraphics[scale=0.51]{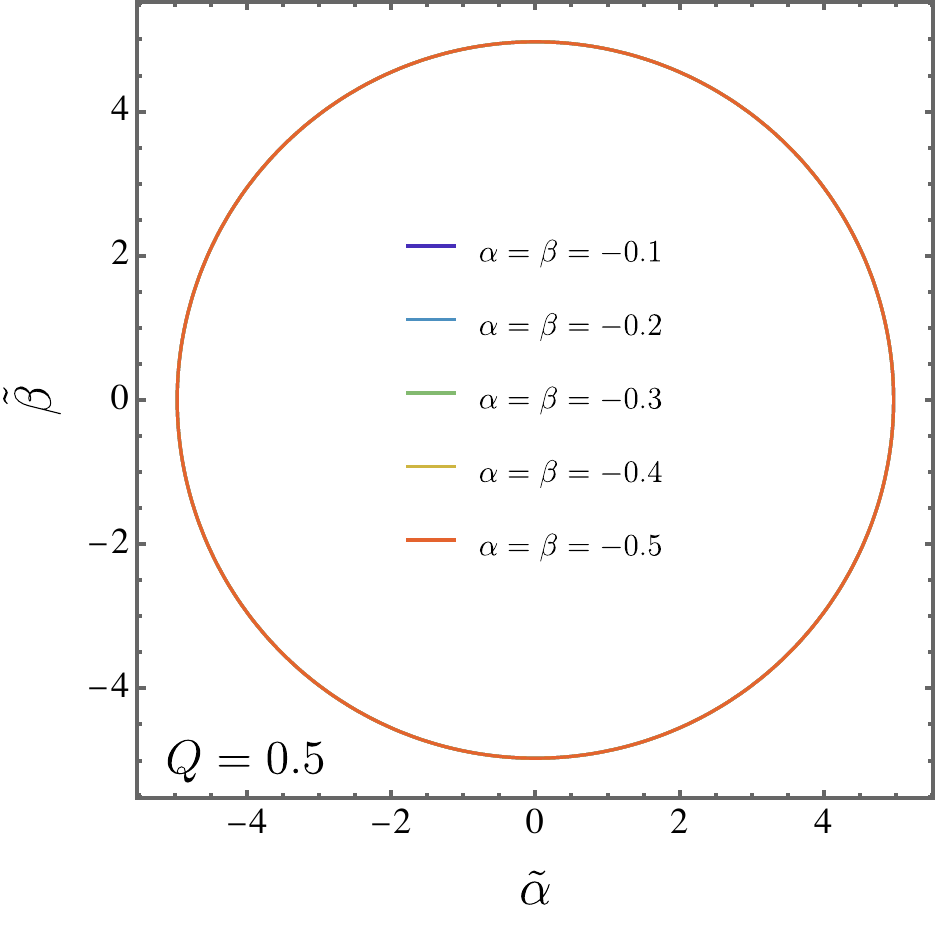}
    \caption{Parametric plots of the shadow radii are presented for different values of $Q$, $\alpha$, and $\beta$.}
    \label{shadowsradius}
\end{figure}

Adopting a widely used procedure in recent studies, the shadow contours are represented through parametric plots in terms of the celestial coordinates $\Tilde{\alpha}$ and $\Tilde{\beta}$ \cite{afrin2024testing,caeala1,caeala2,caeala3,caeala4}. In Fig. \ref{shadowsradius}, the silhouettes of the black hole are illustrated for different combinations of $Q$, $\alpha$, and $\beta$. The left panel demonstrates that increasing the magnetic charge $Q$ leads to a noticeable reduction in the shadow’s radius. A comparable outcome has been reported in recent studies exploring black holes within alternative formulations of nonlinear electrodynamics \cite{AraujoFilho:2024xhm,AraujoFilho:2024lsi}. Conversely, the right panel shows that lowering the values of $\alpha$ and $\beta$—although causing only minimal variation—slightly enlarges the shadow boundary, a trend that is numerically confirmed in Table \ref{shadowstable}.

\begin{table}[!h]
\begin{center}
\begin{tabular}{c c c|| c c c } 
 \hline\hline\hline
 $Q$ & $\alpha=\beta$ &  $\mathcal{R}$ & $Q$  & \, $\alpha=\beta$ &  $\mathcal{R}$  \\ [0.2ex] 
 \hline
 0.6  & -0.01 & 4.858696 & 0.5  & -0.01 & 4.967925  \\ 

 0.7  & -0.01 & 4.720688 & 0.5  & -0.02 & 4.967940   \\
 
 0.8  & -0.01 & 4.545999 & 0.5  & -0.03 & 4.967959  \\
 
 0.9  & -0.01 & 4.319262  & 0.5  & -0.04 & 4.967982  \\
 
 0.99  & -0.01 & 4.038971 & 0.5  & -0.05 & 4.968009   \\ 
 [0.2ex] 
 \hline \hline \hline
\end{tabular}
\caption{\label{shadowstable} The quantitative values of the shadow radii are presented for various values of $Q$, $\alpha$, and $\beta$.}
\end{center}
\end{table}

\subsection{Bounds to shadows based on EHT data}

Up to now, we have established the full framework for analyzing the behavior of light, covering critical photon orbits, shadow formation, and geodesic motion. We now shift our attention to the observational relevance of these results. In particular, we evaluate the phenomenological implications of our model by comparing it with phenomenological data from the EHT observations of $SgrA^{*}$ \cite{vagnozzi2022horizon,akiyama2022first}. In this context, by adopting a $2\sigma$ confidence interval, two distinct bounds on the shadow radius have been established
\cite{vagnozzi2022horizon,akiyama2022first}
\ie
\label{const1}
4.55 < \frac{\mathcal{R}}{M} < 5.22,
\fe
and
\ie
\label{const2}
4.21 < \frac{\mathcal{R}}{M} < 5.56.
\fe

\begin{figure}
    \centering
     \includegraphics[scale=0.51]{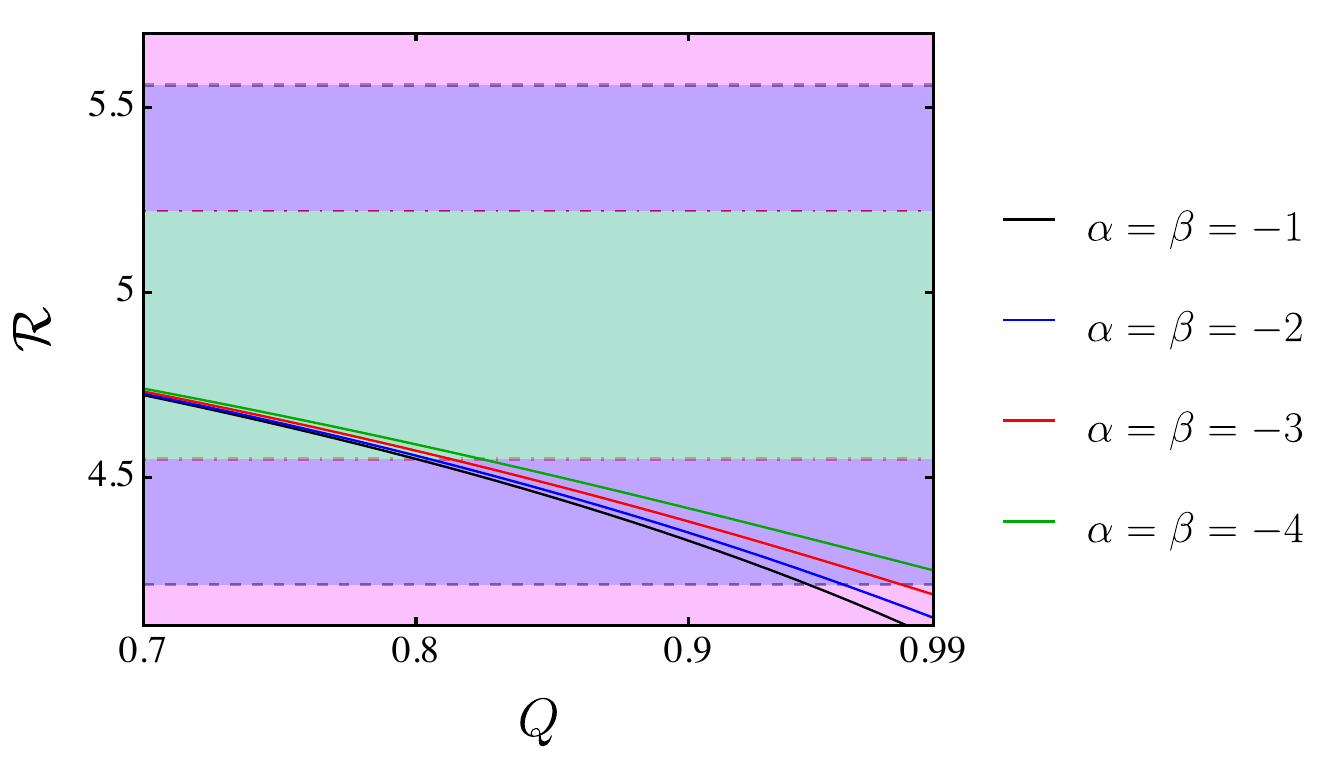}
    \caption{Constraints on the shadow radius as a function of the charge $Q$ are shown for various values of $\alpha$ and $\beta$, based on the observational data from the Event Horizon Telescope for $SgrA^{*}$ \cite{vagnozzi2022horizon,akiyama2022first}.}
    \label{sgrAcontrainst}
\end{figure}

Fig. \ref{sgrAcontrainst} presents the graphical constraints on the charge $Q$ for various fixed values of the parameters $\alpha$ and $\beta$. It is worth noting that the reverse procedure—varying $\alpha$ and $\beta$ for different fixed values of $Q$—was not carried out in this analysis, as the resulting curves become visually indistinguishable, making interpretation challenging in that configuration. In addition, this is why we also have considered ``huge'' values for $\alpha$ and $\beta$. To complement this analysis, Table \ref{boundstab} is included to offer a detailed quantitative assessment of the bounds derived in this work.

\begin{table}[h!]
\centering
\caption{The bounds are evaluated for the charge $Q$ for several choices of $\alpha$ and $\beta$, using observational data from the Event Horizon Telescope for $SgrA^{*}$ \cite{vagnozzi2022horizon,akiyama2022first}.}
\label{boundstab}
\begin{tabular}{lc}
\hline\hline
\textbf{Parameter} & Bounds  \\
\hline
\quad  $\alpha = \beta = -1$  & \, $0.800 \lesssim Q \lesssim 0.945$ \\
\quad  $\alpha = \beta = -2$     &  \, $0.805 \lesssim Q \lesssim 0.955$ \\
\quad  $\alpha = \beta = -3$   & \, $0.810 \lesssim Q \lesssim 0.985$  \\
\quad $\alpha = \beta = -4$  & \, $0.820 \lesssim Q$ \\
\hline\hline
\end{tabular}
\end{table}


\section{Thermal aspects}

To complement the previous analysis, we now turn our attention to the thermodynamic aspects of the nonlinear electrodynamics within the framework of $f(R,T)$ gravity. Specifically, we examine the behavior of the Hawking temperature, entropy, and heat capacity. All thermodynamic quantities will be expressed as functions of the event horizon radius $r_{h}$, as it is a standard procedure in the literature. In particular, special emphasis will be placed on the Hawking temperature, which will also be studied as a function of the black hole mass $M$ to assess the possible existence of a remnant mass.

\subsection{Hawking temperature}

In this subsection, we shall be devoted to investigate the Hawking temperature. Using the surface gravity procedure, it reads
\ie
\begin{split}
\label{htemp}
T_{H} & =  \frac{1}{4 \pi} \frac{1}{\sqrt{\mathrm{A}(r)\mathrm{B}(r)}} \left. \frac{\mathrm{d}}{\mathrm{d}r} \Big[\mathrm{A}(r) \Big]\right|_{r = r_{h}} \\
& = \frac{M}{2 \pi  r_{h}^2} -\frac{Q^2}{2 \pi  r_{h}^3} -\frac{3 \alpha  Q^4}{20 \pi  r_{h}^7} + \frac{3 \alpha  \beta  Q^4}{10 \pi  r_{h}^7}.
\end{split}
\fe
So, the next step is showing the Hawking temperature as a function of the event horizon only. To do so, we consider the solution of $M$ coming from $f(r) = 0$. In this case, it reads
\ie
\label{massadds}
M = \,  \frac{r_{h}}{2} +\frac{Q^2}{2 r_{h}} +\frac{\alpha  Q^4}{20 r_{h}^5} -\frac{\alpha  \beta  Q^4}{10 r_{h}^5}.
\fe
Therefore, after substituting Eq. (\ref{massadds}) in Eq. (\ref{htemp}), we obtain
\ie
\label{ronly}
T_{H} = \frac{1}{4 \pi  r_{h}} -\frac{Q^2}{4 \pi  r_{h}^3} + \frac{\alpha  \beta  Q^4}{4 \pi  r_{h}^7}-\frac{\alpha  Q^4}{8 \pi  r_{h}^7}.
\fe

\begin{figure}
    \centering
     \includegraphics[scale=0.51]{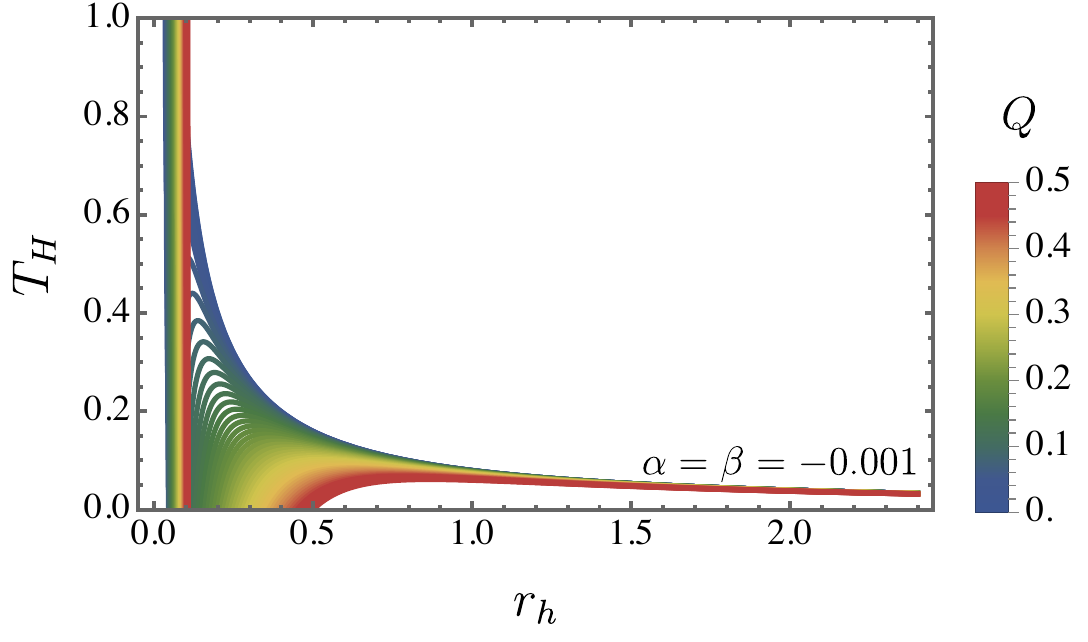}
    \caption{The Hawking temperature is plotted as a function of the event horizon radius $r_{h}$ for various values of the charge $Q$, while keeping the parameters fixed at $\alpha = \beta = -0.001$.}
    \label{hawkingtemperature}
\end{figure}

In Fig. \ref{hawkingtemperature}, the behavior of the Hawking temperature as a function of the event horizon radius $r_{h}$ is depicted for several values of the charge $Q$, assuming fixed values $\alpha = \beta = -0.001$. Furthermore, it is evident that increasing the charge $Q$ leads to a reduction in the Hawking temperature for this specific configuration.

In addition, it would be important to investigate such a thermal quantity as a function of mass $M$. To do so, we have to use the expression of the event horizon present in Eq. (\ref{eventhhh}) and put it in the Hawking temperature expression in Eq. (\ref{ronly}). After that, and considering up to the first order of $\alpha$ and $\beta$, it leads to
\ie
\begin{split}
\label{massssss}
T_{H} \approx  &  \, \, \frac{M \sqrt{M^2-Q^2}+M^2-Q^2}{2 \pi  \left(\sqrt{M^2-Q^2}+M\right)^3} \\
& -\frac{\alpha  \left(-8 \beta  M^2 Q^4+4 M^2 Q^4-8 \beta  M Q^4 \sqrt{M^2-Q^2}+4 M Q^4 \sqrt{M^2-Q^2}+6 \beta  Q^6-3 Q^6\right)}{40 \left[\pi  \left(\sqrt{M^2-Q^2}+M\right)^7 \left(M \sqrt{M^2-Q^2}+M^2-Q^2\right)\right]}.
\end{split}
\fe

\begin{figure}
    \centering
     \includegraphics[scale=0.51]{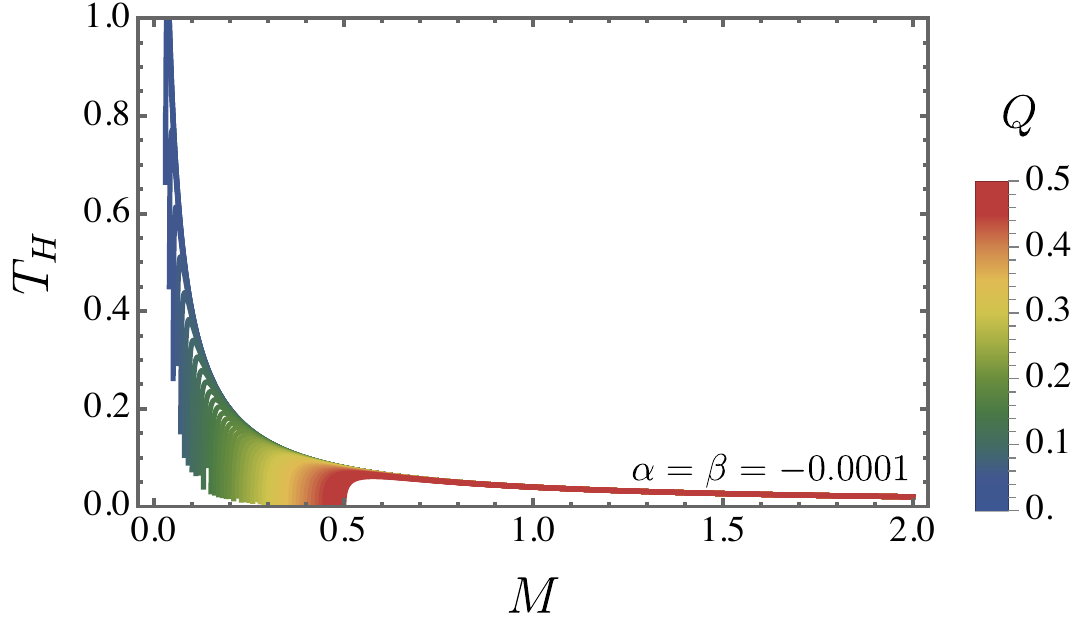}
    \caption{The Hawking temperature is shown as a function of the mass $M$ for different values of the charge $Q$, with the parameters held fixed at $\alpha = \beta = -0.0001$.}
    \label{hawkingtemperaturemass}
\end{figure}

From Fig. \ref{hawkingtemperaturemass}, the existence of remnant masses becomes apparent, as the Hawking temperature approaches zero while the mass remains finite, i.e., $T_{H} \to 0$ as $M \to M_{\text{rem}} \neq 0$. To determine an explicit expression for it, we consider Eq. (\ref{massssss}) in the regime of small $Q$, which yields the following approximation:
\ie
T_{H} \approx \frac{1}{8 \pi  M} -\frac{Q^4}{128 \pi  M^5} + \frac{\alpha  \beta  Q^4}{640 \pi  M^7}-\frac{\alpha  Q^4}{1280 \pi  M^7}.
\fe
Therefore, we impose the condition $T_H \to 0$, which leads to the following result for the mass:
\ie
\begin{split}
M_{rem} & =  \frac{\sqrt{\frac{5 \sqrt[3]{3} Q^4+\sqrt[3]{5} \left(9 \alpha  (1-2 \beta ) Q^4+\sqrt{3} \sqrt{Q^8 \left(27 \alpha ^2 (1-2 \beta )^2-25 Q^4\right)}\right)^{2/3}}{\sqrt[3]{9 \alpha  (1-2 \beta ) Q^4+\sqrt{3} \sqrt{Q^8 \left(27 \alpha ^2 (1-2 \beta )^2-25 Q^4\right)}}}}}{2 \sqrt[3]{15}} \\
& \approx \, \frac{Q}{2} + \frac{\alpha  (1-2 \beta )}{20 Q}. 
\end{split}
\fe

In other words, the above expression indicates that the black hole does not undergo complete evaporation; instead, it leaves behind a nonzero remnant mass, $M_{\text{rem}} \neq 0$. While a detailed analysis of the evaporation process and particle creation in gravitational backgrounds is indeed of significant interest, it falls outside the scope of the present work. However, these aspects will be explored in a forthcoming study, as briefly discussed in the concluding section of this paper.

\subsection{Entropy}

\begin{figure}
    \centering
     \includegraphics[scale=0.51]{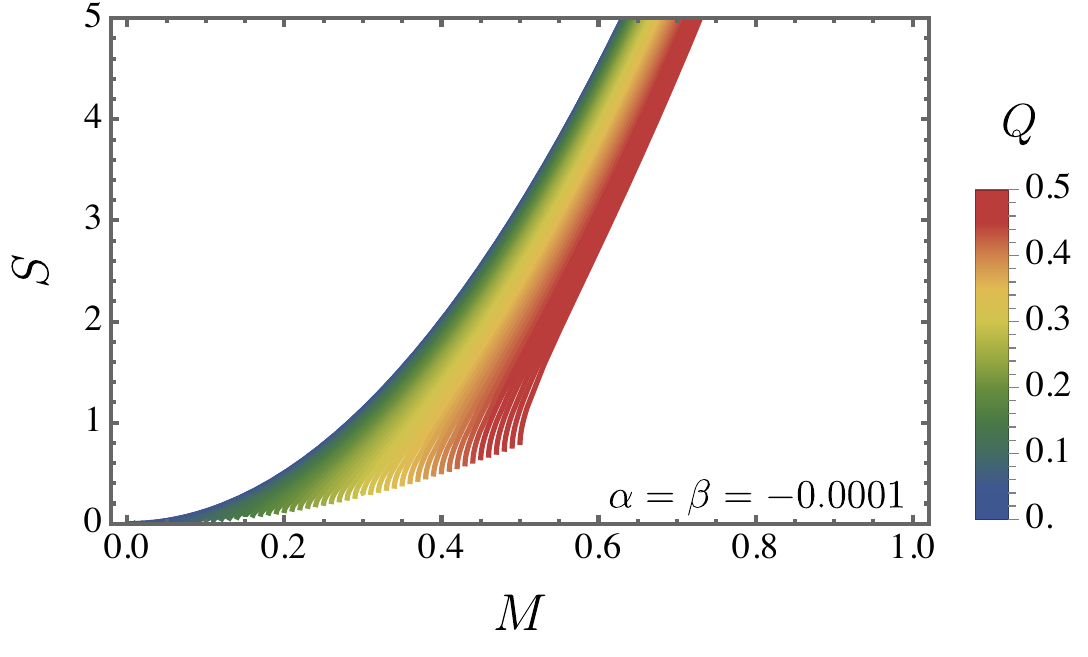}
    \caption{The entropy $S$ is shown as a function of mass $M$ for different values of the charge $Q$, with the parameters held fixed at $\alpha = \beta = -0.0001$.}
    \label{entropy}
\end{figure}

As is well established in the literature, the Bekenstein--Hawking entropy is given by the expression:
\ie
S = \pi r_{h}^{2}.
\fe
In Fig. \ref{entropy}, the entropy is plotted as a function of the horizon radius $r_{h}$ for various values of the charge $Q$, with the parameters fixed at $\alpha = \beta = -0.0001$. As observed for the Hawking temperature, increasing $Q$ results in a reduction of the entropy $S$.

\subsection{Heat capacity}

\begin{figure}
    \centering
     \includegraphics[scale=0.53]{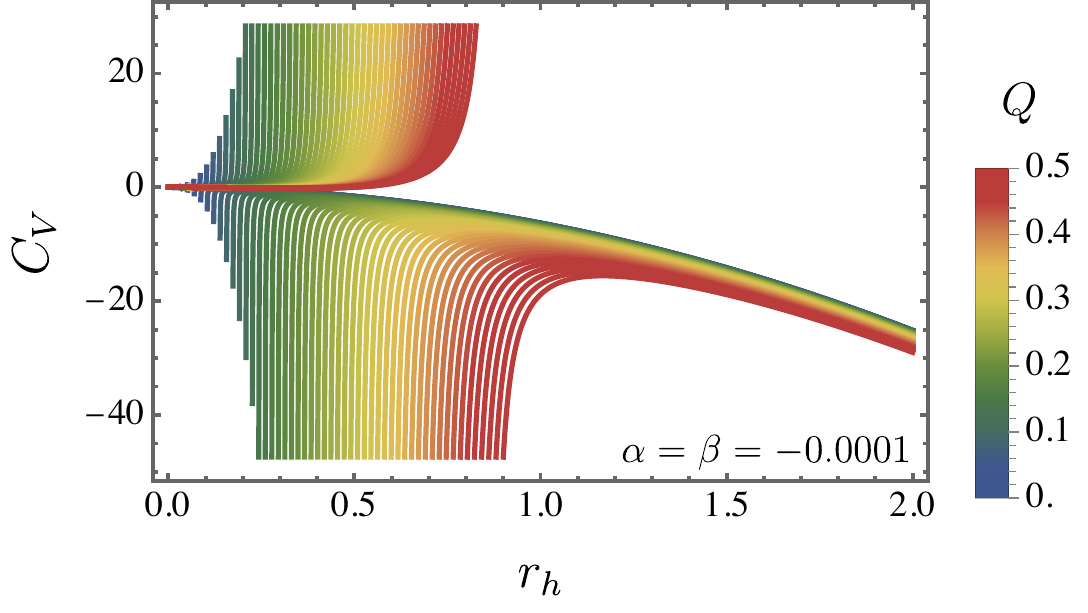}
    \caption{The heat capacity $C_{V}$ is shown as a function of the event horizon $r_{h}$ for different values of the charge $Q$, with the parameters held fixed at $\alpha = \beta = -0.0001$.}
    \label{heatcapp}
\end{figure}

In this subsection, we conclude our analysis by examining the behavior of the heat capacity
\ie
C_{V} = T_H \left( \frac{\partial S}{\partial T_H} \right) = \frac{2 \pi  r_{h}^2 \left[Q^4 (\alpha -2 \alpha  \beta )+2 Q^2 r_{h}^4-2 r_{h}^6\right]}{7 \alpha  (2 \beta -1) Q^4-6 Q^2 r_{h}^4+2 r_{h}^6}.
\fe
In Fig. \ref{heatcapp}, the heat capacity $C_{V}$ is displayed as a function of the event horizon radius $r_{h}$ for different values of the charge $Q$, with the parameters fixed at $\alpha = \beta = -0.0001$. This plot highlights the occurrence of phase transitions, as well as the regions where $C_{V}$ assumes positive or negative values.

\section{Quasinormal modes: bosonic case}

After a black hole undergoes a disturbance, its return to equilibrium is marked by damped gravitational oscillations—referred to as quasinormal modes—that reflect the spacetime geometry rather than the nature of the initial perturbation \cite{Konoplya:2013rxa,Konoplya:2019hlu,Kokkotas:2010zd,Konoplya:2007zx,Konoplya:2011qq}. These modes characterize how the black hole dissipates energy and are considered signatures of its fundamental structure.

Unlike conventional normal modes that describe undamped oscillations in isolated systems, quasinormal modes emerge in scenarios where the system radiates energy, typically through gravitational waves. Their mathematical description involves solving the wave equation under specific boundary conditions imposed by the black hole spacetime, with the solutions corresponding to the complex poles of the associated Green's function.

Extracting the quasinormal spectrum is generally a difficult task, as the equations are rarely solvable in closed form. Consequently, a variety of numerical and semi--analytical techniques have been devised to estimate the frequencies in the context of a given background metric $\mathrm{g}_{\mu\nu}$ \cite{araujo2023analysis,heidari2023gravitational,araujo2024gravitational,MoraisGraca:2017hrf}.

\subsection{Scalar perturbations}

A widely employed approach for estimating quasinormal frequencies is the Wentzel--Kramers--Brillouin (WKB) approximation, originally formulated by Will and Iyer \cite{iyer1987black,iyer1987black1} and subsequently extended to higher orders by Konoplya \cite{konoplya2003quasinormal}. In this study, attention is directed toward scalar perturbations, where the Klein--Gordon equation is solved in the presence of a non--flat gravitational background to determine the corresponding complex frequencies
\ie
\frac{1}{\sqrt{-\mathrm{g}}}\partial_{\mu}(\mathrm{g}^{\mu\nu}\sqrt{-\mathrm{g}}\partial_{\nu}\Tilde{\Phi}) = 0.\label{kkleeinnss}
\fe
Although investigating the influence of backreaction presents a compelling direction, such effects are not addressed in the present work. The analysis is instead restricted to treating the scalar field as a test perturbation on a fixed gravitational background. Within this approximation, Eq. (\ref{kkleeinnss}) emerges as the governing equation for the scalar dynamics under consideration
\ie
\label{jekjsk}
\begin{split}
-& \frac{1}{f(r)} \frac{\partial^{2} \Tilde{\Phi}}{\partial t^{2}} + \frac{1}{r^{2}} \left[  \frac{\partial}{\partial r} \left(  f(r) \, r^{2}  \frac{\partial \Tilde{\Phi}}{\partial r}  \right)  \right] \\  + & \frac{1}{r^{2} \sin \theta}  \left[  \frac{\partial }{\partial \theta} \left( \sin \theta \frac{\partial}{\partial \theta} \Tilde{\Phi}   \right)        \right] 
 +  \frac{1}{r^{2} \sin^{2}}  \frac{\partial^{2} \Tilde{\Phi}}{\partial \varphi^{2}} = 0.
\end{split}
\fe

Given the spherical symmetry of the background geometry, the scalar field is decomposed using a separation of variables approach. The metric determinant takes the form $\sqrt{-\mathrm{g}} = r^{2}\sin\theta$, which facilitates the expansion of the scalar field into angular and radial components
\ie
\label{deffevvd}
\Tilde{\Phi}(t, r, \theta, \varphi) = \sum_{l=0}^{\infty} \sum_{m=-l}^{l}  \mathrm{Y}_{lm}(\theta, \varphi) \frac{\Tilde{\Psi}(t,r)}{r}.
\fe
Notice that, utilizing this decomposition, in which $\mathrm{Y}_{lm}(\theta, \varphi)$ represents the set of spherical harmonic functions, one can isolate the radial part of Eq. (\ref{jekjsk}) and rewrite it in a reduced form that governs the radial dynamics of the scalar perturbation
\ie
\frac{\partial^{2}\Tilde{\Psi}(t,r)}{\partial t^{2}}  + \frac{f(r)}{r} \left\{ \frac{\partial }{\partial r}  \left[ f(r) r^{2} \frac{\partial}{\partial r}  \left( \frac{\Tilde{\Psi}(t,r)}{r} \right)   \right]     \right\} - f(r) \frac{\ell(\ell + 1)}{r^{2}}\Tilde{\Psi}(t,r) = 0 .
\fe

Here, the angular dependence is naturally shown through the spherical harmonics $\mathrm{Y}_{lm}(\theta, \varphi)$. Inserting the decomposed form of the scalar field from Eq. (\ref{deffevvd}) into the original equation (\ref{kkleeinnss}) results it into a Schrödinger--type differential equation. This reformulation brings out its wave-like structure, making it suitable for investigating the quasinormal spectrum. Thereby, the equation becomes
\ie
-\frac{\partial^{2} \Tilde{\Psi}}{\partial t^{2}}+\frac{\partial^{2} \Tilde{\Psi}}{\partial r^{*2}} + V_{\text{S}}(r)\Tilde{\Psi} = 0.
\fe

It is important to highlight a remarkable aspect at this stage: the effective potential $V_{\text{S}}(r)$—often identified as the Regge–Wheeler potential—encodes essential geometric features of the black hole spacetime. For accomplishing the analysis, the radial coordinate is reparametrized using the so--called tortoise coordinate $r^{*}$, which maps the domain smoothly from the horizon to spatial infinity, satisfying the relation $\mathrm{d}r^{*} = \frac{1}{\sqrt{f(r)^2}} \mathrm{d}r$. In other words, it explicitly reads
\ie
\begin{split}
& r^{*} =  \\
& r +  \frac{r_{10}^{10} \ln (r - r_{10})}{(r_{10} - r_{2}) (r_{10}-r_{3}) (r_{10} - r_{4}) (r_{10}-r_{5}) (r_{10}-r_{6}) (r_{10}-r_{7}) (r_{10}-r_{8}) (r_{10}-r_{9}) (r_{10}-r_{h})} \\
& -\frac{r_{2}^{10} \ln (r-r_{2})}{(r_{10}-r_{2}) (r_{2}-r_{3}) (r_{2}-r_{4}) (r_{2}-r_{5}) (r_{2}-r_{6}) (r_{2}-r_{7}) (r_{2}-r_{8}) (r_{2}-r_{9}) (r_{2}-r_{h})} \\
& +\frac{r_{3}^{10} \ln (r-r_{3})}{(r_{10}-r_{3}) (r_{2}-r_{3}) (r_{3}-r_{4}) (r_{3}-r_{5}) (r_{3}-r_{6}) (r_{3}-r_{7}) (r_{3}-r_{8}) (r_{3}-r_{9}) (r_{3}-r_{h})} \\
& -\frac{r_{4}^{10} \ln (r-r_{4})}{(r_{10}-r_{4}) (r_{2}-r_{4}) (r_{3}-r_{4}) (r_{4}-r_{5}) (r_{4}-r_{6}) (r_{4}-r_{7}) (r_{4}-r_{8}) (r_{4}-r_{9}) (r_{4}-r_{h})} \\
& +\frac{r_{5}^{10} \ln (r-r_{5})}{(r_{10}-r_{5}) (r_{2}-r_{5}) (r_{3}-r_{5}) (r_{4}-r_{5}) (r_{5}-r_{6}) (r_{5}-r_{7}) (r_{5}-r_{8}) (r_{5}-r_{9}) (r_{5}-r_{h})}\\
& -\frac{r_{6}^{10} \ln (r-r_{6})}{(r_{10}-r_{6}) (r_{2}-r_{6}) (r_{3}-r_{6}) (r_{4}-r_{6}) (r_{5}-r_{6}) (r_{6}-r_{7}) (r_{6}-r_{8}) (r_{6}-r_{9}) (r_{6}-r_{h})} \\
& +\frac{r_{7}^{10} \ln (r-r_{7})}{(r_{10}-r_{7}) (r_{2}-r_{7}) (r_{3}-r_{7}) (r_{4}-r_{7}) (r_{5}-r_{7}) (r_{6}-r_{7}) (r_{7}-r_{8}) (r_{7}-r_{9}) (r_{7}-r_{h})} \\
& -\frac{r_{8}^{10} \ln (r-r_{8})}{(r_{10}-r_{8}) (r_{2}-r_{8}) (r_{3}-r_{8}) (r_{4}-r_{8}) (r_{5}-r_{8}) (r_{6}-r_{8}) (r_{7}-r_{8}) (r_{8}-r_{9}) (r_{8}-r_{h})} \\
& +\frac{r_{9}^{10} \ln (r-r_{9})}{(r_{10}-r_{9}) (r_{2}-r_{9}) (r_{3}-r_{9}) (r_{4}-r_{9}) (r_{5}-r_{9}) (r_{6}-r_{9}) (r_{7}-r_{9}) (r_{8}-r_{9}) (r_{9}-r_{h})} \\
& -\frac{r_{h}^{10} \ln (r-r_{h})}{(r_{10}-r_{h}) (r_{2}-r_{h}) (r_{3}-r_{h}) (r_{4}-r_{h}) (r_{5}-r_{h}) (r_{6}-r_{h}) (r_{7}-r_{h}) (r_{8}-r_{h}) (r_{9}-r_{h})},
\end{split}
\fe
where $r_2$ through $r_{10}$ denote the remaining solutions of $f(r) = 0$, excluding $r_h$. It should be emphasized that, for certain choices of the parameters $Q$, $M$, $\alpha$, and $\beta$, some of the horizons may no longer correspond to physically meaningful solutions. After carrying out the necessary algebraic procedures, one arrives at the following expression for the effective potential:
\ie
\begin{split}
V_{\text{S}}(r,\alpha,\beta,Q) = f(r) \left(\frac{\ell (\ell+1)}{r^2} + \frac{2 M}{r^3} -\frac{2 Q^2}{r^4} + \frac{3 \alpha  (2 \beta -1) Q^4}{5 r^8} \right)
\end{split}.
\fe

\begin{figure}
    \centering
    \includegraphics[scale=0.55]{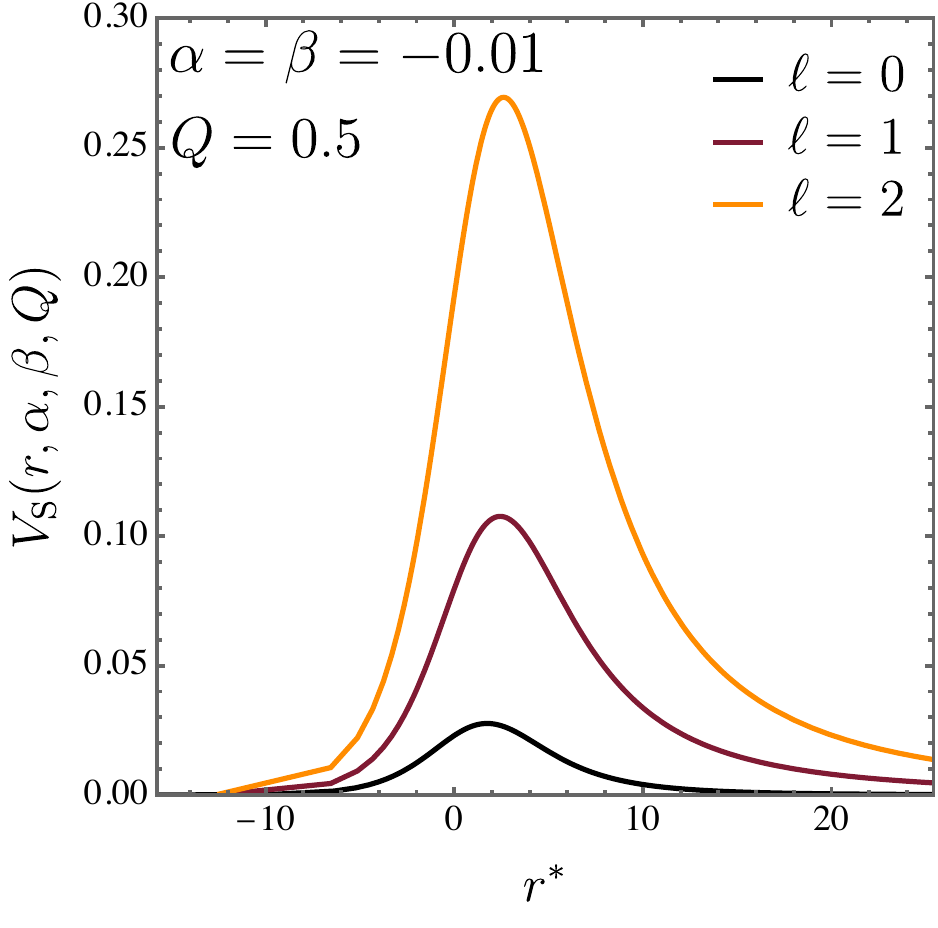}
    \caption{The effective potential $V_{\text{S}}(r,\alpha,\beta,Q)$ for the scalar perturbations is depicted as a function of the tortoise coordinate $r^{*}$, specifically considering different values of $\ell$.}
    \label{veffscalarfiled}
\end{figure}

Notice that the first three terms inside the parentheses correspond to the scalar perturbations in the Reissner--Nordström case. The last term, as expected, arises from the presence of the modified electrodynamics. In the limit $Q \to 0$, the effective potential reduces to that of the Schwarzschild case. Furthermoe, Fig. \ref{veffscalarfiled} illustrates the effective potential $V_{\text{S}}(r,\alpha,\beta,Q)$ as a function of the tortoise coordinate $r^{*}$ for different values of $\ell$.

\begin{table}[!h]
\begin{center}
\caption{\label{qnmtac1} The table displays the quasinormal modes, regarding scalar perturbations for $\ell = 1$ as a function of the parameters $\alpha$, $\beta$ and $Q$.}
\begin{tabular}{c| c | c | c} 
 \hline\hline\hline 
 $\alpha = \beta$ \,\,  $Q$  & $\omega_{0}$ & $\omega_{1}$ & $\omega_{2}$  \\ [0.2ex] 
 \hline 
 \,  -0.01, \,  0.99  & 0.375868 - 0.0904883$i$ & 0.35676 - 0.271701$i$ & 0.332827 - 0.43795$i$ \\
 
\,  -0.02, \,  0.99  & 0.374667 - 0.0908126$i$ & 0.345840 - 0.280433$i$  & 0.292635 - 0.498499$i$  \\
 
 \, -0.03, \,  0.99  & 0.375443 - 0.0906598$i$  & 0.352793 - 0.275052$i$ &  0.316790 - 0.460810$i$   \\
 
\, -0.04, \,  0.99  & 0.374445 - 0.0909383$i$ & 0.34384 - 0.282375$i$ & 0.285973 - 0.510866$i$  \\
 
\, -0.05, \,  0.99  &  0.375374 - 0.0907438$i$  & 0.352096 - 0.275831$i$ & 0.313690 - 0.465575$i$   \\
   [0.2ex] 
 \hline\hline\hline 
  $\alpha = \beta$ \,\,  $Q$  & $\omega_{0}$ & $\omega_{1}$ & $\omega_{2}$  \\ [0.2ex] 
 \hline 
 -0.01, \,   0.6  & 0.313521 - 0.0992213$i$ &  0.287589 - 0.309522$i$  &  0.256853 - 0.543749$i$   \\
 
 -0.01, \,  0.7  & 0.322771 - 0.0994142$i$ & 0.298097 - 0.309376$i$  &  0.268564 - 0.541807$i$  \\

 -0.01, \,   0.8  & 0.335212 - 0.0991068$i$ & 0.312158 - 0.307364$i$ & 0.283644 - 0.536141$i$  \\
 
 -0.01, \,  0.9  & 0.352608 - 0.0972234$i$ & 0.331150 - 0.299522$i$ & 0.301235 - 0.518438$i$  \\
 
 -0.01, \,  0.99  & 0.375868 - 0.0904883$i$ & 0.356760 - 0.271701$i$ & 0.332827 - 0.437950$i$ \\
   [0.2ex] 
 \hline \hline \hline 
\end{tabular}
\end{center}
\end{table}

\begin{table}[!h]
\begin{center}
\caption{\label{qnmtac2} The table displays the quasinormal modes, regarding scalar perturbations for $\ell = 2$ as a function of the parameters $\alpha$, $\beta$ and $Q$.}
\begin{tabular}{c| c | c | c} 
 \hline\hline\hline 
 $\alpha = \beta$ \,\,  $Q$  & $\omega_{0}$ & $\omega_{1}$ & $\omega_{2}$  \\ [0.2ex] 
 \hline 
 \,  -0.01, \,  0.99  & 0.620993 - 0.0902883$i$ & 0.603809 - 0.273446$i$  & 0.569605 - 0.465296$i$ \\
 
\,  -0.02, \,  0.99  & 0.621006 - 0.0903147$i$ & 0.604026 - 0.273443$i$  & 0.570526 - 0.464729$i$  \\
 
 \, -0.03, \,  0.99  & 0.621034 - 0.090340$i$  & 0.604408 - 0.273368$i$ &  0.572273 - 0.463492$i$  \\
 
\, -0.04, \,  0.99  & 0.621057 - 0.0903673$i$ & 0.604739 - 0.273323$i$ & 0.573755 - 0.462502$i$ \\
 
\, -0.05, \,  0.99  &  0.621049 - 0.0904001$i$ & 0.604713 - 0.273441$i$ & 0.573440 - 0.462957$i$  \\
   [0.2ex] 
 \hline\hline\hline 
  $\alpha = \beta$ \,\,  $Q$  & $\omega_{0}$ & $\omega_{1}$ & $\omega_{2}$  \\ [0.2ex] 
 \hline 
 -0.01, \,   0.6  & 0.517386 - 0.0983321$i$ &  0.499370 - 0.299793$i$  &  0.468890 - 0.513974$i$    \\
 
 -0.01, \,  0.7  & 0.532560 - 0.0985752$i$ & 0.515452 - 0.300229$i$  &  0.486450 - 0.513796$i$   \\

 -0.01, \,   0.8  & 0.553050 - 0.0983468$i$ & 0.537189 - 0.299078$i$ & 0.510112 - 0.510476$i$  \\
 
 -0.01, \,  0.9  & 0.581950 - 0.0966402$i$ & 0.567515 - 0.293110$i$  & 0.542159 - 0.498019$i$  \\
 
 -0.01, \,  0.99  & 0.620993 - 0.0902883$i$ & 0.603809 - 0.273446$i$  & 0.569605 - 0.465296$i$  \\
   [0.2ex] 
 \hline \hline \hline 
\end{tabular}
\end{center}
\end{table}

To proceed with the analysis, one rewrites the wave function in a form that isolates its temporal oscillations: $\Tilde{\Psi}(t, r) = e^{-i\omega t} \psi(r)$, with $\omega$ denoting the mode frequency. This choice effectively decouples the time variable, converting the original equation into a static one that governs the spatial behavior. As a result, the problem reduces to solving a time--independent differential equation of the form:
\ie
\frac{\partial^{2} \psi}{\partial r^{*2}} - \left[  \omega^{2} - V_{\text{S}}(r,\alpha,\beta,Q)\right]\psi = 0.\label{sserffs}
\fe

A proper treatment of Eq. (\ref{sserffs}) requires a precise specification of boundary conditions to guarantee meaningful solutions. In the context considered here, physical validity near the event horizon demands that the wave function behaves as a purely ingoing mode at that boundary
\[
    \psi^{in}(r^{*}) \sim 
\begin{cases}
    \Tilde{\beta}_{\ell}(\omega) e^{-i\omega r^{*}} & ( r^{*}\rightarrow - \infty)\\
    \Tilde{\alpha}^{(1)}_{\ell}(\omega) e^{-i\omega r^{*}} + \Tilde{\alpha}^{(2)}_{\ell}(\omega) e^{+i\omega r^{*}} & (r^{*}\rightarrow + \infty).
\end{cases}
\]

The identification of quasinormal frequencies in this framework relies on the behavior of specific complex functions: $\Tilde{\beta}_\ell(\omega)$, $\Tilde{\alpha}^{(1)}\ell(\omega)$, and $\Tilde{\alpha}^{(2)}\ell(\omega)$. These coefficients characterizes the wave dynamics under black hole perturbations. The allowed frequencies $\omega_{n\ell}$ correspond to the roots of $\Tilde{\alpha}^{(1)}_\ell(\omega)$, ensuring that the wave exhibits purely ingoing motion at the event horizon and purely outgoing propagation at spatial infinity. Here, $n$ indicates the overtone number, while $\ell$ denotes the multipole number. Also, in order to extract the spectrum, one must solve Eq. (\ref{sserffs}) as an eigenvalue problem subject to physically motivated boundary conditions. Due to the challenging form of the potential, we employ the WKB approximation—a semiclassical method that offers a relatively simple manner to estimate the quasinormal frequencies.

The WKB approach, first introduced in the context of black hole perturbations by Schutz and Will \cite{schutz1985black}, has since undergone significant refinement. Konoplya later extended the method to incorporate higher--order corrections, enhancing its precision and applicability \cite{konoplya2003quasinormal, konoplya2004quasinormal}. This semiclassical technique is particularly effective when the effective potential exhibits a peak structure and levels off at spatial boundaries, i.e., as $r^{*} \to \pm \infty$.

To determine the quasinormal spectrum, the solution is constructed through a series expansion near the peak of the effective potential—often referred to as the classical turning point. This approach yields accurate approximations for the complex frequencies associated with black hole oscillations. Konoplya’s refinement of the method leads to the following expression for the quasinormal frequencies:
\ie
\frac{i(\omega^{2}_{n}-V_{0})}{\sqrt{-2 V^{''}_{0}}} - \sum^{6}_{j=2} \Lambda_{j} = n + \frac{1}{2}.
\fe
In this formulation, an essential quantity is $V''_{0}$, the second derivative of the effective potential evaluated at its peak position $r_{0}$. The correction terms, denoted by $\Lambda_{j}$, depend on both the potential and its higher--order derivatives at $r_{0}$, is fundamental to  obtaining the accuracy in calculating the quasinormal oscillatory behavior.

As it is straightforward to verify, in the absence of charge ($Q \to 0$), the effective potential describing scalar fluctuations reverts to that of a Schwarzschild black hole. Once this limiting behavior is clarified, one can proceed with evaluating the quasinormal mode spectrum. The potential $V_{\text{S}}(r,\alpha,\beta,Q)$, plotted in Fig. \ref{veeffds} as a function of the tortoise coordinate $r^*$, reveals a sine--barrier-like structure, supporting the use of the WKB formalism for mode computation.

Tabs. \ref{qnmtac1} and \ref{qnmtac2} compile the complex quasinormal frequencies for multipole numbers $\ell = 1$ and $\ell = 2$, considering various parameter choices. When the charge is fixed at $Q = 0.99$, reducing $\alpha = \beta$ results in alternating behavior between weaker and stronger damping across all modes $\omega$ for both $\ell = 1$ and $\ell = 2$. A similar pattern emerges as $Q$ increases with $\alpha = \beta$ held constant at $-0.01$, producing an ``irregular'' sequence of less and more damped frequencies, with the exception of $\omega_{2}$, which does not follow this trend. In other words, this specific mode exhibits reduced damping for the latter configuration (when $Q$ runs for fixed values of $\alpha$ and $\beta$).


\subsection{Vector perturbations}

The analysis of electromagnetic fluctuations is carried out by reformulating the problem in a locally flat frame, constructed through a set of tetrad vectors. This approach, inspired by the procedures outlined in \cite{chandrasekhar1998mathematical, Bouhmadi-Lopez:2020oia, Gogoi:2023kjt}, replaces direct manipulation of the curved metric $\mathrm{g}_{\mu\nu}$ with an equivalent orthonormal basis $\mathrm{e}_{\mu}^{a}$. These vectors are chosen to reproduce the spacetime geometry through the relation imposed by the tetrad formalism:
\ie
\begin{split}
& \mathrm{e}^{a}_\mu \mathrm{e}^\mu_{b} = \delta^{a}_{b}, \, \, \, \,
\mathrm{e}^{a}_\mu \mathrm{e}^\nu_{a} = \delta^{\nu}_{\mu}, \, \, \, \,\\
& \mathrm{e}^{a}_\mu = \mathrm{g}_{\mu\nu} \eta^{a b} \mathrm{e}^\nu_{b}, \, \, \, \,
\mathrm{g}_{\mu\nu} = \eta_{a b}\mathrm{e}^{a}_\mu \mathrm{e}^{b}_\nu = \mathrm{e}_{a\mu} \mathrm{e}^{a}_\nu.
\end{split}
\fe

When reformulating electromagnetic disturbances through the tetrad approach, the antisymmetric structure of the field strength tensor imposes the Bianchi identity, written as $\mathcal{F}_{[ab|c]} = 0$. Enforcing this constraint yields the following expression:
\begin{align}
\left( r \sqrt{\mathrm{A}(r)}\, \mathcal{F}_{t \phi}\right)_{,r} + r \sqrt{\mathrm{B}(r)}\,
\mathcal{F}_{\phi r, t} &=0,  \label{edem1} \\
\left( r \sqrt{\mathrm{A}(r)}\, \mathcal{F}_{ t \phi}\sin\theta\right)_{,\theta} + r^2
\sin\theta\, \mathcal{F}_{\phi r, t} &=0.  \label{dfccdss}
\end{align}

As a direct consequence, the conservation law takes the form:
\ie
\eta^{b c}\! \left( \mathcal{F}_{a b} \right)_{|c} =0.
\fe

It is worth noting that, when expressed in spherical polar coordinates, this equation can be reformulated as:
\ie  \label{edem3}
\left( r \sqrt{\mathrm{A}(r)}\, \mathcal{F}_{\phi r}\right)_{,r} + \sqrt{\mathrm{A}(r) \mathrm{B}(r)}%
\, \mathcal{F}_{\phi \theta,\theta} + r \sqrt{\mathrm{B}(r)}\, \mathcal{F}_{t \phi, t} = 0.
\fe

Within this framework, intrinsic differentiation along the tetrad basis is denoted by a vertical bar, while a comma represents a standard directional derivative. By applying a temporal derivative to Eq.\eqref{edem3} and combining it with the identities provided in Eqs.\eqref{edem1} and \eqref{dfccdss}, one ultimately derives the expression given below:
\ie  \label{edem4}
\begin{split}
& \left[ \sqrt{\mathrm{A}(r) \mathrm{B}(r)^{-1}} \left( r \sqrt{\mathrm{A}(r)}\, \mathcal{F}
\right)_{,r} \right]_{,r} \\
& + \dfrac{\mathrm{A}(r) \sqrt{\mathrm{B}(r)}}{r} \left( \dfrac{%
\mathcal{F}_{,\theta}}{\sin\theta} \right)_{,\theta}\!\! \sin\theta - r 
\sqrt{\mathrm{B}(r)}\, \mathcal{F}_{,tt} = 0.
\end{split}
\fe

Let $F = \mathcal{F}_{t\phi} \sin\theta$ serve as the starting definition. Implementing a Fourier transform via $\partial_t \mapsto -i\omega$ and redefining the field as $F(r,\theta) = F(r), \Tilde{Y}{,\theta} / \sin\theta$, where $\Tilde{Y}(\theta)$ denotes a Gegenbauer function \cite{g3,g2,g1,g6,g5}, one reformulates Eq.~\eqref{edem4} into the following structure:
\ie  
\begin{split}
\label{edem5}
& \left[ \sqrt{\mathrm{A}(r) \mathrm{B}(r)^{-1}} \left( r \sqrt{\mathrm{A}(r)}\, F
\right)_{,r} \right]_{,r} \\
& + \omega^2 r \sqrt{\mathrm{B}(r)}\, F -
\mathrm{A}(r) \sqrt{\mathrm{B}(r)} r^{-1} \ell (\ell + 1)\, F = 0.
\end{split}
\fe

Now, taking into account the redefinition $\psi^{, v} \equiv r \sqrt{\mathrm{A}(r)} , F$, Eq.~\eqref{edem5} transforms into
\ie
\partial^2_{r_{*}} \psi^{\, v} + \omega^2 \psi^{\, v} = V_{\text{V}}(r) \psi^{\, v},
\fe
where the effective potential associated with the vector--type perturbation takes the form
\ie  
V_{\text{V}}(r,\alpha,\beta,Q) = \mathrm{A}(r) \left(\frac{\ell (\ell+1)}{r^2} \right).
\fe

\begin{figure}
    \centering
    \includegraphics[scale=0.55]{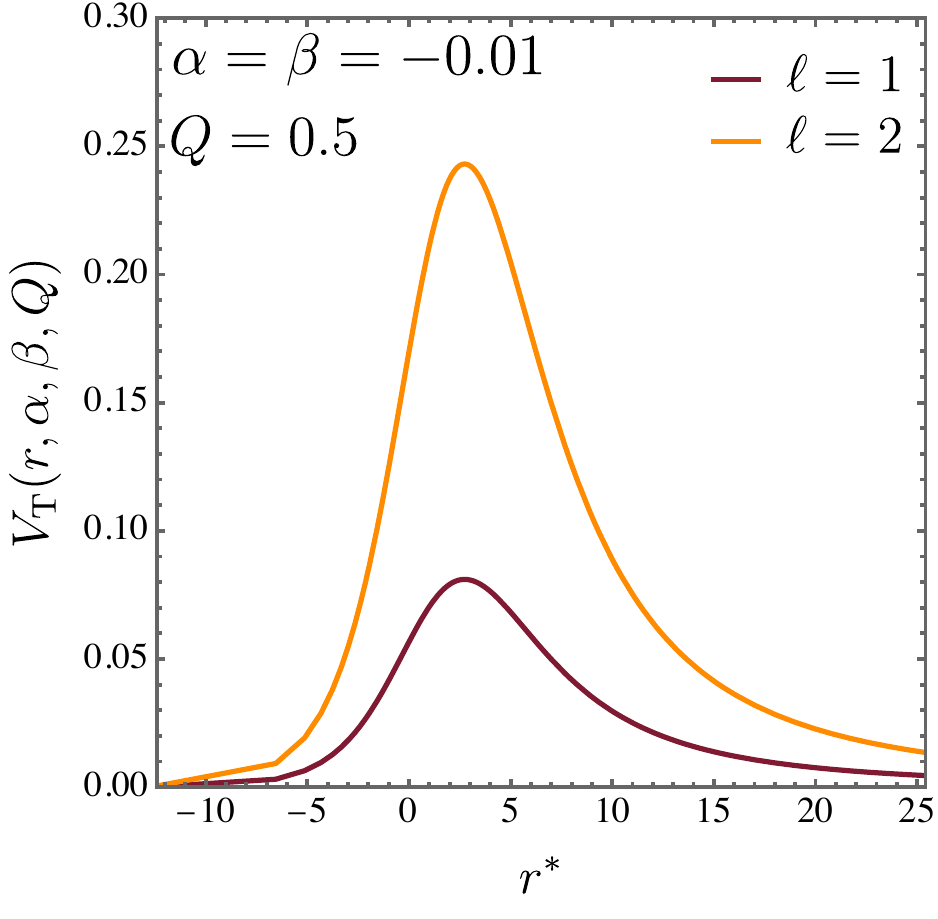}
    \caption{The behavior of the effective potential associated with vector perturbations, $V_{\text{T}}(r,\alpha,\beta,Q)$, is illustrated in terms of the tortoise coordinate $r^*$, with particular attention to variations across different multipole numbers $\ell$.}
    \label{veeffds}
\end{figure}

In the limit where the charge vanishes ($Q \to 0$), the potential linked to vector perturbations simplifies to the Schwarzschild configuration. On the other hand, by approaching $\alpha = \beta \to 0$, one recovers the Reissner--Nordström case, consistent with standard expectations. Fig. \ref{veeffds} displays the behavior of the potential $V_{\text{V}}(r, \alpha, \beta, Q)$ plotted against the tortoise coordinate $r^*$.

Once we have accomplished all these analyzes, the investigation of vector quasinormal frequencies can properly be undertaken. As previously encountered in the scalar analysis, the form of the potential exhibits a sine--like shape, so that it is possible to use the WKB approximation scheme to estimate the quasinormal spectrum, as we did the previous section concerning the scalar perturbations.

Tabs. \ref{qnmtac2vectorial1} and \ref{2qnmtac1vectorial2} present the quasinormal frequencies for vector perturbations with multipole numbers $\ell = 1$ and $\ell = 2$, spanning a range of values for the parameters $Q$, $\alpha$, and $\beta$. When fixing either the electric charge $Q$ or $\alpha$ and $\beta$, the attenuation pattern mirrors the behavior found in the scalar sector: the frequencies alternate between more strongly and more weakly damped modes, depending particularly on the specific choices of $\alpha$ and $\beta$. It should be emphasized that, for all parameter combinations considered in this analysis, the system consistently exhibits stability.

\begin{table}[!h]
\begin{center}
\caption{\label{qnmtac2vectorial1} The table presents the quasinormal mode spectrum for vector--type perturbations with $\ell = 1$, highlighting its dependence on the parameters $\alpha$, $\beta$, and $Q$.}
\begin{tabular}{c| c | c | c} 
 \hline\hline\hline 
 $\alpha = \beta$ \,\,  $Q$  & $\omega_{0}$ & $\omega_{1}$ & $\omega_{2}$  \\ [0.2ex] 
 \hline 
 \,  -0.01, \,  0.99  & 0.332682 - 0.086193$i$ & 0.307812 - 0.261821$i$ & 0.265915 - 0.439619$i$ \\
 
\,  -0.02, \,  0.99  & 0.332658 - 0.0862464$i$ & 0.307624 - 0.262162$i$  & 0.264998 - 0.441408$i$   \\
 
 \, -0.03, \,  0.99  & 0.332965 - 0.0862204$i$  & 0.310462 - 0.260003$i$ &  0.275284 - 0.425505$i$  \\
 
\, -0.04, \,  0.99  & 0.331341 - 0.0866972$i$ & 0.296612 - 0.272385$i$ & 0.231436 - 0.506734$i$ \\
 
\, -0.05, \,  0.99  &  0.333069 - 0.0863111$i$ & 0.311587 - 0.259626$i$ & 0.279357 - 0.420908$i$  \\
   [0.2ex] 
 \hline\hline\hline 
  $\alpha = \beta$ \,\,  $Q$  & $\omega_{0}$ & $\omega_{1}$ & $\omega_{2}$  \\ [0.2ex] 
 \hline 
 -0.01, \,   0.6  & 0.268126 - 0.0946644$i$ &  0.237357 - 0.298635$i$  & 0.201214 - 0.532676$i$   \\
 
 -0.01, \,  0.7  & 0.277264 - 0.0951084$i$  & 0.248126 - 0.299109$i$  &  0.213931 - 0.531056$i$  \\

 -0.01, \,   0.8  & 0.289795 - 0.0951104$i$ & 0.263015 - 0.297680$i$ & 0.23124 - 0.524821$i$   \\
 
 -0.01, \,  0.9  & 0.307902 - 0.0935047$i$ & 0.284487 - 0.289602$i$ & 0.254614 - 0.502195$i$  \\
 
 -0.01, \,  0.99  & 0.332682 - 0.086193$i$ & 0.307812 - 0.261821$i$  & 0.265915 - 0.439619$i$ \\
   [0.2ex] 
 \hline \hline \hline 
\end{tabular}
\end{center}
\end{table}

\begin{table}[!h]
\begin{center}
\caption{\label{2qnmtac1vectorial2} The table shows the quasinormal mode frequencies corresponding to vector perturbations with $\ell = 2$, examined as functions of the parameters $\alpha$, $\beta$, and $Q$.}
\begin{tabular}{c| c | c | c} 
 \hline\hline\hline 
 $\alpha = \beta$ \,\,  $Q$  & $\omega_{0}$ & $\omega_{1}$ & $\omega_{2}$  \\ [0.2ex] 
 \hline 
 \,  -0.01, \,  0.99  & 0.595601 - 0.0886477$i$  & 0.577573 - 0.268751$i$  & 0.540663 - 0.459178$i$ \\
 
\,  -0.02, \,  0.99  & 0.595498 - 0.0886965$i$  & 0.57651 - 0.269363$i$  & 0.535281 - 0.464043$i$   \\
 
 \, -0.03, \,  0.99  & 0.595657 - 0.0887067$i$  & 0.578393 - 0.268600$i$ &  0.544341 - 0.456527$i$  \\
 
\, -0.04, \,  0.99  & 0.595767 - 0.0887268$i$ & 0.579748 - 0.268103$i$ & 0.550971 - 0.451327$i$ \\
 
\, -0.05, \,  0.99  &  0.595750 - 0.0887658$i$ & 0.579654 - 0.268268$i$ & 0.550294 - 0.452106$i$  \\
   [0.2ex] 
 \hline\hline\hline 
  $\alpha = \beta$ \,\,  $Q$  & $\omega_{0}$ & $\omega_{1}$ & $\omega_{2}$  \\ [0.2ex] 
 \hline 
 -0.01, \,   0.6  & 0.490904 - 0.0967198$i$ &  0.471752 - 0.295298$i$  & 0.439375 - 0.507546$i$   \\
 
 -0.01, \,  0.7  & 0.505982 - 0.0970274$i$  & 0.487816 - 0.295912$i$  &  0.457066 - 0.507606$i$  \\

 -0.01, \,   0.8  & 0.526457 - 0.0968786$i$  & 0.509671 - 0.294973$i$ & 0.481088 - 0.504549$i$  \\
 
 -0.01, \,  0.9  & 0.555605 - 0.0952366$i$ & 0.540542 - 0.289114$i$ & 0.514280 - 0.491927$i$  \\
 
 -0.01, \,  0.99  & 0.595601 - 0.0886477$i$ & 0.577573 - 0.268751$i$ & 0.540663 - 0.459178$i$ \\
   [0.2ex] 
 \hline \hline \hline 
\end{tabular}
\end{center}
\end{table}


\subsection{Tensor perturbations}

Rather than starting from a specific fundamental framework, the derivation of the master equations was carried out by assuming that the Klein--Gordon and Maxwell equations remain applicable. It should be stressed, however, that these field equations do not inherently ensure the conservation of stress--energy in extended gravitational models, unless the matter fields are minimally coupled to the background geometry defined by $\mathrm{g}_{\mu\nu}$.

To investigate odd--parity (axial) gravitational fluctuations, one must introduce perturbations not only in the metric but also in the stress--energy distribution. Since the analysis is not grounded in a complete underlying theory, a different route is adopted. Here, the system is explored using the Einstein equations supplemented with an effective source term. This treatment aligns with approaches found in earlier studies \cite{ashtekar2018quantum, ashtekar2018quantum2,baruah2025quasinormal}. From a modeling standpoint, the stress--energy content driving the black hole configuration is represented by
\ie
T_{\mu\nu} = \left(\rho + p_{2}\right)u_{\mu} u_{\nu} + \left(p_{1} - p_{2} \right)x_{\mu} x_{\nu} + p_{2} \,\mathrm{g}_{\mu\nu}.\label{strreessss}
\fe

In this context, the parameter $\rho$ denotes the energy density measured by an observer moving with the fluid. The four--vector $u^{\mu}$ identifies the fluid’s temporal flow, while $x^{\mu}$ defines a spacelike direction orthogonal to both $u^\mu$ and the angular components. The expression in Eq.~\eqref{strreessss} features $p_1$ and $p_2$, representing the pressures along the radial and transverse directions, respectively. Furthermore, the vectors $u^{\mu}$ and $x^{\mu}$ are constrained by the following normalization and orthogonality relations:
\ie
u_{\mu} u^{\mu} = -1\,,\qquad x_{\mu} x^{\mu} = 1\,.\label{ffvello}
\fe
Notice that, in the reference frame moving with the fluid, the timelike four--velocity and the orthogonal radial unit vector take the forms $u^{\mu} = (u^t, 0, 0, 0)$ and $x^\mu = (0, x^r, 0, 0)$. Applying the condition given in Eq.~(\ref{ffvello}), one obtains the following relation:
\ie
u_t^2 = \mathrm{A}(r )u_tu^t = - \mathrm{A}(r)\,,\qquad x_r^2 = \mathrm{B}(r)x_rx^r = \mathrm{B}(r)\,.
\fe

Furthermore, at the background level, the components of the stress–energy tensor take the form:
\begin{align}
T_{tt}&= - \mathrm{A}(r)\rho\,,\qquad T_{t}^{t}=-\rho\,,\\
T_{rr}& = \mathrm{B}(r)p_{1}\,,\qquad T_{r}^{r} = p_{1}\,,\\
&T_{\theta}^{\theta} = T_{\varphi}^{\varphi} = p_{2}\,.
\end{align}
It is important to emphasize that the functions $\rho$, $p_1$, and $p_2$ vary with the radial coordinate $r$ and can be explicitly obtained by evaluating the corresponding components of the Einstein tensor for the given geometry.

The study of quasinormal oscillations for a static, spherically symmetric black hole begins by introducing deviations from the background geometry, leading to a time--dependent, axisymmetric perturbation. Under such a deformation, the spacetime metric is altered and assumes the following form \cite{chen2019gravitational}:
\begin{align}
\mathrm{d}s^2=&-e^{2\Tilde{\nu}}\left(\mathrm{d}x^0\right)^2+e^{2\Tilde{\psi}}\left(\mathrm{d}x^1 - \Tilde{\sigma} \mathrm{d}x^0 - \Tilde{q}_2\mathrm{d}x^2 -\Tilde{q}_3\mathrm{d}x^3\right)^2\nonumber\\&+e^{2\Tilde{\mu}_2}\left(\mathrm{d}x^2\right)^2+e^{2\Tilde{\mu}_3}\left(\mathrm{d}x^3\right)^2\,.\label{mtccgg}
\end{align}

Parameters $\Tilde{\nu}$, $\Tilde{\psi}$, $\Tilde{\mu}_2$, $\Tilde{\mu}_3$, $\Tilde{\sigma}$, $\Tilde{q}_2$, and $\Tilde{q}_3$ are treated as functions of three variables: the temporal coordinate $t$ ($x^0$), the radial position $r$ ($x^2$), and the polar angle $\theta$ ($x^3$). Since the configuration under consideration is axisymmetric, all these functions are assumed to be independent of the azimuthal variable $\varphi$ ($x^1$). The labeling and formalism adopted here are based on the framework developed in Ref.~\cite{chen2019gravitational}. For a purely static and spherically symmetric background, the fields $\Tilde{q}_2$, $\Tilde{q}_3$, and $\Tilde{\sigma}$ are identically zero; therefore, when perturbations are introduced, these terms are regarded as first--order contributions and treated accordingly in the linearized theory.

Advancing the investigation requires reformulating the spacetime structure through an orthonormal tetrad basis tailored to the geometry described by Eq.~\eqref{mtccgg}. This method simplifies, as we did to the vector perturbations, the handling of perturbations by translating the problem into a locally flat frame, where the relevant quantities can be more conveniently analyzed
\begin{align}
\mathrm{e}^{\mu}_{0}&=\left(\mathrm{e}^{-\Tilde{\nu}},\Tilde{\sigma} \mathrm{e}^{-\Tilde{\nu}},0,0\right)\,,\nonumber\\
\mathrm{e}^{\mu}_{1}&=\left(0, \mathrm{e}^{-\Tilde{\psi}}, 0,0\right)\,,\nonumber\\
\mathrm{e}^{\mu}_{2}&=\left(0, \Tilde{q}_2 \mathrm{e}^{-\Tilde{\mu}_2}, \mathrm{e}^{-\Tilde{\mu}_2},0\right)\,,\nonumber\\
\mathrm{e}^{\mu}_{3}&=\left(0, \Tilde{q}_3 \mathrm{e}^{-\Tilde{\mu}_3}, 0, \mathrm{e}^{-\Tilde{\mu}_3}\right)\,.
\end{align}

To distinguish between coordinate and frame components, indices associated with the tetrad are enclosed in parentheses. The formalism operates by mapping all geometrical and physical quantities from the coordinate basis defined by $\mathrm{g}_{\mu\nu}$ onto a locally inertial frame characterized by $\eta_{ab}$, using the appropriate tetrad vectors.  Within this setup, any vector or tensor field is rewritten in terms of its projections onto the tetrad basis
\begin{align}
A_{\mu}& = \mathrm{e}_{\mu}^{a}A_{a}\,,\quad A_{a} = \mathrm{e}_{a}^{\mu}A_{\mu}\,,\nonumber\\
B_{\mu\nu}& = \mathrm{e}_{\mu}^{a} \mathrm{e}_{\nu}^{b}B_{ab}\,,\quad B_{ab} = \mathrm{e}_{a}^{\mu}\mathrm{e}_{b}^{\nu}B_{\mu\nu}\,.
\end{align}

Therefore, the modified (perturbated) version of the stress--energy tensor, by taking into account all these previous features, is
\begin{align}
\delta T_{ab}=&\,(\rho+p_2)\delta(u_{a}u_{b})+(\delta\rho+\delta p_2)u_{a}u_{b}\nonumber\\
&+(p_1-p_2)\delta(x_{a}x_{b})+(\delta p_1-\delta p_2)x_{a}x_{b}\nonumber\\&+\delta p_2\eta_{ab}.
\end{align}
By implementing the normalization condition from Eq.~\eqref{ffvello} together with the orthogonality constraint $u^\mu x_\mu = 0$, one remarkably finds that the axial components of the perturbed stress--energy tensor vanish identically in the tetrad frame:
\ie
\delta T_{10} = \delta T_{12}=\delta T_{13}=0\,.
\fe

Therefore, the Einstein equation is rewritten as
\ie
R_{ab}-\frac{1}{2}\eta_{ab}R=8\pi T_{ab}\,.\label{eineq}
\fe
Because the axial components of the perturbed stress--energy tensor identically vanish, the dynamics of axial gravitational perturbations are governed solely by the condition $R_{ab}|_{\text{axial}} = 0$. After carrying out the corresponding manipulations and tensorial reductions—detailed extensively in the Appendix of Ref.~\cite{chen2019gravitational}—one ultimately derives a master equation whose structure leads to the following form for the effective potential associated with gravitational perturbations \cite{baruah2025quasinormal}:
\ie
V_{\text{T}}(r,\alpha,\beta,Q) = \mathrm{A}(r) \left\{ \dfrac{2}{r^2} \left[ 
\frac{1}{\mathrm{B}(r)} - 1 \right] + \dfrac{\ell(\ell+1)}{r^2} - \dfrac{1}{r \sqrt{\mathrm{A}(r) B(r)}} \left( \dfrac{\mathrm{d}}{\mathrm{d}r} \sqrt{\mathrm{A}(r) \mathrm{B}^{-1}(r)} \right) \right\},
\fe
or
\ie
\begin{split}
& V_{\text{T}}(r,\alpha,\beta,Q) =  \, \mathrm{A}(r) \left[ \frac{\ell (\ell+1)}{r^2} -\frac{4 M}{r^3} + \frac{2 Q^2}{r^4} +\frac{\alpha  (1-2 \beta ) Q^4}{5 r^8}  \right. \\
& \left.  + \frac{\left(10 M r^5+3 \alpha  (2 \beta -1) Q^4-10 Q^2 r^4\right) \left(10 r^5 (r-2 M)+Q^4 (\alpha -2 \alpha  \beta )+10 Q^2 r^4\right)}{5 r^8 \sqrt{\left(10 r^5 (r-2 M)+Q^4 (\alpha -2 \alpha  \beta )+10 Q^2 r^4\right)^2}}   \right] .
\end{split}
\fe

In line with expectations, taking the limit $\alpha = \beta \to 0$ restores the effective potential corresponding to tensorial perturbations in the Reissner--Nordström geometry. Similarly, setting $Q \to 0$ yields the Schwarzschild case for the tensor sector.

\begin{figure}
    \centering
    \includegraphics[scale=0.55]{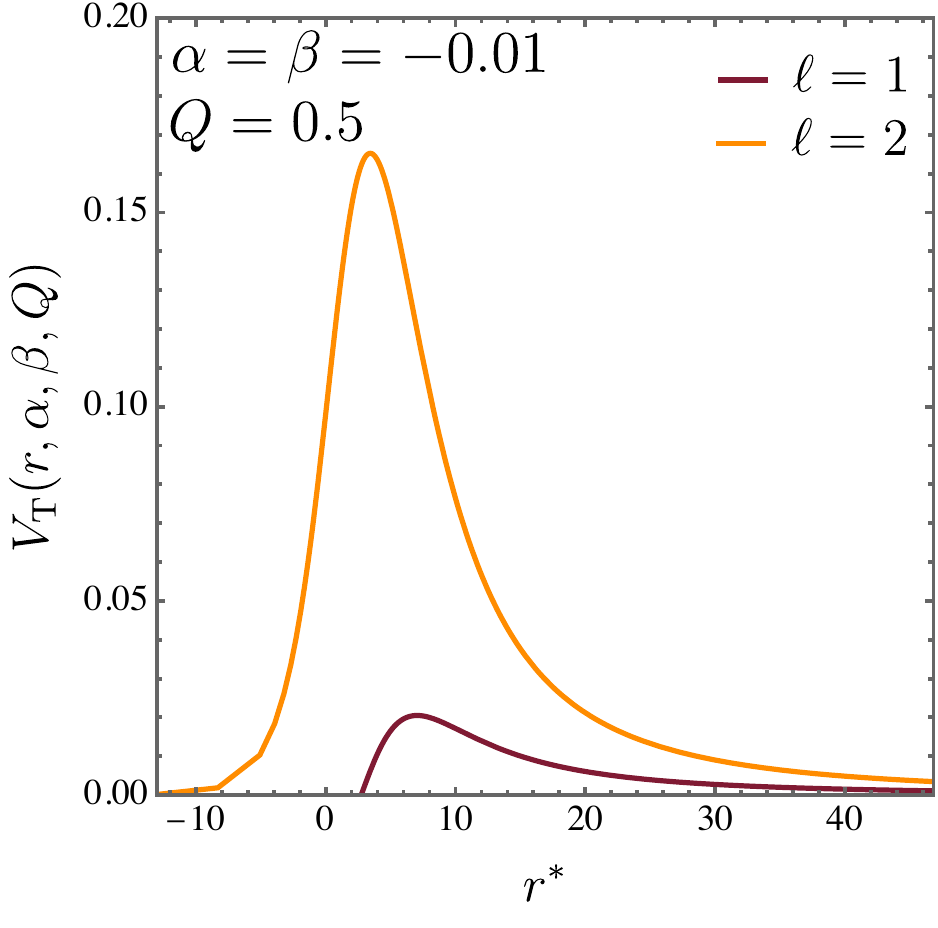}
    \caption{The behavior of the effective potential associated with tensorial perturbations, $V_{\text{T}}(r,\alpha,\beta,Q)$, is shown as a function of the tortoise coordinate $r^*$, with particular emphasis on its variation for different multipole numbers $\ell$.}
    \label{vftns}
\end{figure}

The potential $V_{\text{T}}(r,\alpha,\beta,Q)$ is displayed in Fig.~\ref{vftns} as a function of the tortoise coordinate $r^*$. As seen in the analysis of previous perturbation types, this potential maintains a sinusoidal--like structure, which justifies the application of the WKB approximation for determining the quasinormal spectrum.

Tabs. \ref{qnmmtensoor} and \ref{qnmtac2tensor} report the quasinormal frequencies for tensor perturbations with $\ell = 1$ and $\ell = 2$, covering a range of values for $Q$, $\alpha$, and $\beta$. As with earlier cases, the damping rates alternate depending on the chosen parameters, exhibiting either stronger or weaker attenuation. A distinct behavior emerges here compared to the scalar and vector sectors: for the mode $\omega_2$, keeping $\alpha$ and $\beta$ fixed while varying $Q$ may lead to instabilities in the computed frequencies—at least within the parameter explored here. It is also important to point out that, specifically for the case $\ell = 2$, the third--order WKB approximation was adopted, in contrast to the sixth--order scheme utilized earlier for scalar and vector perturbations. Recent claims on the detection of QNMs beyond the $\ell=m=2$ have been a matter of recent discussion and we expect that this window could be used to constrain alternative gravity proposals in the future \cite{Franchini:2023eda}.

\begin{table}[!h]
\begin{center}
\caption{\label{qnmmtensoor} The table presents the quasinormal mode frequencies for tensor--type perturbations with $\ell = 1$, highlighting their dependence on the parameters $\alpha$, $\beta$, and $Q$.}
\begin{tabular}{c| c | c | c} 
 \hline\hline\hline 
 $\alpha = \beta$ \,\,  $Q$  & $\omega_{0}$ & $\omega_{1}$ & $\omega_{2}$  \\ [0.2ex] 
 \hline 
 \,  -0.01, \,  0.6  & 0.843177 - 0.0134701$i$ & 2.94714 - 0.007348$i$ & 7.44771 - 0.00240293$i$ \\
 
\,  -0.02, \,  0.6  & 0.784042 - 0.0144285$i$ & 2.84423 - 0.00739006$i$  & 7.3801 - 0.00187955$i$  \\
 
 \, -0.03, \,  0.6  & 0.233795 - 0.0480162$i$  & 0.476666 - 0.0429147$i$ &  0.73061 - 0.0165599$i$  \\
 
\, -0.04, \,  0.6  & 0.225702 - 0.0493197$i$  & 0.468161 - 0.0408386$i$  & 0.78037 - 0.00468963$i$ \\
 
\, -0.05, \,  0.6  &  0.623823 - 0.0178113$i$ & 2.18230 - 0.00856006$i$ & 5.55378 - 0.0000833042$i$  \\
   [0.2ex] 
 \hline\hline\hline 
\,  $\alpha = \beta$ \,\,  $Q$  & $\omega_{0}$ & $\omega_{1}$ & $\omega_{2}$  \\ [0.2ex] 
 \hline 
\, -0.01, \,   0.1  & 0.104524 - 0.0997893$i$ &  0.0334908 - 0.479693$i$  & \text{Unstable}   \\
 
\, -0.01, \,  0.2  & 0.0916284 - 0.115183$i$ & 0.0373754 - 0.456598$i$  &  \text{Unstable}   \\

\, -0.01, \,   0.3  & 0.345455 - 0.0306078$i$ & 1.1842 - 0.0140684$i$ & \text{Unstable}   \\
 
\, -0.01, \,  0.4  & 0.036154 - 0.295617$i$  & 0.0133302 - 1.266820$i$ & \text{Unstable}   \\
 
\, -0.01, \,  0.5  & 0.333668 - 0.032605$i$ & 1.30788 - 0.0135937$i$  & \text{Unstable}  \\
   [0.2ex] 
 \hline \hline \hline 
\end{tabular}
\end{center}
\end{table}

\begin{table}[!h]
\begin{center}
\caption{\label{qnmtac2tensor} The table reports the quasinormal mode frequencies corresponding to tensor perturbations with $\ell = 2$, explored as functions of the parameters $\alpha$, $\beta$, and $Q$.}
\begin{tabular}{c| c | c | c} 
 \hline\hline\hline 
 $\alpha = \beta$ \,\,  $Q$  & $\omega_{0}$ & $\omega_{1}$ & $\omega_{2}$  \\ [0.2ex] 
 \hline 
 \,  -0.01, \,  0.6  & 0.400219 - 0.0907304$i$ & 0.375352 - 0.278875$i$ & 0.336032 - 0.476918$i$  \\
 
\,  -0.02, \,  0.6  & 0.400220 - 0.0907334$i$ & 0.375362 - 0.278889$i$  & 0.336072 - 0.476947$i$  \\
 
 \, -0.03, \,  0.6  & 0.400219 - 0.090730$i$  & 0.375348 - 0.278871$i$ &  0.336017 - 0.476910$i$  \\
 
\, -0.04, \,  0.6  & 0.400219 - 0.0907336$i$ & 0.375360 - 0.278888$i$ & 0.336065 - 0.476945$i$ \\
 
\, -0.05, \,  0.6  &  0.400219 - 0.0907329$i$ & 0.375356 - 0.278884$i$ & 0.336050 - 0.476935$i$  \\
   [0.2ex] 
 \hline\hline\hline 
  $\alpha = \beta$ \,\,  $Q$  & $\omega_{0}$ & $\omega_{1}$ & $\omega_{2}$  \\ [0.2ex] 
 \hline 
\, -0.01, \,   0.1  & 0.373812 - 0.0892675$i$ &  0.346716 - 0.275054$i$  & 0.303716 - 0.471280$i$   \\
 
\, -0.01, \,  0.2  & 0.375789 - 0.0894155$i$  & 0.348843 - 0.275460$i$  &  0.306092 - 0.471909$i$  \\

\, -0.01, \,   0.3  & 0.379188 - 0.0896565$i$ & 0.352506 - 0.276118$i$ & 0.310199 - 0.472921$i$  \\
 
\, -0.01, \,  0.4  & 0.384178 - 0.0899769$i$ & 0.357897 - 0.276976$i$ & 0.316251 - 0.474219$i$  \\
 
\, -0.01, \,  0.5  & 0.391039 - 0.0903529$i$ & 0.365341 - 0.277955$i$ & 0.324659 - 0.475661$i$ \\
   [0.2ex] 
 \hline \hline \hline 
\end{tabular}
\end{center}
\end{table}


\section{Quasinormal modes: fermionic case}

This part of the study is devoted to examining the dynamics of massless Dirac fields in the background of a static and spherically symmetric black hole. The analysis is carried out using the Newman--Penrose formalism, a powerful framework tailored for handling spin--$1/2$ fields in curved geometries. Within this approach, the Dirac equations that describe the evolution of fermionic perturbations are formulated as follows \cite{newman1962approach, chandrasekhar1984mathematical, Albuquerque:2023lhm}:
\begin{align}
(D + \epsilon - \rho) F_1 +( \bar{\delta} + \pi - \alpha) F_2 &= 0, \\
(\Delta + \mu - \gamma) F_2 + (\delta + \beta - \tau) F_1 &= 0.
\end{align}
Here, $F_1$ and $F_2$ represent the spinor components of the Dirac field, while the differential operators $D = l^\mu \partial_\mu$, $\Delta = n^\mu \partial_\mu$, $\delta = m^\mu \partial_\mu$, and $\bar{\delta} = \bar{m}^\mu \partial_\mu$ correspond to derivatives taken along the null tetrad vectors that define the chosen Newman--Penrose basis.

To carry out this investigation, the null tetrad frame is explicitly defined using the components of the underlying spacetime metric. The basis vectors are expressed in terms of these metric functions as follows:
\begin{align}
l^\mu &= \left(\frac{1}{\mathrm{A}(r)}, \sqrt{\frac{1}{\mathrm{B}(r) \mathrm{A}(r)}}, 0, 0\right), \quad \quad
n^\mu = \frac{1}{2} \left(1, -\sqrt{\frac{\mathrm{A}(r)}{\mathrm{B}(r)}}, 0, 0\right), \\
m^\mu &= \frac{1}{\sqrt{2} r} \left(0, 0, 1, \frac{i}{\sin \theta}\right), \quad \quad
\bar{m}^\mu = \frac{1}{\sqrt{2} r} \left(0, 0, 1, \frac{-i}{\sin \theta}\right).
\end{align}

In this manner, the non--vanishing spin coefficients can be computed and are given by the following expressions:
\ie
 \rho = -\frac{1}{r} \frac{1}{\mathrm{A}(r) \mathrm{B}(r)}, \quad 
\mu = -\frac{\sqrt{\frac{\mathrm{A}(r)}{\mathrm{B}(r)} }}{2r},  \quad 
\gamma = \frac{\mathrm{A}(r)'}{4}\sqrt{\frac{1}{\mathrm{A}(r) \mathrm{B}(r)}}, \quad 
\beta = -\alpha = \frac{\cot{\theta}}{2\sqrt{2}\, r} .
\fe

Through the decoupling procedure applied to the system describing the massless Dirac field, one arrives at a single differential equation governing the behavior of the spinor component $F_1$:
\begin{align}
\left[(D - 2\rho)(\Delta + \mu - \gamma) - (\delta + \beta) (\bar{\delta}+\beta)\right] F_1 = 0.
\end{align}

Substituting the earlier expressions for the directional derivatives and spin coefficients into the decoupled equation allows it to be rewritten in an explicit manner
\begin{align}
&\left[ \frac{1}{2\mathrm{A}(r)} \partial_t^2 - \left( \frac{\sqrt{\frac{\mathrm{A}(r)}{\mathrm{B}(r)}}}{2r} +\frac{\mathrm{A}(r)'}{4}\sqrt{\frac{1}{\mathrm{A}(r) \mathrm{B}(r)}}\right)\frac{1}{\mathrm{A}(r)}\partial_t \right. \\ \nonumber 
& \left. - \frac{\sqrt{\frac{\mathrm{A}(r)}{\mathrm{B}(r)}}}{2} \sqrt{\frac{1}{\mathrm{A}(r) \mathrm{B}(r)}}\partial_r^2 -\sqrt{\frac{1}{\mathrm{A}(r) \mathrm{B}(r)}} \partial_r \left( \frac{\sqrt{\frac{\mathrm{A}(r)}{\mathrm{B}(r)}}}{2} + \frac{\mathrm{A}(r)'}{4}{\sqrt{\frac{1}{\mathrm{A}(r) \mathrm{B}(r)}}} \right) \right] F_1  \\ \nonumber
& + \left[ \frac{1}{\sin^2\theta} \partial_\varphi^2 + i \frac{\cot \theta}{\sin \theta}\partial_\varphi + \frac{1}{\sin \theta}\partial_\theta \left( \sin \theta \partial_\theta \right) - \frac{1}{4} \cot^2 \theta + \frac{1}{2} \right] F_1 = 0.
\end{align}

To facilitate the separation of variables into radial and angular parts, the spinor wave function is expressed in the following form:
\ie
F_1 = \mathrm{R}(r) S_{\ell,m}(\theta, \varphi) e^{-i \omega t},
\fe
where the radial part is 
\begin{align}
&\left[\frac{-\omega^2}{2\mathrm{A}(r)} - \left(\frac{\sqrt{\frac{\mathrm{A}(r)}{\mathrm{B}(r)}}}{2r}+\frac{\mathrm{A}(r)'}{4} + \sqrt{\frac{1}{\mathrm{A}(r) \mathrm{B}(r)}}\right)\frac{- i\omega}{\mathrm{A}(r)} - \frac{\sqrt{\frac{\mathrm{A}(r)}{\mathrm{B}(r)}}}{2} \sqrt{\frac{1}{\mathrm{A}(r) \mathrm{B}(r)}}\partial_r^2 \right. \\ \nonumber
& \left. - \sqrt{\frac{1}{\mathrm{A}(r) \mathrm{B}(r)}}\partial_r \left(\frac{\sqrt{\frac{\mathrm{A}(r)}{\mathrm{B}(r)}}}{2r} + \frac{\mathrm{A}(r)'}{4}\sqrt{\frac{1}{\mathrm{A}(r) \mathrm{B}(r)}}\right)-\lambda_{\ell}\right] \mathrm{R}(r) = 0.
\end{align}

Here, $\lambda_{\ell}$ gives rise to the separation constant arising from the variable decomposition. Upon introducing the generalized tortoise coordinate $r^{*}$ in place of the standard radial coordinate, the radial part of the Dirac equation assumes the structure of a Schrödinger--like differential equation (similar to the other perturbations studied in this manuscript), written as:
\begin{align}
\left[\frac{\mathrm{d}^2 }{\mathrm{d}r_*^2} +( \omega^2 - V^{\uparrow \downarrow}) \right]U_{\uparrow \downarrow} = 0.
\end{align}
Therefore, the potentials $V^{\uparrow \downarrow}$ shown above are given by \cite{Albuquerque:2023lhm, al2024massless, arbey2021hawking}:
\ie
V^{\uparrow \downarrow} = \frac{(\ell + \frac{1}{2})^2}{r^2} \mathrm{A}(r) \pm \left(\ell + \frac{1}{2}\right) \sqrt{\frac{\mathrm{A}(r)}{ \mathrm{B}(r)}} \partial_r \left(\frac{\sqrt{\mathrm{A}(r)}}{r}\right).
\fe

For definiteness, the analysis is carried out using the potential $V^{\uparrow}$, as its qualitative features are nearly identical to those of $V^{\downarrow}$ \cite{Albuquerque:2023lhm, devi2020quasinormal}. While both potentials yield comparable physical features, the focus is placed on $V^{\uparrow}$ to avoid redundancy. The explicit expression is given by:
\ie
\begin{split}
V^{\uparrow}(r,\alpha,\beta,Q) = & \, \, \mathrm{A}(r) \left[ \frac{(2 \ell+1) \left(5 r^5 (r-3 M)+\alpha  (2-4 \beta ) Q^4+10 Q^2 r^4\right)}{10 \sqrt{10} \, r^{11} \sqrt{\frac{1}{10 r^5 (r-2 M)+Q^4 (\alpha -2 \alpha  \beta )+10 Q^2 r^4}}} \right. \\
& \left. + \frac{(2 \ell+1)^2 \left(10 r^5 (r-2 M)+Q^4 (\alpha -2 \alpha  \beta )+10 Q^2 r^4\right)}{40 \, r^8}   \right]. \end{split}
\fe

\begin{figure}
    \centering
      \includegraphics[scale=0.55]{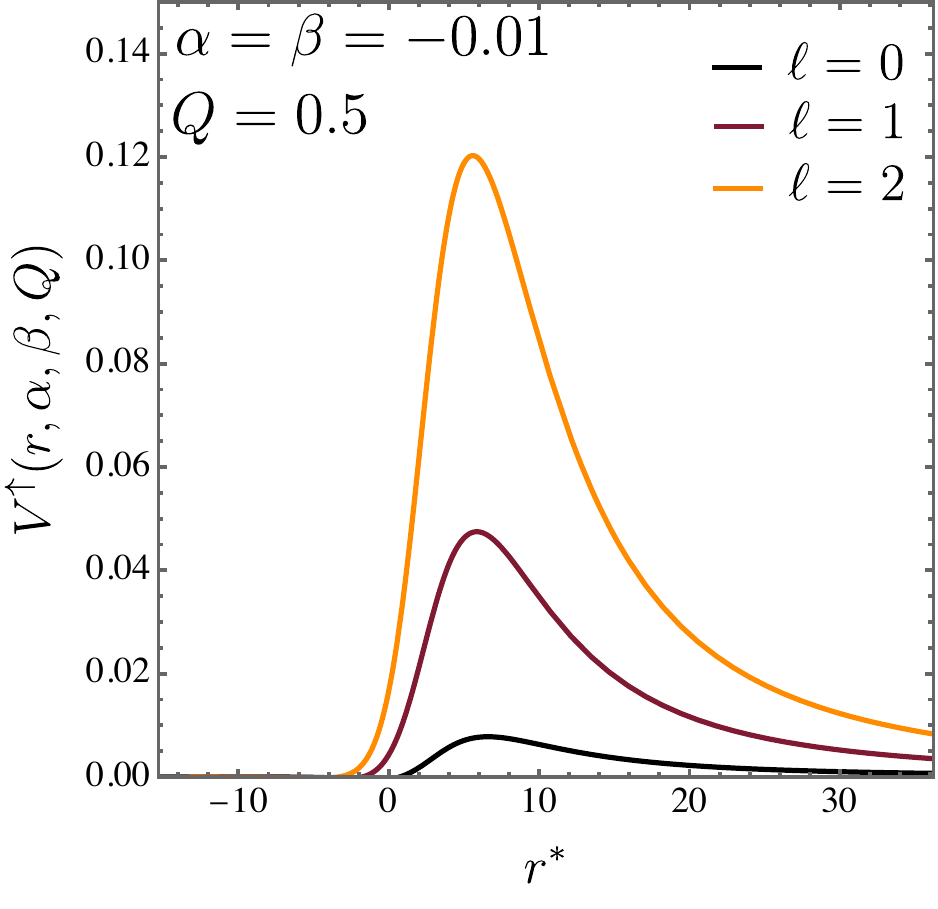}
    \caption{The effective potential $V^{\uparrow}(r,\alpha,\beta,Q)$, corresponding to the fermionic sector, is plotted for several values of the multipole number $\ell$, while keeping the parameters $Q$, $\alpha$, and $\beta$ fixed.}
    \label{vfermions}
\end{figure}

To provide a clearer understanding of the above expression, Fig.~\ref{vfermions} displays the profile of $V^{\uparrow}(r,\alpha,\beta,Q)$ as a function of the tortoise coordinate $r^{*}$, considering various values of the angular momentum number $\ell$. As evident from the plot, the potential exhibits a characteristic oscillatory, sine--like structure. Unlike the treatment adopted for bosonic perturbations, this fermionic sector analysis is carried out by setting $M = 0.5$ rather than $M = 1$. Additionally, the quasinormal mode frequencies are computed using the third--order WKB approximation, as we did for the tensor perturbations as well.

\begin{table}[!h]
\begin{center}
\caption{\label{spinorialperturbations1} The table presents the quasinormal mode frequencies associated with spin--$1/2$ (fermionic) perturbations for $\ell = 1$, showing their dependence on the parameters $\alpha$, $\beta$, and $Q$.}
\begin{tabular}{c| c | c | c} 
 \hline\hline\hline 
 $\alpha = \beta$ \,\,  $Q$  & $\omega_{0}$ & $\omega_{1}$ & $\omega_{2}$  \\ [0.2ex] 
 \hline 
 \,  -0.01, \,  0.5  & 0.453671 - 0.174431$i$ & 0.375909 - 0.547421$i$ & 0.251908 - 0.943040$i$ \\
 
\,  -0.02, \,  0.5  & 0.453694 - 0.174417$i$  & 0.375953 - 0.547388$i$  & 0.252011 - 0.942984$i$  \\
 
 \, -0.03, \,  0.5  & 0.453719 - 0.174407$i$   & 0.376020 - 0.547368$i$  &  0.252177 - 0.942943$i$  \\
 
\, -0.04, \,  0.5  & 0.453744 - 0.174393$i$ & 0.376070 - 0.547335$i$ & 0.252296 - 0.942887$i$ \\
 
\, -0.05, \,  0.5  &  0.252426 - 0.942831$i$ & 0.376124 - 0.547303$i$ & 0.252426 - 0.942831$i$  \\
   [0.2ex] 
 \hline\hline\hline 
  $\alpha = \beta$ \,\,  $Q$  & $\omega_{0}$ & $\omega_{1}$ & $\omega_{2}$  \\ [0.2ex] 
 \hline 
 \, -0.01, \,   0.1  & 0.384826 - 0.174107$i$ &  0.300527 - 0.542882$i$  & 0.155979 - 0.934397$i$   \\
 
\,  -0.01, \,  0.2  & 0.391142 - 0.175166$i$ & 0.307691 - 0.546246$i$  &  0.165190 - 0.939702$i$  \\

\, -0.01, \,   0.3  & 0.402772 - 0.176721$i$ & 0.321084 - 0.551308$i$ & 0.182879 - 0.947612$i$  \\
 
\, -0.01, \,  0.4  & 0.422008 - 0.177927$i$  & 0.343417 - 0.555766$i$ & 0.213122 - 0.954521$i$  \\
 
\,  -0.01, \,  0.5  & 0.453671 - 0.174431$i$ & 0.375909 - 0.547421$i$ & 0.251908 - 0.943040$i$\\
   [0.2ex] 
 \hline \hline \hline 
\end{tabular}
\end{center}
\end{table}

\begin{table}[!h]
\begin{center}
\caption{\label{spinorialperturbations2} The table provides the quasinormal mode frequencies for spinorial perturbations with $\ell = 2$, highlighting how they vary with the parameters $\alpha$, $\beta$, and $Q$.}
\begin{tabular}{c| c | c | c} 
 \hline\hline\hline 
 $\alpha = \beta$ \,\,  $Q$  & $\omega_{0}$ & $\omega_{1}$ & $\omega_{2}$  \\ [0.2ex] 
 \hline 
 \,  -0.01, \,  0.5  & 0.764755 - 0.175544$i$ & 0.710881 - 0.538795$i$ & 0.621425 - 0.921456$i$ \\
 
\,  -0.02, \,  0.5  & 0.764770 - 0.175552$i$ & 0.710937 - 0.538804$i$  & 0.621539 - 0.921441$i$  \\
 
 \, -0.03, \,  0.5  & 0.764786 - 0.175564$i$  & 0.711010 - 0.538833$i$ &  0.621715 - 0.921465$i$   \\
 
\, -0.04, \,  0.5  & 0.764802 - 0.175574$i$ & 0.711076 - 0.538852$i$ & 0.621864 - 0.921468$i$ \\
 
\, -0.05, \,  0.5  &  0.764818 - 0.175585$i$ & 0.711144 - 0.538872$i$ & 0.622017 - 0.921470$i$  \\
   [0.2ex] 
 \hline\hline\hline 
  $\alpha = \beta$ \,\,  $Q$  & $\omega_{0}$ & $\omega_{1}$ & $\omega_{2}$  \\ [0.2ex] 
 \hline 
  \, -0.01, \,   0.1  & 0.648516 - 0.173018$i$ &  0.587867 - 0.532099$i$  & 0.485294 - 0.910721$i$   \\
 
\, -0.01, \,  0.2  & 0.658917 - 0.174201$i$  & 0.598892 - 0.535658$i$  &  0.497686 - 0.916579$i$  \\

\, -0.01, \,   0.3  & 0.678097 - 0.176024$i$  & 0.619409 - 0.541120$i$ & 0.521065 - 0.925526$i$  \\
 
 \, -0.01, \,  0.4  & 0.710040 - 0.177817$i$ & 0.653918 - 0.546357$i$ & 0.560825 - 0.933967$i$   \\
 
 \, -0.01, \,  0.5  & 0.764755 - 0.175544$i$ & 0.710881 - 0.538795$i$ & 0.621425 - 0.921456$i$ \\
   [0.2ex] 
 \hline \hline \hline 
\end{tabular}
\end{center}
\end{table}


\section{Time--domain solution: bosonic case}

A thorough understanding of the temporal dynamics of scalar, vector, and tensor perturbations requires examining their evolution through time--dependent analysis, especially to evaluate the impact of quasinormal spectra on scattering processes. Owing to the complex structure of the governing potentials, this task demands a robust and precise numerical method. To overcome this challenge, the characteristic integration technique originally formulated by Gundlach and collaborators \cite{Gundlach:1993tp} is adopted.

The numerical procedure, as developed in \cite{Baruah:2023rhd, Bolokhov:2024ixe, Guo:2023nkd, Yang:2024rms, Gundlach:1993tp, Skvortsova:2024wly, Shao:2023qlt}, utilizes a transformation to null coordinates via the definitions $\Tilde{u} = t - r^{*}$ and $\Tilde{v} = t + r^{*}$. This reparameterization reduces the wave equation to a more manageable form. In this framework, the wave equation is rewritten as:
\ie
\left(4 \frac{\partial^{2}}{\partial \Tilde{u} \partial \Tilde{v}} + V(\Tilde{u},\Tilde{v})\right) \Tilde{\psi} (\Tilde{u},\Tilde{v}) = 0.
\fe

A consistent numerical solution to the equation can be achieved by discretizing the system using finite--difference schemes
\ie
\Tilde{\psi}(\Tilde{N}) = -\Tilde{\psi}(\Tilde{S}) + \Tilde{\psi}(\Tilde{W}) + \Tilde{\psi}(\Tilde{E}) - \frac{\Tilde{h}^{2}}{8}\Tilde{V}(\Tilde{S})[\Tilde{\psi}(\Tilde{W}) + \Tilde{\psi}(\Tilde{E})] + \mathcal{O}(\Tilde{h}^{4}).
\fe

To construct the numerical grid, points are designated within the null coordinate plane as follows: the base point is $\Tilde{S} = (\Tilde{u}, \Tilde{v})$, while neighboring positions are labeled $\Tilde{W} = (\Tilde{u} + \Tilde{h}, \Tilde{v})$, $\Tilde{E} = (\Tilde{u}, \Tilde{v} + \Tilde{h})$, and $\Tilde{N} = (\Tilde{u} + \Tilde{h}, \Tilde{v} + \Tilde{h})$, where $\Tilde{h}$ denotes the step size used for discretization. The computational domain is initialized along the characteristic lines $\Tilde{u} = \Tilde{u}_0$ and $\Tilde{v} = \Tilde{v}_0$, which serve as the foundation for evolving the system forward. As a starting profile, a Gaussian function centered at $\Tilde{v} = \Tilde{v}_c$ with width parameter $\sigma$ is imposed along the initial slice $\Tilde{u} = \Tilde{u}_0$, specifying the initial wave configuration for the evolution algorithm
\ie
\Tilde{\psi}(\Tilde{u} = \Tilde{u}_{0},\Tilde{v}) = A e^{-(\Tilde{v}-\Tilde{v}_{0})^{2}}/2\sigma^{2}, \,\,\,\,\,\, \Tilde{\psi}(\Tilde{u},\Tilde{v}_{0}) = \Tilde{\psi}_{0}.
\fe

The computational process begins by prescribing the field value along $\Tilde{v} = \Tilde{v}_0$ through the condition $\Tilde{\psi}(\Tilde{u}, \Tilde{v}_0) = \Tilde{\psi}_0$, where $\Tilde{\psi}_0$ is assigned a value of zero for convenience. The evolution is carried out iteratively along lines of constant $\Tilde{u}$, with $\Tilde{v}$ progressing in accordance with the null grid configuration. For simplicity and computational clarity, the test case involves massless perturbations (as already derived so far) in a spacetime with mass parameter fixed at $M = 1$. A Gaussian pulse, centered at $\Tilde{v} = 0$, is used as the initial wave packet, characterized by a width parameter $\sigma = 1$ and zero initial amplitude. The simulation employs a uniform lattice over the domain $\Tilde{u}, \Tilde{v} \in [0, 1000]$, with each step on the grid separated by an interval $\Tilde{h} = 0.1$.

\subsection{Scalar perturbations}

\begin{figure}
    \centering
    \includegraphics[scale=0.51]{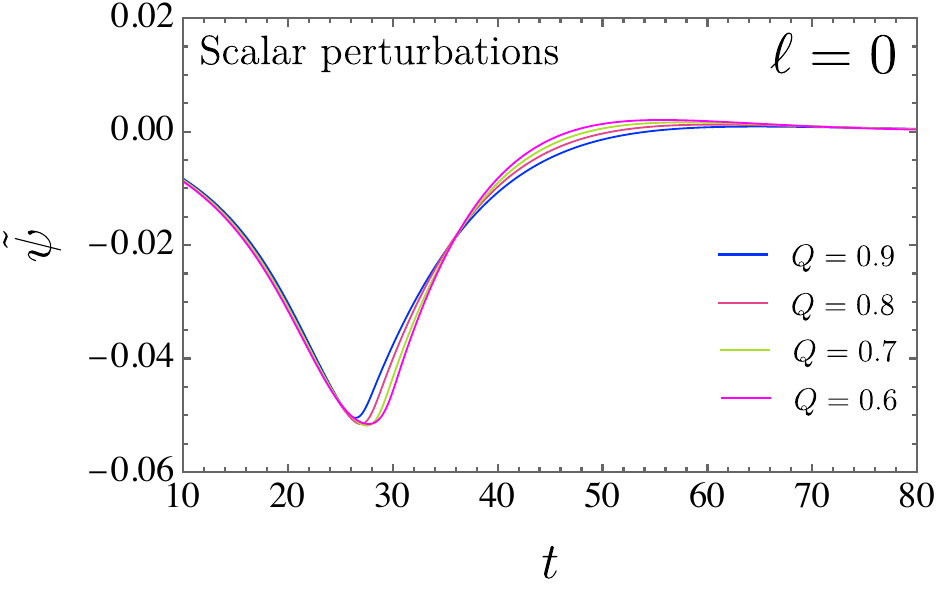}
    \includegraphics[scale=0.51]{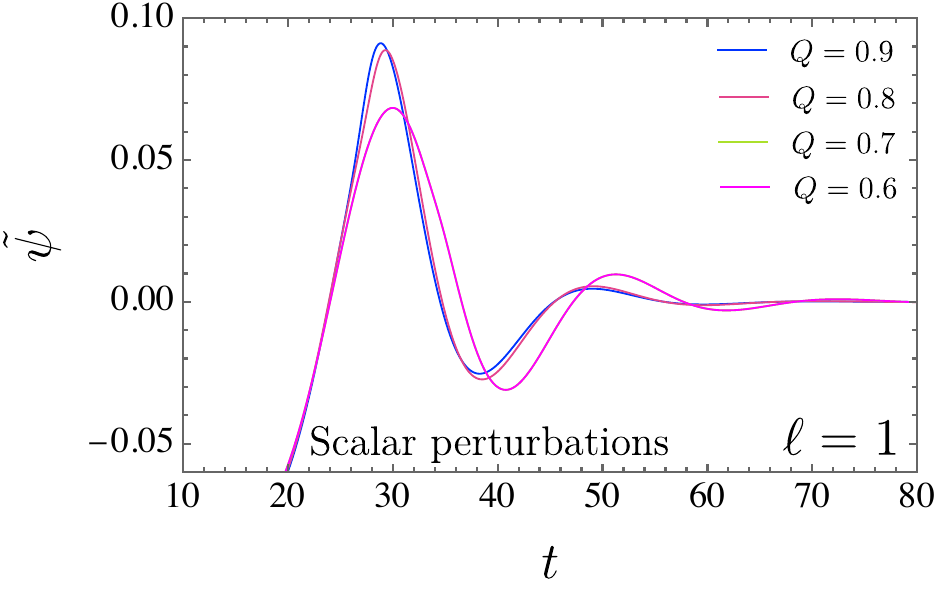}
    \includegraphics[scale=0.51]{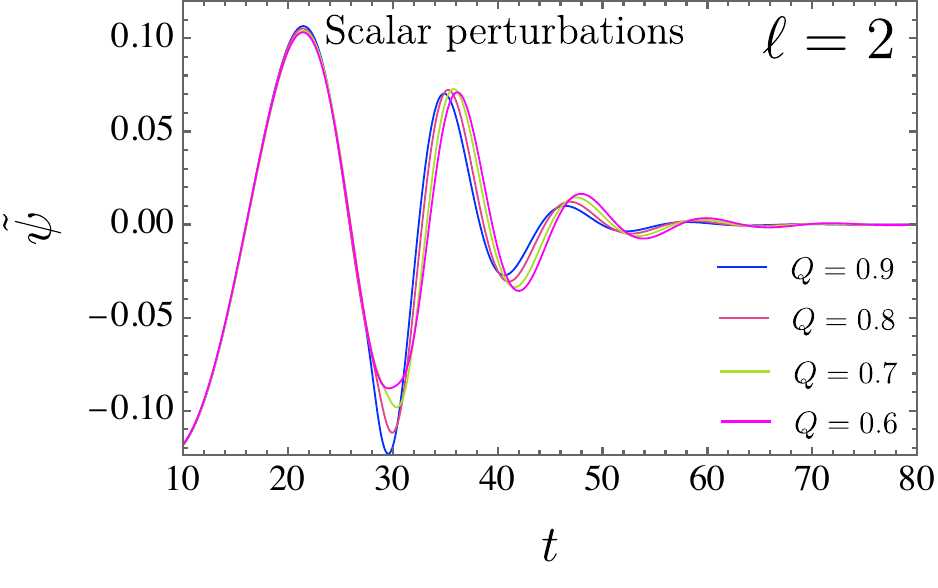}
    \caption{Scalar field perturbations are examined by plotting the time evolution of the waveform $\Tilde{\psi}$ for several values of the charge parameter $Q$, while keeping the coupling constant fixed at $\alpha =\beta = -0.01$. The chosen values for $Q$ include $0.6$, $0.7$, $0.8$, and $0.9$. The corresponding wave dynamics are displayed for: $\ell = 0$ in the upper--left panel, $\ell = 1$ in the upper--right, and $\ell = 2$ in the bottom panel.}
    \label{psiscalar}
\end{figure}

This part of the analysis focuses on the temporal profile of scalar perturbations within the black hole background. In Fig.~\ref{psiscalar}, the field $\Tilde{\psi}$ is shown evolving over time for a fixed choice of coupling parameters, $\alpha = \beta = -0.01$, while varying the charge $Q$ across values $0.6$, $0.7$, $0.8$, and $0.9$. The corresponding plots are organized by angular mode: $\ell = 0$ (upper left), $\ell = 1$ (upper right), and $\ell = 2$ (bottom). Naturally, the time series clearly demonstrate decaying oscillations, which is an indicative of quasinormal ringing.

\begin{figure}
    \centering
    \includegraphics[scale=0.51]{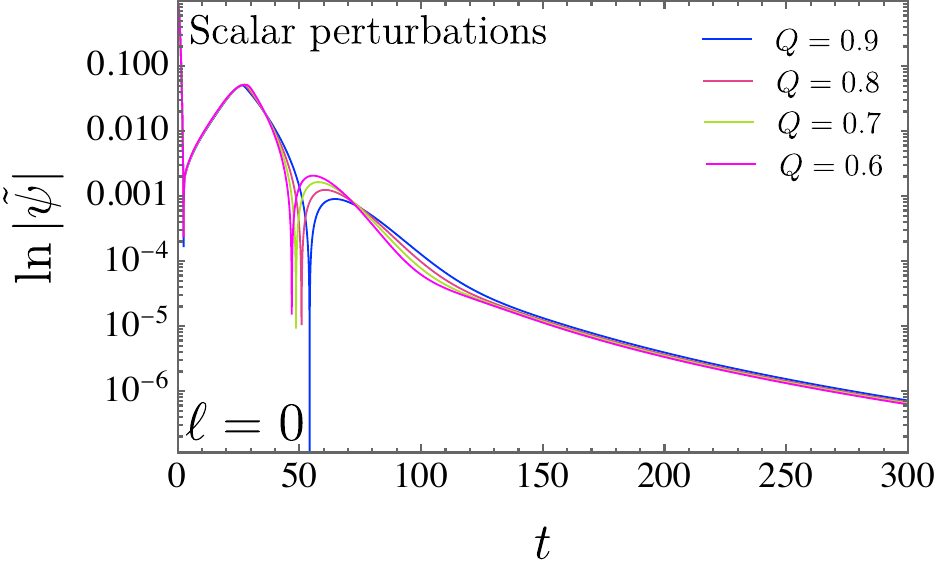}
    \includegraphics[scale=0.51]{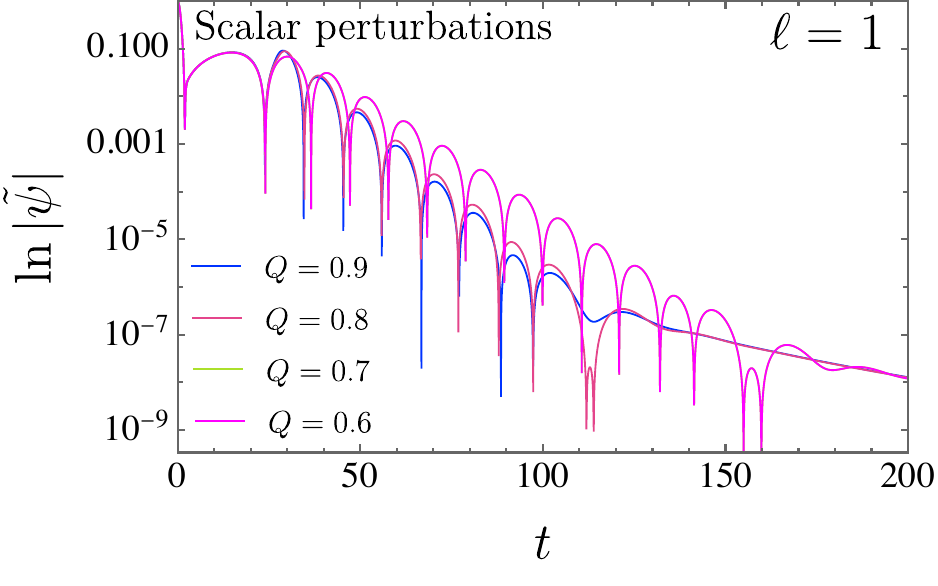}
    \includegraphics[scale=0.51]{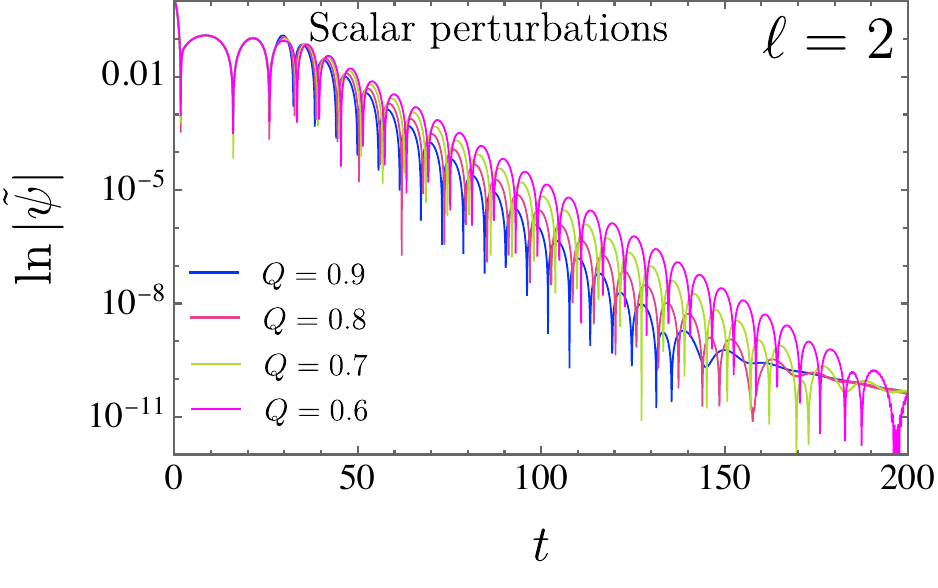}
    \caption{Logarithmic profile of the scalar field amplitude, $\ln|\Tilde{\psi}|$, as a function of time $t$ for fixed $\alpha = \beta = -0.01$ and varying charge values $Q = 0.6$, $0.7$, $0.8$, and $0.9$. Results are shown for angular modes $\ell = 0$ (top left), $\ell = 1$ (top right), and $\ell = 2$ (bottom).}
    \label{lnpsiscalar}
\end{figure}

To examine the attenuation more precisely, Fig.~\ref{lnpsiscalar} plots the logarithm of the absolute value, $\ln|\Tilde{\psi}|$, for the same set of charges and multipole indices. The damping pattern is preserved, and a transition to late--time power--law decay becomes evident—consistent with the expected tail behavior following the quasinormal phase.

\begin{figure}
    \centering
    \includegraphics[scale=0.51]{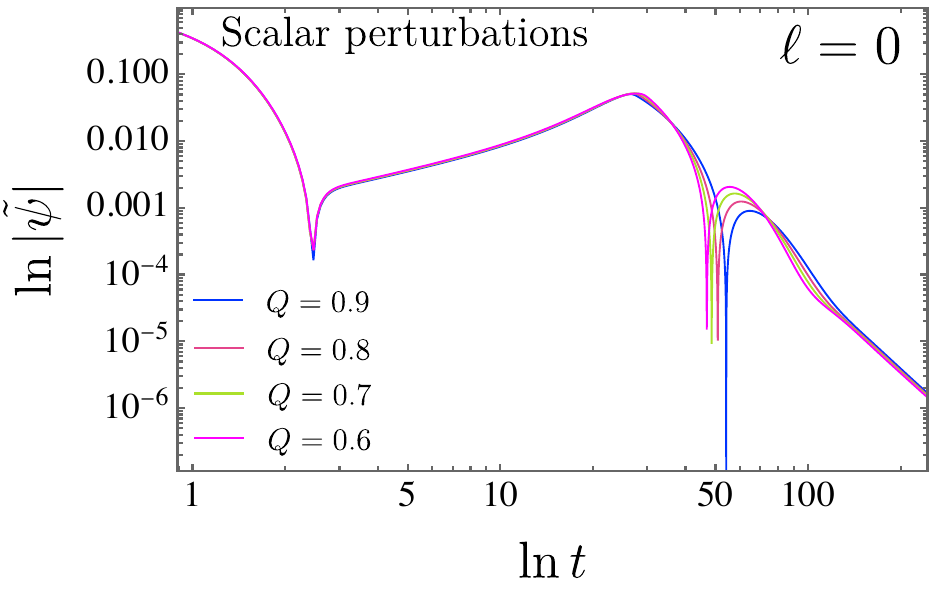}
    \includegraphics[scale=0.51]{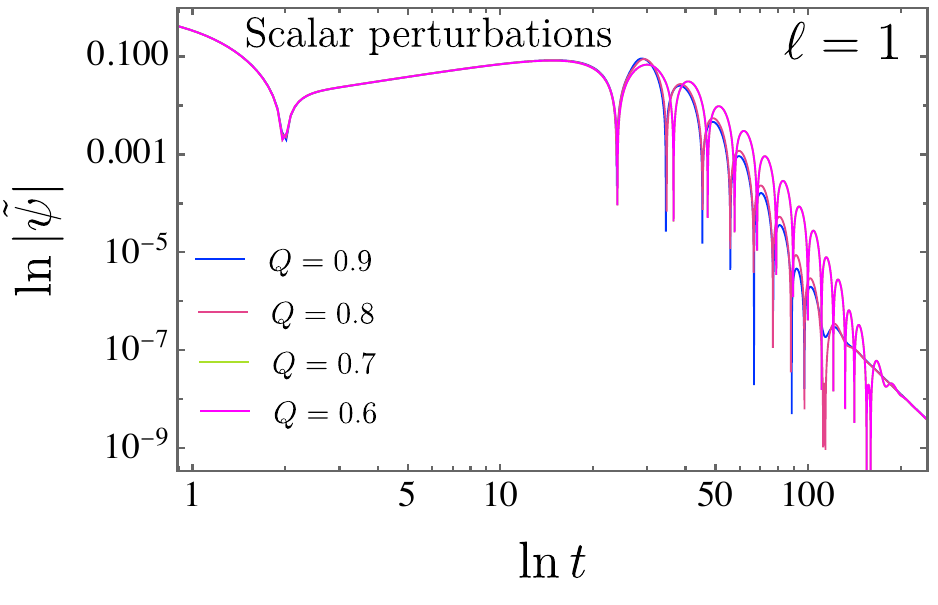}
    \includegraphics[scale=0.51]{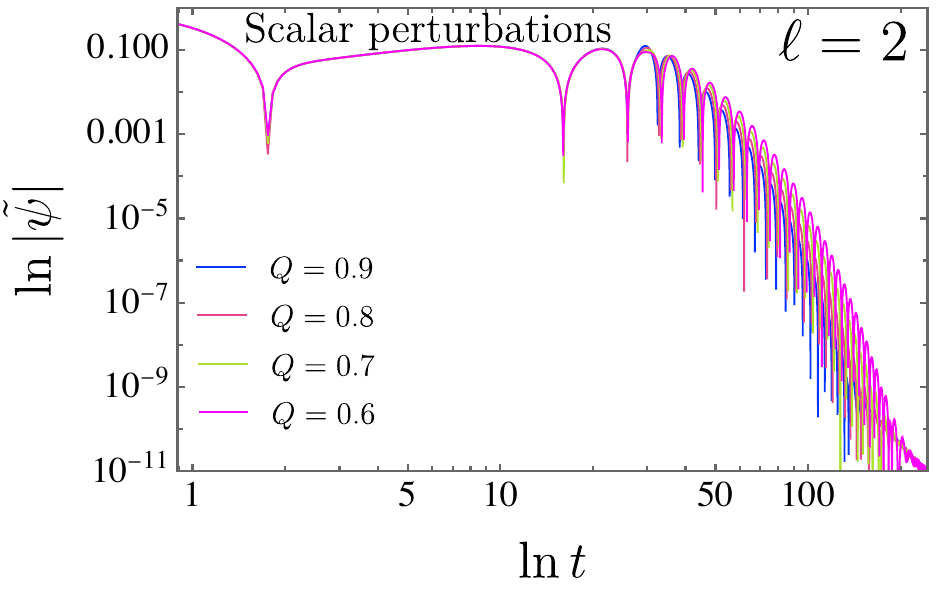}
    \caption{The figure presents the evolution of the scalar field in a ln–ln scale, where $\ln|\Tilde{\psi}|$ is plotted against $\ln|t|$ for several values of the charge parameter $Q$, while the parameters $\alpha = \beta$ remain fixed at $-0.01$. The charge values examined are $Q = 0.6$, $0.7$, $0.8$, and $0.9$. Each subplot corresponds to a distinct multipole number: $\ell = 0$ in the upper--left frame, $\ell = 1$ in the upper--right, and $\ell = 2$ in the lower panel.}
    \label{lnpsilnscalar}
\end{figure}

Further attention is devoted to Fig.~\ref{lnpsilnscalar}, which displays $\Tilde{\psi}$ against time $t$ in a ln--ln format. Using the same values for $Q$ and $\ell$, the panels are structured identically to the previous figures: top--left for $\ell = 0$, top--right for $\ell = 1$, and bottom for $\ell = 2$. This representation emphasizes the long--time decay and confirms the emergence of characteristic power--law tails.

\subsection{Vector perturbations}

\begin{figure}
    \centering
    \includegraphics[scale=0.51]{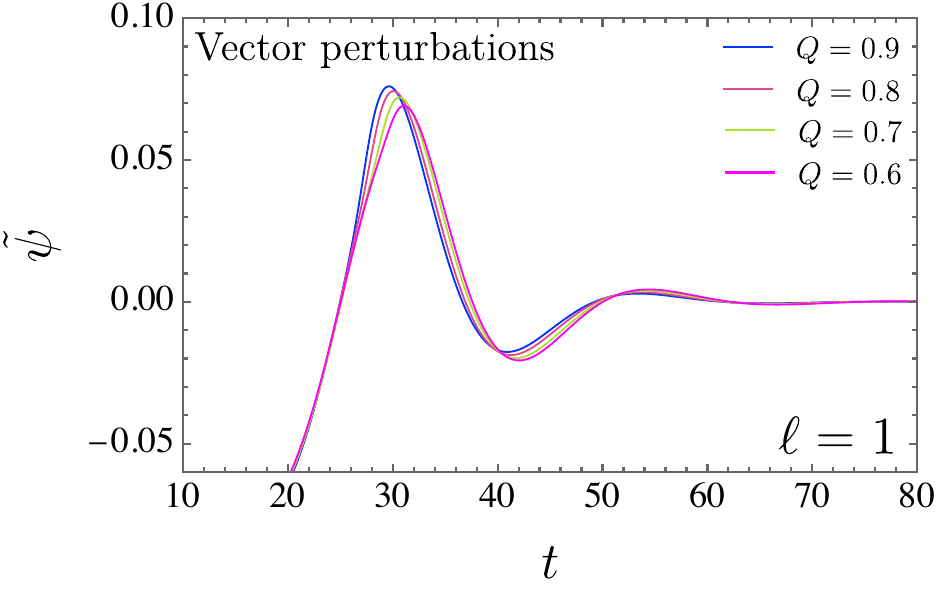}
    \includegraphics[scale=0.51]{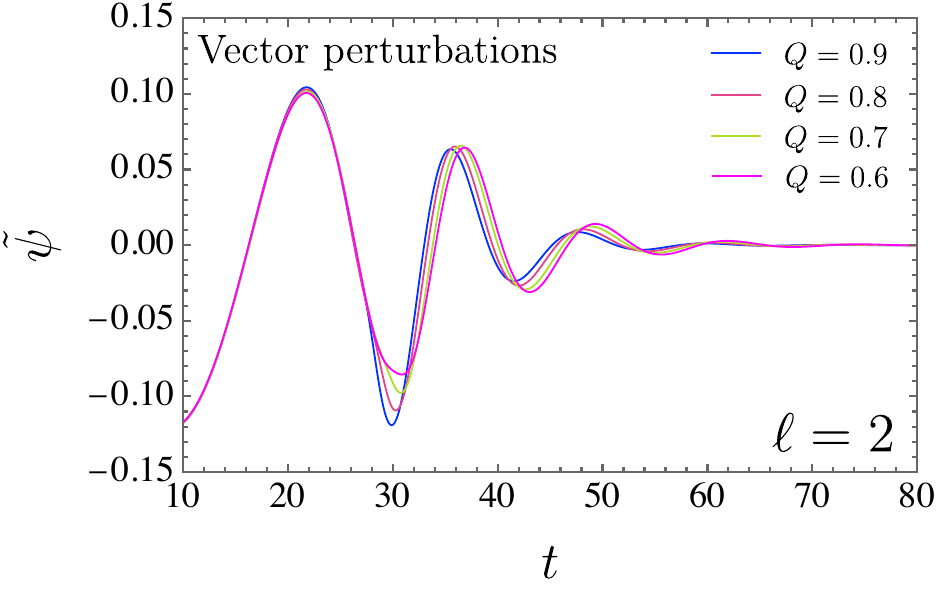}
    \caption{Time evolution of the vector perturbation waveform $\Tilde{\psi}$ for varying charge values $Q = 0.6$, $0.7$, $0.8$, and $0.9$, with fixed parameters $\alpha = \beta = -0.01$. The left and right panels correspond to angular modes $\ell = 1$ and $\ell = 2$, respectively.}
    \label{psivector}
\end{figure}

A comprehensive investigation of the temporal evolution of vector field disturbances is conducted by examining how the signal $\Tilde{\psi}$ behaves under variations of the electric charge parameter $Q$, while maintaining fixed coupling values $\alpha = \beta = -0.01$. As illustrated in Fig. \ref{psivector}, the analysis spans four distinct values of $Q$—namely $0.6$, $0.7$, $0.8$, and $0.9$—and considers angular momentum numbers , $\ell = 1$, and $2$, depicted in the left, and right panels, respectively. In all configurations, the waveform undergoes damped oscillations, gradually losing amplitude as time progresses.

\begin{figure}
    \centering
    \includegraphics[scale=0.51]{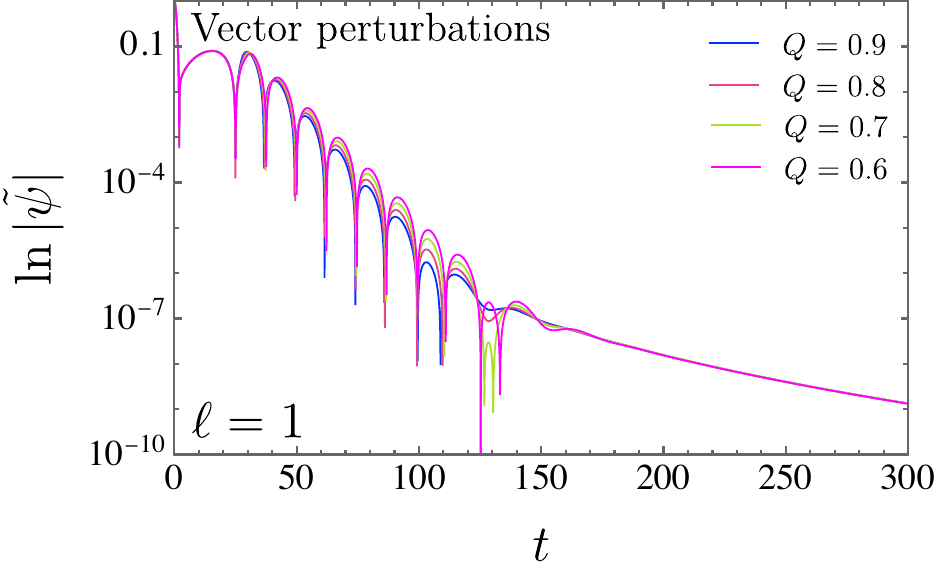}
    \includegraphics[scale=0.51]{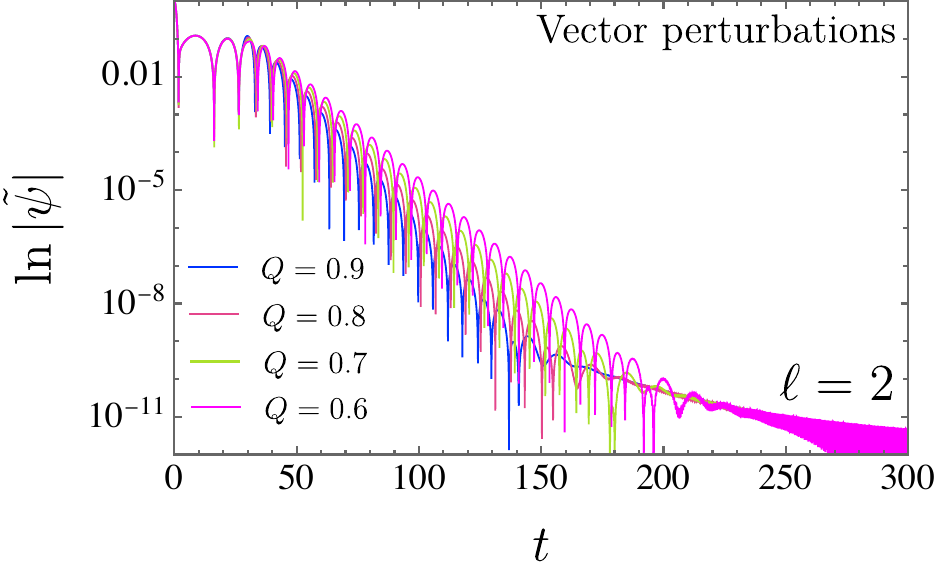}
    \caption{Logarithmic profile $\ln|\Tilde{\psi}|$ for vector--type perturbations plotted against time $t$, considering $Q = 0.6$, $0.7$, $0.8$, and $0.9$, with coupling parameters set to $\alpha = \beta = -0.01$. The left and right panels display the results for $\ell = 1$ and $\ell = 2$, respectively.}
    \label{lnpsivector}
\end{figure}

To highlight the attenuation pattern more effectively, Fig. \ref{lnpsivector} presents a logarithmic plot where the vertical axis displays $\ln|\Tilde{\psi}|$ against time $t$. The decay curves demonstrate that the initial exponential suppression of the signal transitions into a slower falloff at later stages. This shift signals the emergence of a power--law tail—a late--time signature often arising from scattering effects in curved backgrounds.

\begin{figure}
    \centering
    \includegraphics[scale=0.51]{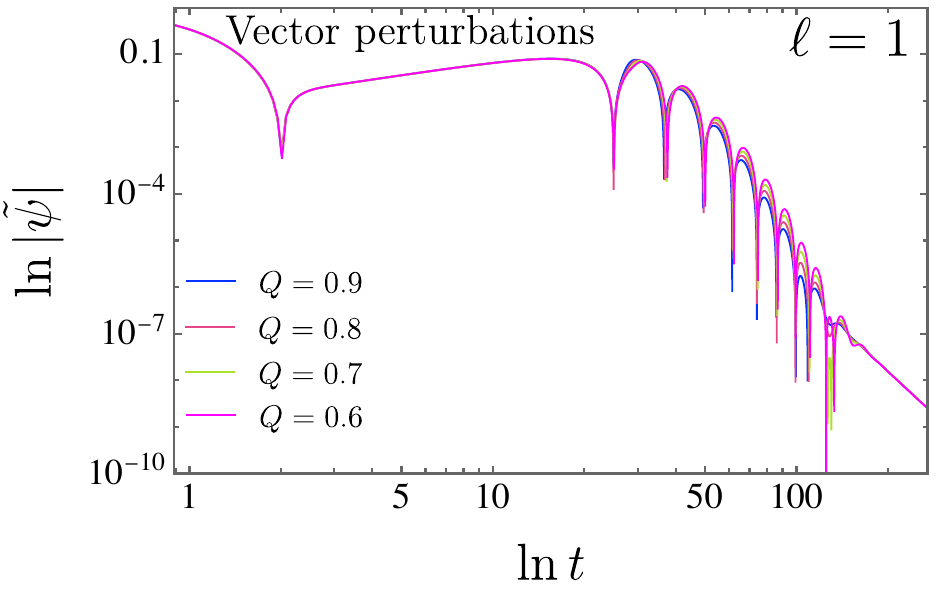}
    \includegraphics[scale=0.51]{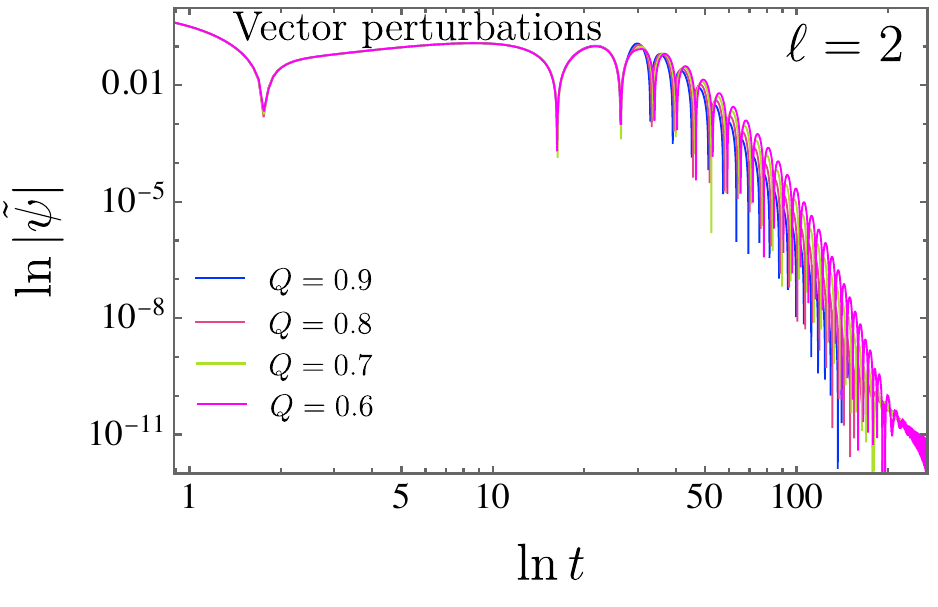}
    \caption{$\ln$--$\ln$ plots of $\ln|\Tilde{\psi}|$ versus $\ln|t|$ for vector perturbations, shown for charge values $Q = 0.6$, $0.7$, $0.8$, and $0.9$, with the parameters $\alpha = \beta$ kept constant at $-0.01$. The left panel corresponds to $\ell = 1$, while the right panel presents results for $\ell = 2$.}
    \label{lnpsilnvector}
\end{figure}

Furthermore, it is also shown in Fig. \ref{lnpsilnvector}, where both axes are logarithmic. This double--logarithmic depiction of $\Tilde{\psi}$ versus $t$ confirms that the late--time dynamics follow a power--law profile, largely independent of the values of $Q$ or the angular index $\ell$.

\subsection{Tensor perturbations}

\begin{figure}
    \centering
    \includegraphics[scale=0.51]{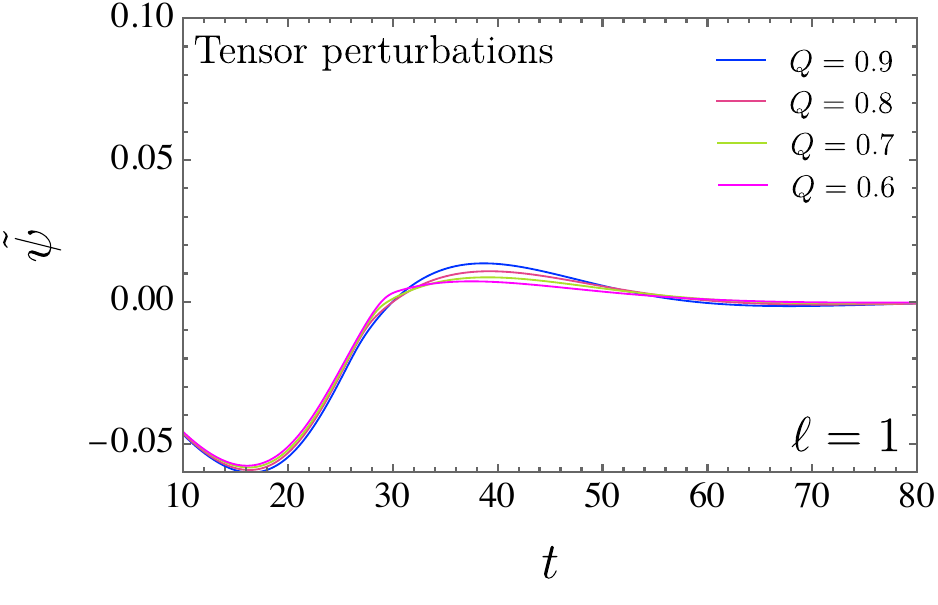}
    \includegraphics[scale=0.51]{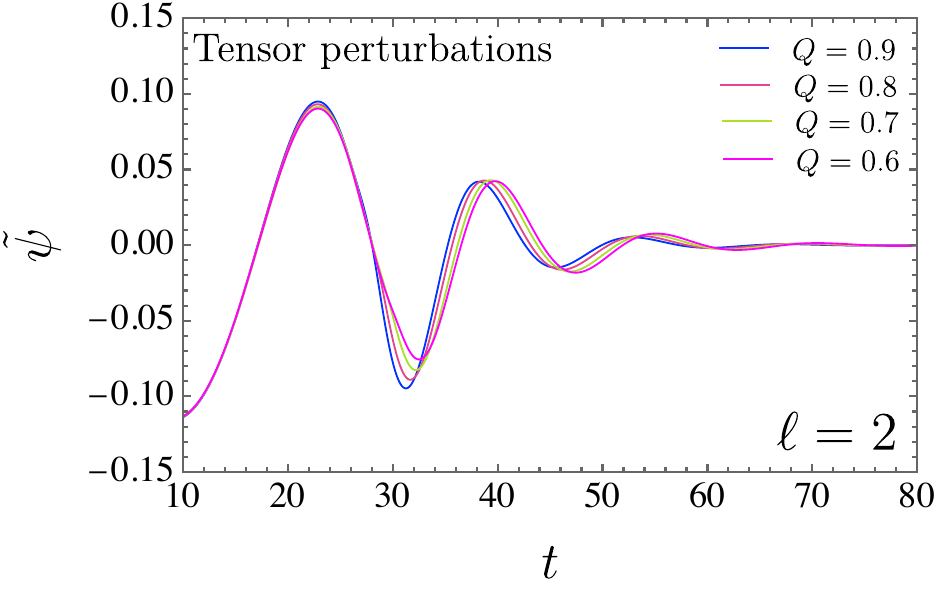}
    \caption{Time evolution of tensor--mode perturbations represented by the waveform $\Tilde{\psi}$ plotted against $t$ for fixed $\alpha = \beta = -0.01$ and varying $Q = 0.6$, $0.7$, $0.8$, and $0.9$. The left and right panels correspond to $\ell = 1$ and $\ell = 2$.}
    \label{psitensor}
\end{figure}

\begin{figure}
    \centering
    \includegraphics[scale=0.51]{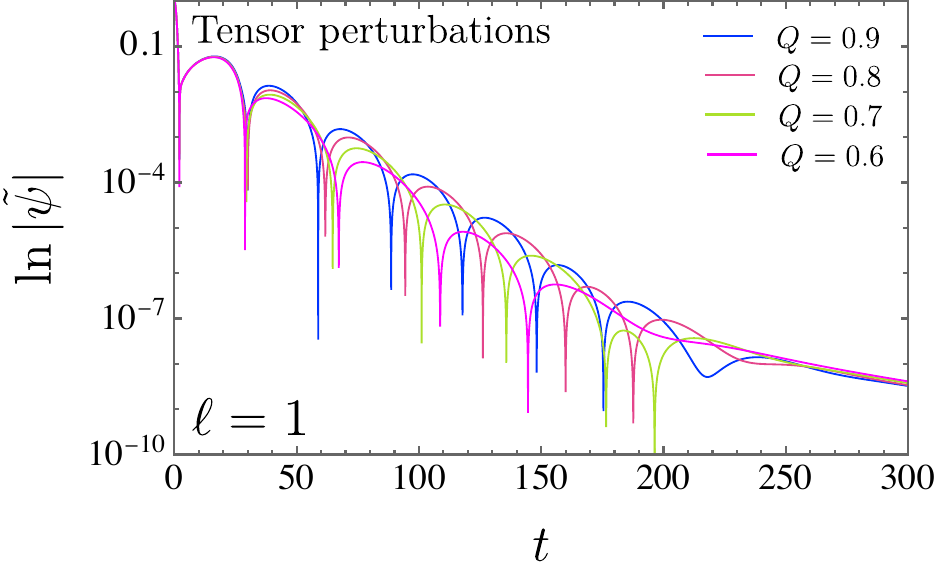}
    \includegraphics[scale=0.51]{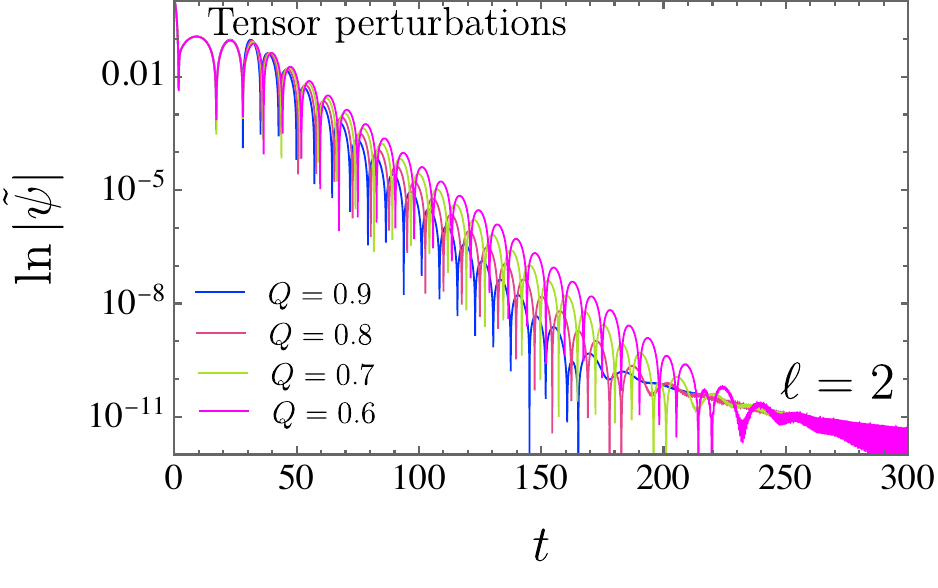}
    \caption{Time evolution of tensor--mode perturbations depicted through the logarithmic profile $\ln|\Tilde{\psi}|$ as a function of $t$, evaluated for charge values $Q = 0.6$, $0.7$, $0.8$, and $0.9$, with fixed parameters $\alpha = \beta = -0.01$. The plots correspond to angular indices $\ell = 1$ (left) and $\ell = 2$ (right).}
    \label{lnpsitensor}
\end{figure}

\begin{figure}
    \centering
    \includegraphics[scale=0.51]{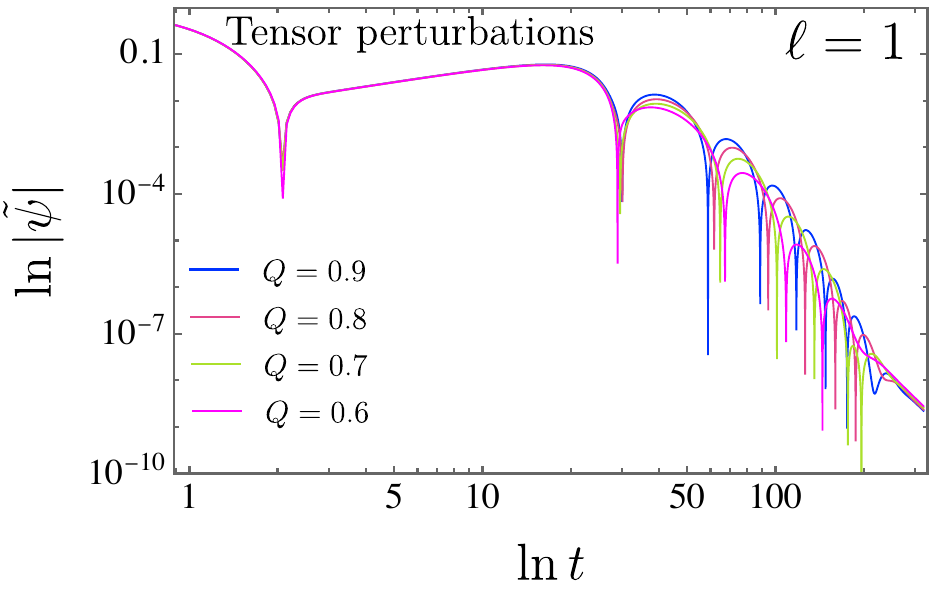}
    \includegraphics[scale=0.51]{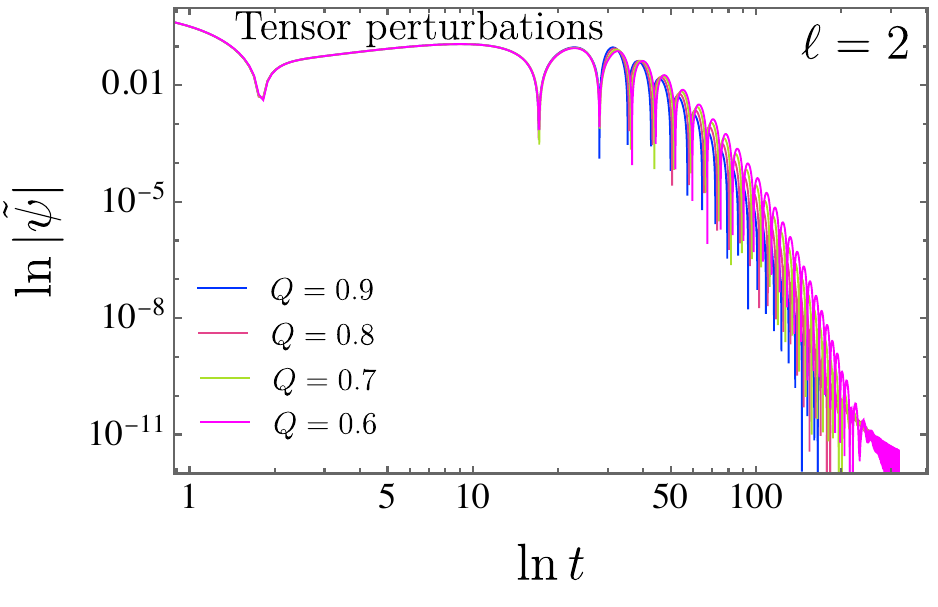}
    \caption{Late--time behavior of tensorial perturbations displayed through a $\ln|\Tilde{\psi}|$ versus $\ln|t|$ plot, where the coupling constants are fixed at $\alpha = \beta -0.01$ and the electric charge parameter varies over $Q = 0.6$, $0.7$, $0.8$, and $0.9$. The left and right panels correspond to angular momentum numbers $\ell = 1$ and $\ell = 2$.}
    \label{lnpsilntensor}
\end{figure}

This part addresses the temporal response of tensorial fluctuations under varying electric charge. The waveform $\Tilde{\psi}$ is exhibited as it evolves with time $t$, as shown in Fig. \ref{psitensor}, where the influence of different $Q$ values—specifically $0.6$, $0.7$, $0.8$, and $0.9$—is explored while the deformation parameters remain fixed at $\alpha = \beta = -0.01$. The plots correspond to $\ell = 1$ (left panel) and $\ell = 2$ (right panel). Furthermore, in order to show another detailed view of this decay process, we show Fig. \ref{lnpsitensor}, which it is displayed $\ln|\Tilde{\psi}|$ versus time $t$. To emphasize the asymptotic regime, Fig. \ref{lnpsilntensor} employs a double--logarithmic scale, plotting $\ln|\Tilde{\psi}|$ against $\ln|t|$ for the same combinations of $Q$ and angular index.

\section{Time--domain solution: fermionic case}

\begin{figure}
    \centering
    \includegraphics[scale=0.51]{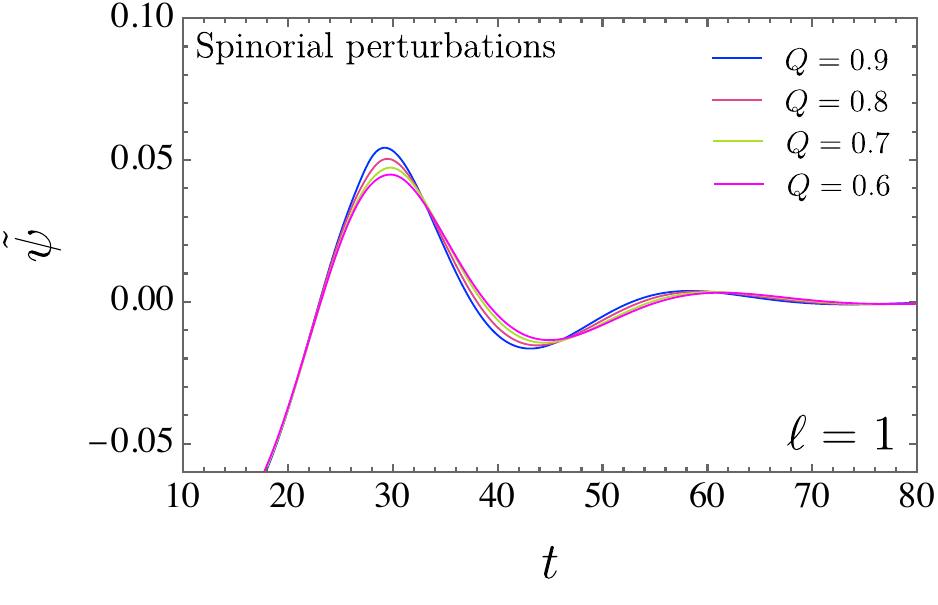}
    \includegraphics[scale=0.51]{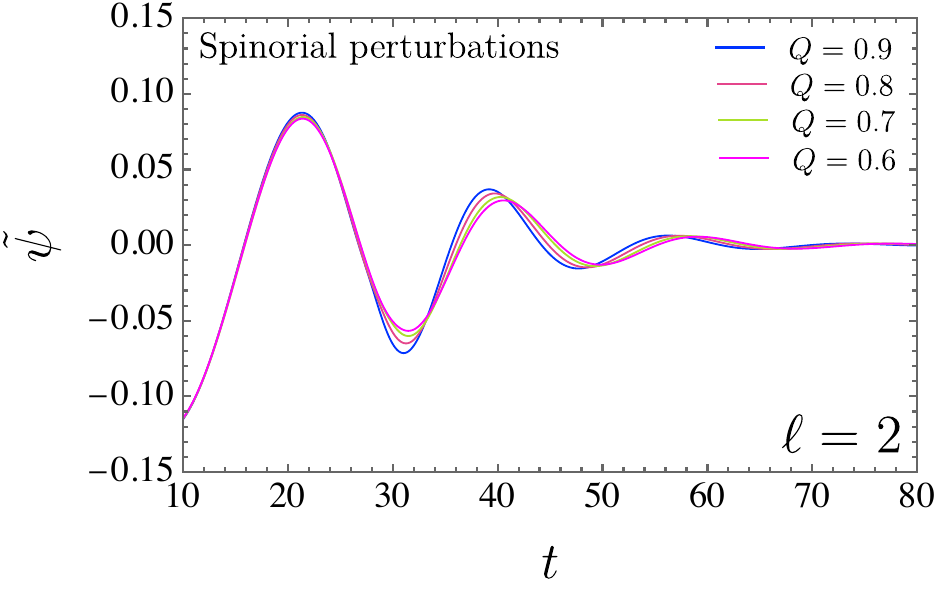}
    \caption{Temporal profiles of spinorial perturbations shown through the waveform $\Tilde{\psi}$ as a function of time $t$, with deformation parameters set to $\alpha = \beta = -0.01$ and charge values ranging from $Q = 0.6$ to $0.9$. Results for angular modes $\ell = 1$ and $\ell = 2$ are displayed in the left and right panels, respectively.}
    \label{psispin}
\end{figure}

\begin{figure}
    \centering
    \includegraphics[scale=0.51]{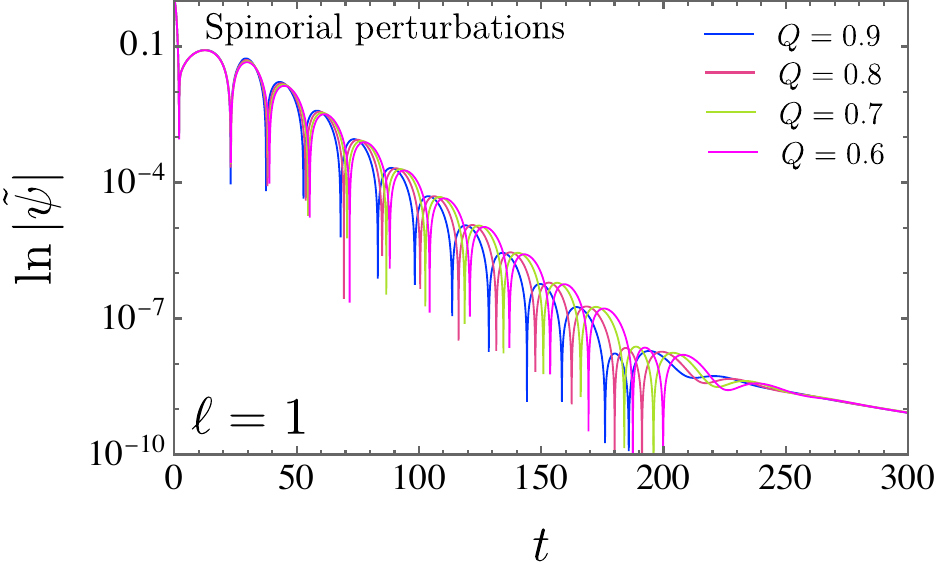}
    \includegraphics[scale=0.51]{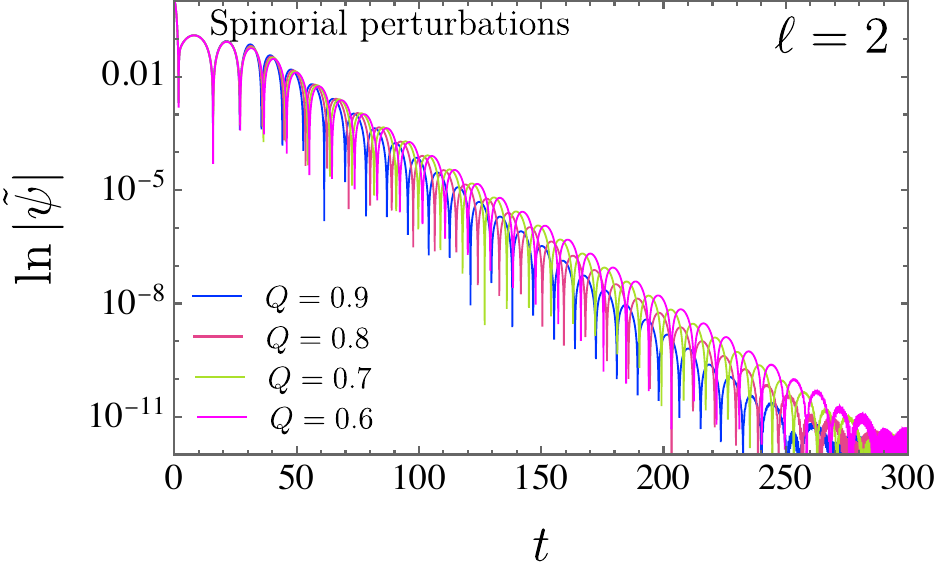}
    \caption{Logarithmic representation of spinorial perturbations, displaying $\ln|\Tilde{\psi}|$ versus time $t$ for fixed $\alpha = \beta = -0.01$ and varying $Q = 0.6$, $0.7$, $0.8$, and $0.9$. The left and right panels illustrate the behavior for angular momentum modes $\ell = 1$ and $\ell = 2$, respectively.}
    \label{lnpsispin}
\end{figure}

\begin{figure}
    \centering
    \includegraphics[scale=0.51]{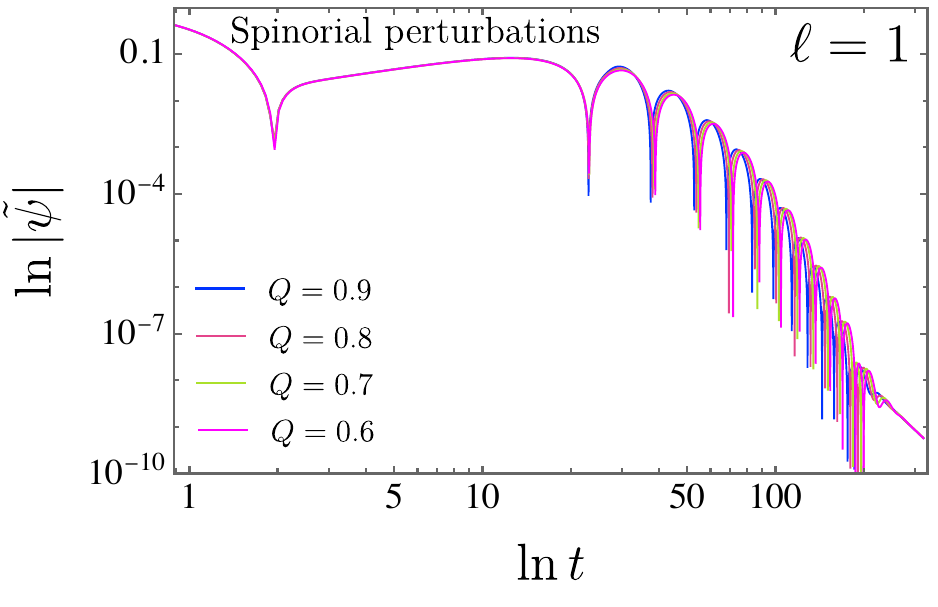}
    \includegraphics[scale=0.51]{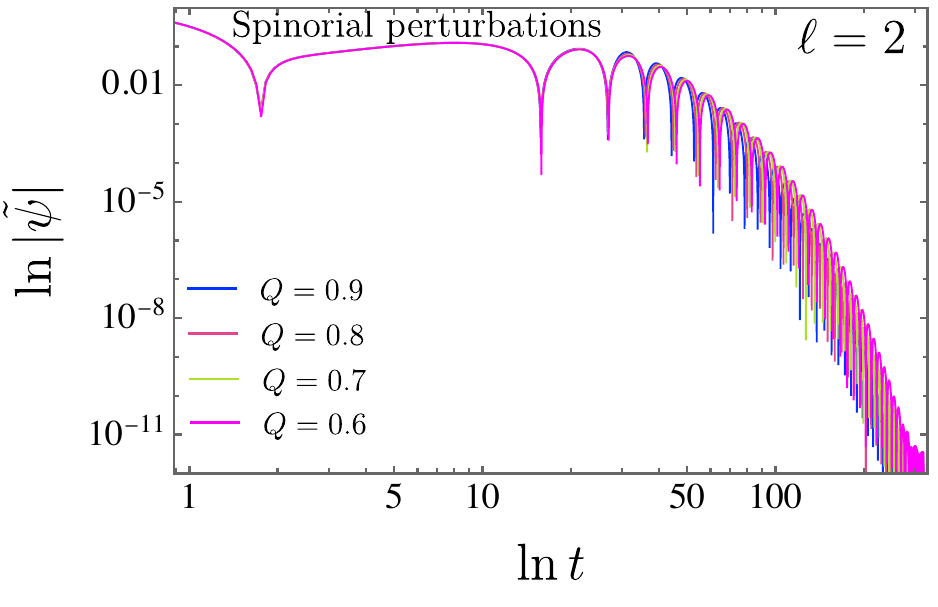}
    \caption{Late--time dynamics of spinor perturbations illustrated using a $\ln$--$\ln$ plot of $\ln|\Tilde{\psi}|$ against $\ln|t|$, with deformation parameters set to $\alpha = \beta = -0.01$ and charge values $Q = 0.6$, $0.7$, $0.8$, and $0.9$. Results for $\ell = 1$ and $\ell = 2$ are shown in the left and right panels, respectively.}
    \label{lnpsilnspin}
\end{figure}

Up to now,  we have examined the time evolution of scalar, vector, and tensor perturbations. To complete this analysis, we now turn our attention to the time--domain behavior of spinorial perturbations.
In this manner, Fig. \ref{psispin} shows the time evolution of the spinor perturbation $\Tilde{\psi}$ for fixed $\alpha = \beta = -0.01$ and varying $Q = 0.6$, $0.7$, $0.8$, and $0.9$. The left and right panels display results for $\ell = 1$ and $\ell = 2$, respectively, highlighting the gradual damping of the signal. In Fig. \ref{lnpsispin}, the decay is further analyzed through a $\ln$ plot of $\ln|\Tilde{\psi}|$ versus $t$. Finally, Fig. \ref{lnpsilnspin} presents a $\ln–\ln$ plot of $\ln|\Tilde{\psi}|$ against $\ln|t|$, confirming the emergence of power--law tails.

\pagebreak

\section{Weak field lensing regime}

In this section, we focus on the analysis of gravitational lensing within the framework of the weak deflection limit, employing the Gauss--Bonnet approach \cite{Gibbons:2008rj} to guide our investigation.

We begin by examining the stability of the photon spheres described in Eq. (\ref{photonsphererph}). To this end, we compute the Gaussian curvature, which is essential in assessing the nature of the critical orbits. As will be shown, the sign of the curvature determines the stability: positive curvature indicates stable orbits, while negative curvature corresponds to instability.

\subsection{Stability of the critical orbits}

The dynamics of photon rings near black holes are intimately tied to the curvature structure of the associated optical manifold, where the presence of conjugate points is essential in determining orbital stability. Rather than remaining indefinitely along circular trajectories, photons subjected to slight perturbations either spiral into the black hole or scatter away, depending on whether the photon orbit is unstable or stable. In stable cases, the photon remains in a localized region, circling near its original path \cite{qiao2022geometric,Heidari:2025iiv,qiao2022curvatures,AraujoFilho:2024rcr}.

This distinction can be analyzed through the lens of differential geometry. The emergence of conjugate points along null paths is directly influenced by the sign of the Gaussian curvature $\mathcal{K}(r)$, which encodes the intrinsic geometry of the effective optical surface. The Cartan--Hadamard theorem asserts that a non--positive curvature prohibits conjugate points, thereby implying instability, whereas a positive curvature allows them, pointing to the possibility of stable configurations \cite{qiao2024existence}. In this context, one investigates null trajectories constrained by the condition $\mathrm{d}s^2 = 0$, which can be reformulated as \cite{AraujoFilho:2024xhm}:
\ie
\mathrm{d}t^2=\gamma_{ij}\mathrm{d}x^i \mathrm{d}x^j = \frac{\mathrm{B}(r)}{\mathrm{A}(r)}\mathrm{d}r^2  +\frac{\Bar{\mathrm{D}}(r)}{\mathrm{A}(r)}\mathrm{d}\varphi^2.
\fe

In this formulation, indices $i$ and $j$ span the spatial coordinates $1$ through $3$, with $\gamma_{ij}$ referring to the components of the induced optical geometry. The function $\Bar{\mathrm{D}}(r)$ is defined as the equatorial restriction of the original metric function, i.e., $\Bar{\mathrm{D}}(r) \equiv \mathrm{D}(r, \theta = \pi/2)$. Thereby, the curvature governing the optical surface is characterized by the Gaussian curvature, given explicitly in \cite{qiao2024existence} as:
\ie
\mathcal{K}(r,\alpha,\beta,Q) = \frac{R}{2} =  -\frac{\mathrm{A}(r)}{\sqrt{\mathrm{B}(r) \,  \Bar{\mathrm{D}}(r)}}  \frac{\partial}{\partial r} \left[  \frac{\mathrm{A}(r)}{2 \sqrt{\mathrm{B}(r) \, \Bar{\mathrm{D}}(r) }}   \frac{\partial}{\partial r} \left(   \frac{\Bar{\mathrm{D}}(r)}{\mathrm{A}(r)}    \right)    \right],
\fe
with $R$ being for the Ricci scalar computed on the two--dimensional optical subspace. Now, taking into account small values of $\alpha$, $\beta$, and $Q$, we have
\ie
\begin{split}
\label{gdadudssdiadncdurdvadtdurde}
& \mathcal{K}(r,\alpha,\beta,Q) =   \frac{3 M^2}{r^4}-\frac{6 M Q^2}{r^5}+\frac{2 Q^4}{r^6} -\frac{2 M}{r^3}+\frac{9 \alpha  Q^6}{5 r^{10}}+\frac{3 Q^2}{r^4} -\frac{19 \alpha  M Q^4}{5 r^9}\\
& + \frac{38 \alpha  \beta  M Q^4}{5 r^9}+\frac{12 \alpha ^2 \beta ^2 Q^8}{25 r^{14}}-\frac{12 \alpha ^2 \beta  Q^8}{25 r^{14}}+\frac{3 \alpha ^2 Q^8}{25 r^{14}}-\frac{18 \alpha  \beta  Q^6}{5 r^{10}}-\frac{21 \alpha  \beta  Q^4}{5 r^8}+\frac{21 \alpha  Q^4}{10 r^8}.
\end{split}
\fe

Several studies \cite{qiao2024existence,Heidari:2025iiv,qiao2022curvatures,AraujoFilho:2024rcr,qiao2022geometric} have shown that the sign of the Gaussian curvature $\mathcal{K}(r, \alpha, \beta, Q)$ determines whether circular photon trajectories are dynamically stable or not. A positive value of $\mathcal{K}$ implies the presence of stable configurations, while a negative curvature indicates instability in the photon motion.

\begin{figure}
    \centering
    \includegraphics[scale=0.61]{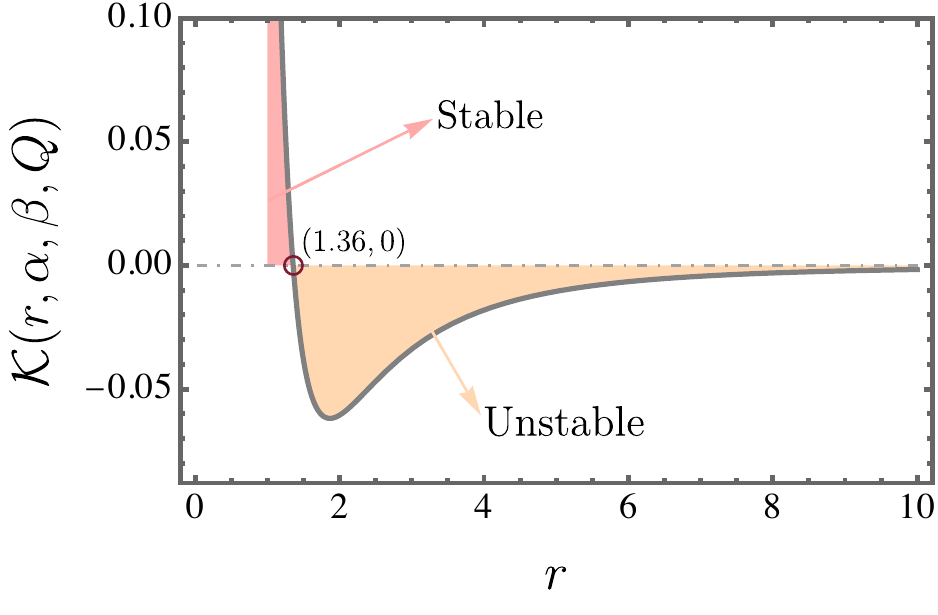}
    \caption{Profile of the Gaussian curvature $\mathcal{K}(r, \alpha, \beta, Q)$ plotted for $M = 1$, $\alpha = \beta = -0.01$, and $Q = 0.5$. Also, the transition between stability and instability regions for photon orbits is represented by the wine circle at $(1.36,0)$.}
    \label{gafugshsjcujrvjatjujrje}
\end{figure}

To examine this behavior in more detail, Fig. \ref{gafugshsjcujrvjatjujrje} presents the profile of $\mathcal{K}(r, \alpha, \beta, Q)$ as a function of the radial coordinate $r$, using the representative values $M = 1$, $\alpha = \beta = -0.01$, and $Q = 0.5$. The curvature curve clearly separates regions of stable (light pink) and unstable (light orange) photon motion. Notably, the graph features a transition point at $r = 1.36$, where the curvature crosses zero. This point marks the boundary between stability and instability: photon paths are confined and stable for $r < 1.36$, but lose stability beyond that. However, since the actual photon sphere lies beyond this transition point, the black hole spacetime under consideration supports only unstable circular photon orbits.

\subsection{Weak deflection angle}

Starting from the expression for the Gaussian curvature obtained in Eq. (\ref{gdadudssdiadncdurdvadtdurde}), we proceed to calculate the light deflection angle in the weak--field regime by applying the Gauss--Bonnet theorem \cite{Gibbons:2008rj}. As part of this procedure, the optical domain is restricted to the equatorial plane, and the corresponding surface element is given by:
\ie
\mathrm{d}S = \sqrt{\gamma} \, \mathrm{d} r \mathrm{d}\varphi = \sqrt{\frac{\mathrm{B}(r)}{\mathrm{A}(r)}  \frac{\mathrm{D}(r)}{\mathrm{A}(r)} } \, \mathrm{d} r \mathrm{d}\varphi.
\fe

Now, based on above expression, the deflection angle can properly be determined
\ie
\begin{split}
\label{lensing}
& \hat{\alpha} (b,\alpha,\beta,Q) =  - \int \int_{D} \mathcal{K} \mathrm{d}S = - \int^{\pi}_{0} \int^{\infty}_{\frac{b}{\sin \varphi}} \mathcal{K} \mathrm{d}S \\
& \simeq  \, \frac{4 M}{b} + \frac{3 \pi  M^2}{4 b^2} + \frac{525 \pi  M^2 Q^4}{256 b^6}+\frac{12 M Q^4}{5 b^5}-\frac{45 \pi  M^2 Q^2}{32 b^4}+\frac{15 \pi  Q^4}{64 b^4}-\frac{8 M Q^2}{3 b^3}-\frac{3 \pi  Q^2}{4 b^2}\\
& +\frac{189 \pi  \alpha  \beta  M^2 Q^4}{512 b^8}-\frac{189 \pi  \alpha  M^2 Q^4}{1024 b^8}+\frac{128 \alpha  \beta  M Q^4}{175 b^7}-\frac{64 \alpha  M Q^4}{175 b^7}+\frac{7 \pi  \alpha  \beta  Q^4}{32 b^6}-\frac{7 \pi  \alpha  Q^4}{64 b^6}.
\end{split}
\fe

The decomposition of Eq. (\ref{lensing}) reveals distinct contributions to the light deflection angle. The first pair of terms on the second line reproduce the bending behavior characteristic of the Schwarzschild solution. When the next set of terms—from the third to the eighth—is included, the result aligns with the deflection predicted in the Reissner--Nordström geometry. Additional corrections ascribed to the nonlinear electrodynamics sector, controlled by the parameters $\alpha$ and $\beta$, appear in the final segment of the expression.

\begin{figure}
    \centering
    \includegraphics[scale=0.51]{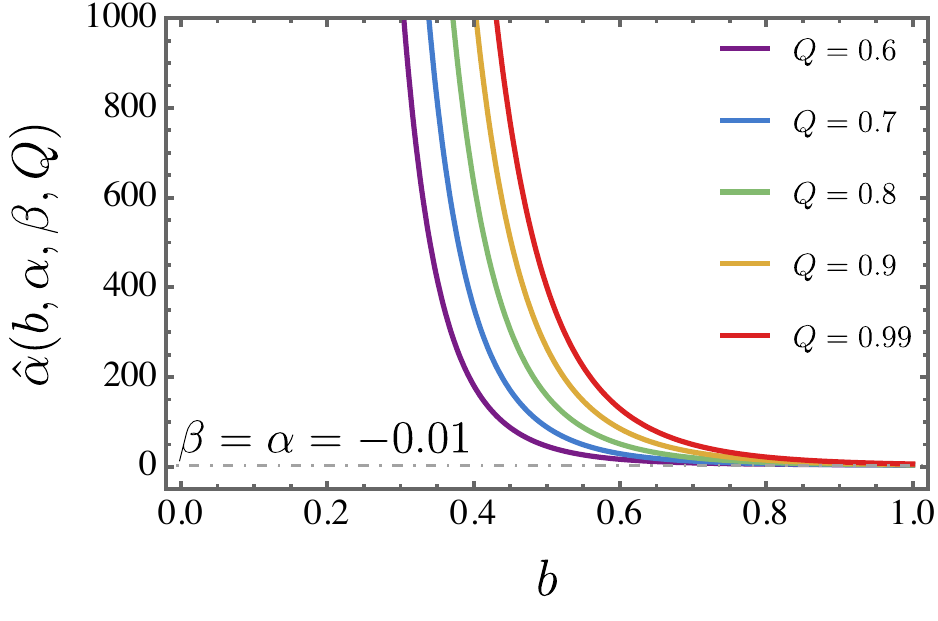}
    \includegraphics[scale=0.51]{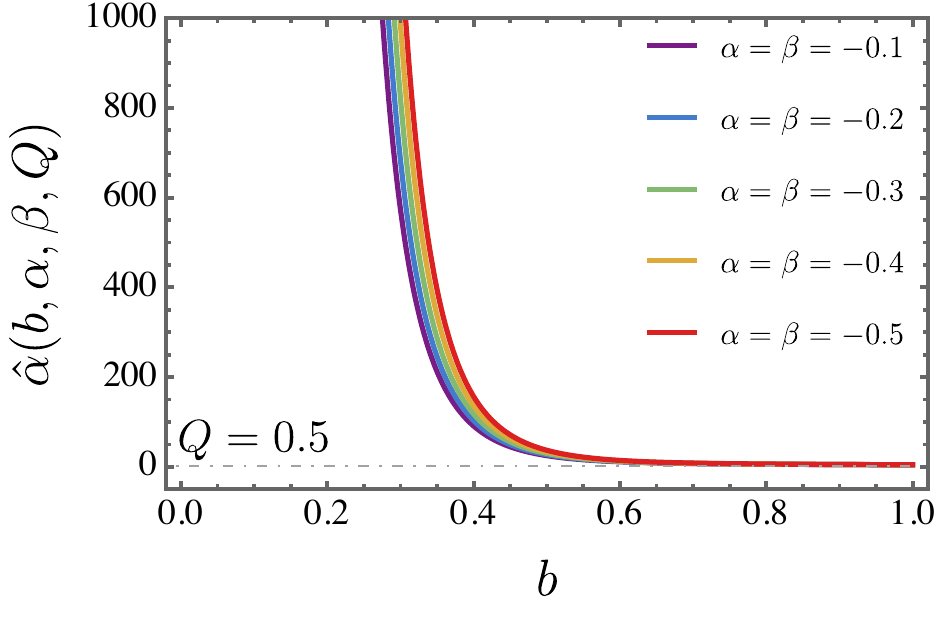}
    \caption{Deflection angle $\hat{\alpha}(b, \alpha, \beta, Q)$ is shown versus the impact parameter $b$, for various choices of the charge $Q$ and nonlinear parameters $\alpha$ and $\beta$.}
    \label{deflecccc}
\end{figure}

In Fig. \ref{deflecccc}, the response of the deflection angle $\hat{\alpha}(b, \alpha, \beta, Q)$ is examined as a function of the black hole parameters. For a constant impact parameter ($b = 0.5$), an increase in electric charge $Q$ leads to a stronger deflection when the nonlinear coefficients are fixed at $\alpha = \beta = -0.01$. Likewise, lowering the values of $\alpha$ and $\beta$ amplifies the bending angle.

\section{Strong field lensing regime}

The derivation of the light deflection angle in the strong field regime is outlined in this part of the analysis. Following the approach adopted in several recent studies (such as \cite{nascimento2024gravitational, heidari2024absorption,araujo2024effects}), the calculation is developed under the assumption of a static, spherically symmetric background that approaches flatness at large distances. Specifically, the geometry is described by \cite{tsukamoto2017deflection}
\ie
\mathrm{d}s^{2} = - \mathrm{A}(r) \mathrm{d}t^{2} + \mathrm{B}(r) \mathrm{d}r^{2} + \mathrm{C}(r)(\mathrm{d}\theta^2 + \sin^{2}\theta\mathrm{d}\phi^2).
\fe

In order to implement the analytical framework outlined by Tsukamoto \cite{tsukamoto2017deflection}, one must ensure the background geometry adheres to the condition of asymptotic flatness. This implies that the metric functions must exhibit specific limits at spatial infinity: $\mathrm{A}(r)$ and $\mathrm{B}(r)$ should both converge to unity, while $\mathrm{C}(r)$ must asymptotically behave as $r^2$, i.e.,
\ie
\lim_{r \to \infty} \mathrm{A}(r) = 1, \quad \lim_{r \to \infty} \mathrm{B}(r) = 1, \quad \lim_{r \to \infty} \mathrm{C}(r) = r^2.
\nonumber
\fe

The process for deriving the deflection angle in the strong field approximation starts by redefining the radial dependence through the introduction of a suitable auxiliary quantity, $\Tilde{\Bar{D}}(r)$, which facilitates the handling of divergences near the photon sphere and streamlines the analytical treatment of the light trajectory
\ie
\Tilde{\Bar{D}}(r) \equiv \frac{\mathrm{C}^{\prime}(r)}{\mathrm{C}(r)} - \frac{\mathrm{A}^{\prime}(r)}{\mathrm{A}(r)},
\fe
in which derivatives with respect to the radial coordinate are denoted by primes. The function $\Tilde{\Bar{D}}(r)$ is assumed to vanish at least once for a positive value of $r$. Among its real positive roots, the photon sphere is identified by the largest one, labeled $r_{ph}$. For the formalism to be consistent in this domain, it is required that the metric components $\mathrm{A}(r)$, $\mathrm{B}(r)$, and $\mathrm{C}(r)$ remain regular and strictly greater than zero for all $r$ equal to or exceeding $r_{ph}$.

Due to the underlying symmetries of the spacetime, specifically time translation and axial invariance, two conserved quantities naturally emerge along null geodesics: the energy, given by $E = \mathrm{A}(r)\Dot{t}$, and the angular momentum, expressed as $L = \mathrm{C}(r)\Dot{\phi}$. Assuming both $E$ and $L$ take non-vanishing values, one can define the impact parameter $b$ as the ratio between them:
\ie
b \equiv \frac{L}{E} = \frac{\mathrm{C}(r)\Dot{\phi}}{\mathrm{A}(r)\Dot{t}}.
\fe

Notice that by considering the rotational symmetry of the spacetime, the analysis can be simplified by confining the trajectory to the equatorial plane, setting $\theta = \pi/2$ without loss of generality. Under this condition, the equation governing the radial motion of photons reduces to the following form:
\ie
\Dot{r}^{2} = V(r).
\fe
Let us consider the effective potential
$V(r) = \frac{L^2\, \mathrm{R}(r)}{\mathrm{B}(r)\, \mathrm{C}(r)},$
with
$\mathrm{R}(r) = \frac{\mathrm{C}(r)}{\mathrm{A}(r)\, b^2} - 1.$
This relation resembles the equation of motion for a massless particle under a radial potential. The condition $V(r) \geq 0$ determines the permitted region of photon motion. Since the spacetime is asymptotically flat, we have $\lim_{r \to \infty} V(r) = E^2 > 0$, allowing the photon to reach infinity. Moreover, it is considered that $\mathrm{R}(r) = 0$ possess one positive real defined (at least) solution.

In gravitational lensing analysis, the photon path under consideration starts at infinity, moves toward the central object, reaches its nearest distance $r_0$, and then returns to infinity. This closest approach $r_0$ must be greater than the photon sphere radius $r_{ph}$, which excludes circular trajectories. The value $r_0$ is the largest positive solution to $\mathrm{R}(r) = 0$, where both $\mathrm{B}(r)$ and $\mathrm{C}(r)$ remain regular. At this point, the effective potential $V(r)$ vanishes, making $\mathrm{R}(r_0) = 0$ a critical condition for the analysis
\ie
\mathrm{A}_{0}\Dot{t}^{2}_{0} = \mathrm{C}_{0}\Dot{\phi}^{2}_{0}.
\fe

From this point onward, all quantities labeled with the subscript “$0$” refer to their values at the radial coordinate $r = r_0$. When examining the propagation of a single photon, it is convenient—and physically justified—to restrict the impact parameter $b$ to positive values. Since $b$ is conserved along the photon's path, it admits the following representation:
\ie
b(r_{0}) = \frac{L}{E} = \frac{\mathrm{C}_{0}\Dot{\phi}_{0}}{\mathrm{A}_{0}\Dot{t}_{0}} = \sqrt{\frac{\mathrm{C}_{0}}{\mathrm{A}_{0}}}.
\fe
Note that an alternative form of $\mathrm{R}(r)$ can be written as
\ie
\mathrm{R}(r)= \frac{\mathrm{A}_{0}\mathrm{C}}{\mathrm{A}\mathrm{C}_{0}} - 1.
\fe

A necessary and sufficient criterion for the existence of a circular null geodesic can be established by following the methodology presented in Ref. \cite{hasse2002gravitational}. Under this framework, the trajectory equation takes the form
\ie
\frac{\mathrm{B}\,\mathrm{C}\, \Dot{r}^{2}}{E^{2}} + b^{2} = \frac{\mathrm{C}}{\mathrm{A}},
\fe
so that
\ie
\ddot{r} + \frac{1}{2}\left( \frac{\mathrm{B}^{\prime}}{\mathrm{B}} + \frac{\mathrm{C}^{\prime}}{\mathrm{C}} \Dot{r}^{2} \right) = \frac{E^{2}\Tilde{\Bar{D}}(r)}{\mathrm{A}\mathrm{B}}. 
\fe
Assuming $r \geq r_{ph}$, the metric functions $\mathrm{A}(r)$, $\mathrm{B}(r)$, and $\mathrm{C}(r)$ remain regular and strictly positive. Since the energy $E$ is also positive, the condition $\Tilde{\Bar{D}}(r) = 0$ characterizes the existence of a stable circular photon trajectory. Moreover, one finds that the derivative $\mathrm{R}'(r)$ evaluated at the photon sphere satisfies
$\mathrm{R}'_{ph} = \frac{\Tilde{\Bar{D}}_{ph}\, \mathrm{C}_{ph}\, \mathrm{A}_{ph}}{b^2} = 0,$
with the index “ph” being the label to describe the evaluation at $r = r_{ph}$.

We next examine the critical impact parameter, represented by $b_c$:
\ie
b_{c}(r_{ph}) \equiv \lim_{r_{0} \to r_{ph}} \sqrt{\frac{\mathrm{C}_{0}}{\mathrm{A}_{0}}}.
\fe
This regime will be referred to as the strong deflection limit. Differentiating the effective potential $V(r)$ with respect to the radial coordinate yields
\ie
V^{\prime}(r) = \frac{L^{2}}{\mathrm{B}\mathrm{C}} \left[ \mathrm{R}^{\prime} + \left( \frac{\mathrm{C}^{\prime}}{\mathrm{C}} - \frac{\mathrm{B}^{\prime}}{\mathrm{B}}   \right)   \mathrm{R}  \right].
\fe
In the strong deflection limit, when $r_0$ tends toward $r_{ph}$, both the potential $V(r_0)$ and its radial derivative $V'(r_0)$ approach zero. Accordingly, the equation governing the trajectory takes the form
\ie
\left(  \frac{\mathrm{d}r}{\mathrm{d}\phi}     \right)^{2} = \frac{\mathrm{R}(r)\mathrm{C}(r)}{\mathrm{B}(r)}.
\fe
In this manner, the corresponding deflection angle, denoted by $\alpha(r_0)$, is therefore expressed as
\ie
\alpha(r_{0}) = I(r_{0}) - \pi,
\fe
in which $I(r_{0})$ reads
\ie
I(r_{0}) \equiv 2 \int^{\infty}_{r_{0}} \frac{\mathrm{d}r}{\sqrt{\frac{\mathrm{R}(r)\mathrm{C}(r)}{\mathrm{B}(r)}}}.
\fe

As a starting point, we must address the integral involved—an analytically demanding task, as highlighted by Tsukamoto in Ref. \cite{tsukamoto2017deflection}. Additionally, we adopt the following definition from \cite{tsukamoto2017deflection}:
\ie
z \equiv 1 - \frac{r_{0}}{r},
\fe
which allows the integral to be reformulated as
\ie
I(r_{0}) = \int^{1}_{0} f(z,r_{0}) \mathrm{d}z,
\fe
where we have 
\ie
f(z,z_{0}) \equiv \frac{2r_{0}}{\sqrt{G(z,r_{0})}}, \,\,\,\,\,\,\,\, \text{and} \,\,\,\,\,\,\,\,  G(z,r_{0}) \equiv \mathrm{R} \frac{\mathrm{C}}{\mathrm{B}}(1-z)^{4}.
\fe

On the other hand, expressed in terms of the variable $z$, the function $\mathrm{R}(r)$ takes the form
\ie
\mathrm{R}(r) = \Tilde{{\Bar{D}}}_{0} \, r_{0} z + \left[ \frac{r_{0}}{2}\left( \frac{\mathrm{C}^{\prime\prime}_{0}}{\mathrm{C}_{0}} - \frac{\mathrm{A}_{0}^{\prime\prime}}{\mathrm{A}_{0}}  \right) + \left( 1 - \frac{\mathrm{A}_{0}^{\prime}r_{0}}{\mathrm{A}_{0}}  \right) \Tilde{{\Bar{D}}}_{0}  \right] r_{0} z^{2} + \mathcal{O}(z^{3})+ ...    \,\,\,\,.
\fe
By performing a series expansion of $G(z, r_0)$ around $z = 0$, it turns out:
\ie
G(z,r_{0}) = \sum^{\infty}_{n=1} c_{n}(r_{0})z^{n},
\fe
with $c_{1}(r)$ and $c_{2}(r)$ being, respectively
\ie
c_{1}(r_{0}) = \frac{\mathrm{C}_{0}\Tilde{\Bar{D}}_{0}r_{0}}{\mathrm{B}_{0}},
\fe
and
\ie
c_{2}(r_{0}) = \frac{\mathrm{C}_{0}r_{0}}{\mathrm{B}_{0}} \left\{ \Tilde{\Bar{D}}_{0} \left[ \left( \Tilde{\Bar{D}}_{0} - \frac{\mathrm{B}^{\prime}_{0}}{\mathrm{B}_{0}}  \right)r_{0} -3       \right] + \frac{r_{0}}{2} \left(  \frac{\mathrm{C}^{\prime\prime}_{0}}{\mathrm{C}_{0}} - \frac{\mathrm{A}^{\prime\prime}_{0}}{\mathrm{A}_{0}}  \right)                 \right\}.
\fe

Furthermore, under the strong deflection limit approximation, one finds that
\ie
c_{1}(r_{ph}) = 0, \,\,\,\,\,\, \text{and} \,\,\,\,\,\, c_{2}(r_{ph}) =  \frac{\mathrm{C}_{ph}r^{2}_{ph}}{2 \mathrm{B}_{ph}}\Tilde{\Bar{D}}^{\prime}_{ph}, \,\,\,\,\,\,\, \text{with} \,\,\,\,\, \Tilde{\Bar{D}}^{\prime}_{ph} = \frac{\mathrm{C}^{\prime\prime}}{\mathrm{C}_{ph}} - \frac{\mathrm{A}^{\prime\prime}}{\mathrm{A}_{ph}},
\fe
where $G(z, r_0)$ is expressed using a more compact notation, given by:
\ie
G_{ph}(z) = c_{2}(r_{ph})z^{2} + \mathcal{O}(z^{3}).
\fe

As $r_0$ approaches the photon sphere radius $r_{ph}$, the function $f(z, r_0)$ becomes singular, with its dominant divergence scaling as $1/z$. This behavior induces a logarithmic blow--up in the integral $I(r_0)$. To systematically address this divergence, the integral is separated into two distinct contributions: one containing the divergent behavior, labeled $I_D(r_0)$, and the other remaining finite, denoted by $I_R(r_0)$. The term capturing the divergence, $I_D(r_0)$, is therefore represented in the form:
\ie
I_{D}(r_{0}) \equiv \int^{1}_{0} f_{D}(z,r_{0}) \mathrm{d}z, \,\,\,\,\,\,\, \text{with} \,\,\,\,\,\,f_{D}(z,r_{0}) \equiv \frac{2 r_{0}}{\sqrt{c_{1}(r_{0})z + c_{2}(r_{0})z^{2}}}.
\fe
Upon performing the integration, the result takes the form
\ie
I_{D} (r_{0}) = \frac{4 r_{0}}{\sqrt{c_{2}(r_{0})}} \ln \left[  \frac{\sqrt{c_{2}(r_{0})} + \sqrt{c_{1}(r_{0}) + c_{2}(r_{0})     }  }{\sqrt{c_{1}(r_{0})}}  \right].
\fe

Expanding $c_1(r_0)$ and $b(r_0)$ in a Taylor series around the point $r_0 = r_{ph}$ yields:
\ie
c_{1}(r_{0}) = \frac{\mathrm{C}_{ph}r_{ph}\Tilde{\Bar{D}}^{\prime}_{ph}}{\mathrm{B}_{ph}} (r_{0}-r_{ph}) + \mathcal{O}((r_{0}-r_{ph})^{2}),
\fe
and
\ie
b(r_{0}) = b_{c}(r_{ph}) + \frac{1}{4} \sqrt{\frac{\mathrm{C}_{ph}}{\mathrm{A}_{ph}}}  \Tilde{\Bar{D}}^{\prime}_{ph}(r_{0}-r_{ph})^{2} + \mathcal{O}((r_{0}-r_{ph})^{3}),
\fe
resulting in the expression below when approaching the strong deflection regime:
\ie
\lim_{r_{0} \to r_{ph}} c_{1}(r_{0})  =  \lim_{b \to b_{c}} \frac{2 \mathrm{C}_{ph} r_{ph} \sqrt{\Tilde{{\Bar{D}}}^{\prime}}}{\mathrm{B}_{ph}} \left(  \frac{b}{b_{c}} -1  \right)^{1/2}.
\fe
In this way, we can properly write $I_{D}(b)$ as shown below
\ie
I_{D}(b) = - \frac{r_{ph}}{\sqrt{c_{2}(r_{ph})}} \ln\left[ \frac{b}{b_{c}} - 1 \right] + \frac{r_{ph}}{\sqrt{c_{2}(r_{ph})}}\ln \left[ r^{2}\Tilde{{\Bar{D}}}^{\prime}_{ph}\right] + \mathcal{O}[(b-b_{c})\ln(b-b_{c})].
\fe

In addition, we define the regular contribution $I_{R}(b)$ as
\ie
I_{R}(b) = \int^{0}_{1} f_{R}(z,b_{c})\mathrm{d}z + \mathcal{O}[(b-b_{c})\ln(b-b_{c})].
\fe
Define $f_{R}$ as the difference $f_{R} = f(z, r_{0}) - f_{D}(z, r_{0})$. Under the strong deflection approximation, the deflection angle takes the form
\ie
\label{sftgrgonadsgsddefdsdledccxccxxxcc}
a(b) = - \Tilde{a} \ln \left[ \frac{b}{b_{c}}-1    \right] + \Tilde{b} + \mathcal{O}[(b-b_{c})\ln(b-b_{c})],
\fe
where the following parameters are considered
\ie
\Tilde{a} = \sqrt{\frac{2 \mathrm{B}_{ph}\mathrm{A}_{ph}}{\mathrm{C}^{\prime\prime}_{ph}\mathrm{A}_{ph} - \mathrm{C}_{ph}\mathrm{A}^{\prime\prime}_{ph}}}, \,\,\,\,\,\,\,\, \text{and} \,\,\,\,\,\,\,\, \Tilde{b} = \Tilde{a} \ln\left[ r^{2}_{ph}\left( \frac{\mathrm{C}^{\prime\prime}}{\mathrm{C}_{ph}}  -  \frac{\mathrm{A}^{\prime\prime}_{ph}}{\mathrm{C}_{ph}} \right)   \right] + I_{R}(r_{ph}) - \pi.
\fe


\subsection{Gravitational lensing of a black hole with a modified electrodynamics}

Having established the general framework, we now apply this procedure to the specific spacetime described by the metric in Eq. (\ref{metricfuntion}). This yields:
\ie
b_{c} = 3 \sqrt{3} M  -\frac{\sqrt{3} Q^2}{2 M} -\frac{7 Q^4}{24 \left(\sqrt{3} M^3\right)} +  \frac{\alpha  \left((2 \beta -1) Q^4\right)}{540 \sqrt{3} M^5}.
\fe
In addition, the quantities $\Tilde{a}$ and $\Tilde{b}$ can be written as
\ie
\Tilde{a} = 1 + \frac{Q^2}{9 M^2} +\frac{11 Q^4}{162 M^4}   +\frac{(1-2 \beta ) Q^4}{729 M^6}\alpha
\fe
Accordingly, we may express it as
\ie
\begin{split}
\Tilde{b} = & \left(1 +\frac{Q^2}{9 M^2} +\frac{11 Q^4}{162 M^4} + \frac{\alpha  (1-2 \beta ) Q^4}{729 M^6}\right) \left(\ln[6] -\frac{Q^2}{9 M^2} -\frac{11 Q^4}{162 M^4} + \frac{\alpha  (2 \beta -1) Q^4}{486 M^6}\right)\\
& + I_{R}(r_{ph}) - \pi.
\end{split}
\fe

Differing from the Schwarzschild case, the parameter $\Tilde{a}$ is predominantly shaped by the contributions arising from the nonlinear electrodynamics. Additionally, the regular integral $I_{R}(r_{ph})$ takes the form
\ie
\begin{split}
 & I_{R}(r_{ph}) =   \int_{0}^{1} \mathrm{d}z \left\{   \frac{2}{\sqrt{1-\frac{2 z}{3}} z} -\frac{2}{z} -\frac{11 Q^4}{81 M^4 z}-\frac{2 Q^2}{9 M^2 z} +\frac{Q^2 \left(3 \sqrt{M z^5}-8 \sqrt{M z^3}+6 \sqrt{M z}\right)}{3 \sqrt{3} M^{5/2} ((3-2 z) z)^{3/2}} \right. \\
 & \left. +\frac{Q^4 \left(27 \sqrt{M z^{11}}-240 \sqrt{M z^9}+700 \sqrt{M z^7}-864 \sqrt{M z^5}+396 \sqrt{M z^3}\right)}{108 \sqrt{3} M^2 (M (3-2 z) z)^{5/2}}   \right. \\
 & \left.  +\frac{2 \alpha  (2 \beta -1) Q^4}{729 M^6 z} +    \frac{\left(\alpha  (2 \beta -1) Q^4 (z (3 z (z (z ((z-8) z+28)-56)+70)-160)+60)\right)}{2430 M^6 \sqrt{9-6 z} \, z (2 z-3)}      \right\} \\
 & = 2 \ln \left[6 \left(2-\sqrt{3}\right)\right]  +\frac{Q^4 \left(19 \sqrt{3}-53+44 \ln \left[3-\sqrt{3}\right]\right)}{162 M^4} \\
 & -\frac{\alpha  (2 \beta -1) Q^4 \left(163 \sqrt{3}-603+420 \ln \left[3-\sqrt{3}\right]\right)}{76545 M^6} \\
 & +\frac{Q^2 \left(2 \sqrt{3}-5+4 \ln[6]-4 \ln \left[\sqrt{3}+3\right]\right)}{9 M^2}.
\end{split}
\fe
Notably, this leads to an exact analytical expression. With this result, the deflection angle in the strong deflection regime can be determined using Eq. (\ref{sftgrgonadsgsddefdsdledccxccxxxcc}), yielding the following form:
\ie
\begin{split}
& \hat{\alpha}^{\text{s}} (b,\alpha,\beta,Q)  =  -\left( 1 + \frac{Q^2}{9 M^2} +\frac{11 Q^4}{162 M^4}   +\frac{(1-2 \beta ) Q^4}{729 M^6}\alpha \right)  \\
& \times \, \ln \left[ \frac{b}{3 \sqrt{3} M  -\frac{\sqrt{3} Q^2}{2 M} -\frac{7 Q^4}{24 \left(\sqrt{3} M^3\right)} +  \frac{\alpha  \left((2 \beta -1) Q^4\right)}{540 \sqrt{3} M^5}} - 1  \right] \\
& +  2 \ln \left[6 \left(2-\sqrt{3}\right)\right]  +\frac{Q^4 \left(19 \sqrt{3}-53+44 \ln \left[3-\sqrt{3}\right]\right)}{162 M^4} \\
 & -\frac{\alpha  (2 \beta -1) Q^4 \left(163 \sqrt{3}-603+420 \ln \left[3-\sqrt{3}\right]\right)}{76545 M^6} \\
 & +\frac{Q^2 \left(2 \sqrt{3}-5+4 \ln[6]-4 \ln \left[\sqrt{3}+3\right]\right)}{9 M^2} - \pi  \\
 & + \left(1 +\frac{Q^2}{9 M^2} +\frac{11 Q^4}{162 M^4} + \frac{\alpha  (1-2 \beta ) Q^4}{729 M^6}\right) \left(\ln[6] -\frac{Q^2}{9 M^2} -\frac{11 Q^4}{162 M^4} + \frac{\alpha  (2 \beta -1) Q^4}{486 M^6}\right)\\
 & + \mathcal{O}\Bigg\{\Bigg[b- \left( 3 \sqrt{3} M  -\frac{\sqrt{3} Q^2}{2 M} -\frac{7 Q^4}{24 \left(\sqrt{3} M^3\right)} +  \frac{\alpha  \left((2 \beta -1) Q^4\right)}{540 \sqrt{3} M^5} \right) \Bigg]  \\
& \times \, \ln\Bigg[b- \left( 3 \sqrt{3} M  -\frac{\sqrt{3} Q^2}{2 M} -\frac{7 Q^4}{24 \left(\sqrt{3} M^3\right)} +  \frac{\alpha  \left((2 \beta -1) Q^4\right)}{540 \sqrt{3} M^5} \right) \Bigg] \Bigg\}.
\end{split}
\fe

\begin{figure}
    \centering
     \includegraphics[scale=0.51]{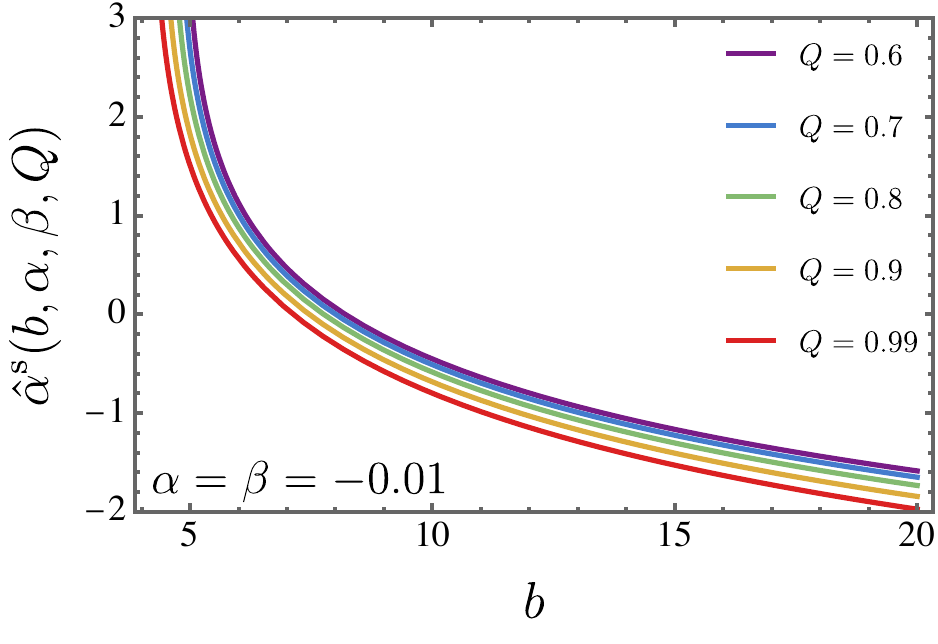}
    \includegraphics[scale=0.51]{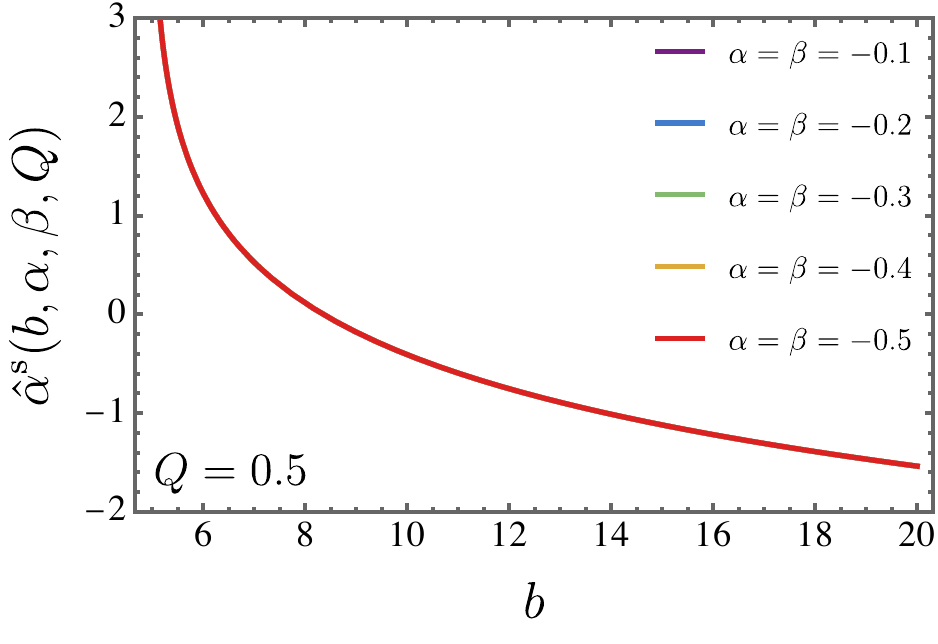}
    \caption{The deflection angle $\hat{\alpha}^{\text{s}} (b,\alpha,\beta,Q)$ as a function of $b$ for different values of $Q$, $\alpha$ and $\beta$.}
    \label{strongdeflec}
\end{figure}

To improve clarity, Fig. \ref{strongdeflec} displays the deflection angle as a function of $b$ for different system configurations. Overall, we observe that increasing the charge $Q$ leads to a decrease in $\hat{\alpha}^{\text{s}} (b,\alpha,\beta,Q)$. Note that this feature is consistent with the analysis accomplished in the geodesic analysis, Fig. \ref{geodesicslight}, where the light trajectory approaches the photon sphere. Furthermore, as one should expect, variations in $\alpha$ and $\beta$ produce only minor changes in the deflection angle.

To support the findings obtained so far within the strong deflection limit, the next subsection will focus on observable features, primarily using data from the EHT telescope.

\section{Topological features}


\subsection{Topological thermodynamics}

Recent advances in black hole thermodynamics have introduced a topological perspective as a powerful framework for analyzing critical points and phase transitions \cite{wei2022black,wu2025novel,yerra2022topology,wu2023topological,wu2023topological1,fang2023revisiting,zhang2023bulk,gogoi2023thermodynamic,fan2023topological,sadeghi2024thermodynamic,Wu:2023sue,Wu:2023fcw,Zhu:2024zcl,Afshar:2024bgi}. The main idea comes from Duan’s $\phi$--mapping topological current theory \cite{Duane1984}, which systematically assigns topological charges to critical points--where the phase transitions occur.

Building on this foundation, Wei et al. have developed the concept of thermodynamic topology \cite{wei2022black}, where in black holes are modeled as topological defects embedded within a thermodynamic parameter space. This method employs a scalar thermodynamic potential and constructs a corresponding two--dimensional vector field whose zero points indicate thermodynamic critical points. The total topological charge and the winding numbers around these points can be identified as both conventional and novel critical points. This topological framework not only extends the classification of black hole thermodynamics but also offers robust tools for studying stability and non--equilibrium phenomena in gravitational systems \cite{gashti2025thermodynamic,rathi2025topology,huang2025interaction,Gashti2025,EslamPanah2024}.

Now, based on the Hawking temperature calculated in Eq. \ref{ronly}, the thermodynamic scalar potential is introduced as 
\begin{equation}
    \Phi=\frac{1}{\sin\theta}T_H=\frac{\csc \theta  \left(\alpha  (2 \beta -1) Q^4-2 Q^2 r_h^4+2 r_h^6\right)}{8 \pi  r_h^7}.
\end{equation}
This scalar potential enables examination of thermodynamic critical points by generating a two-dimensional vector field in the radial and angular directions as $(\phi^{r_h},\phi^\theta)$, which are explicitly determined as follows
\begin{align}
    &\phi^{r_h}=\partial_{r_h}\Phi=\frac{\csc \theta  \left(7 \alpha  (1-2 \beta ) Q^4+6 Q^2 r_h^4-2 r_h^6\right)}{8 \pi  r_h^8},\\
    &\phi^{\theta}=\partial_{\theta}\Phi=\frac{\cot \theta  \csc \theta  \left(Q^4 (\alpha -2 \alpha  \beta )+2 Q^2 r_h^4-2 r_h^6\right)}{8 \pi  r_h^7}.
\end{align}
and the normalized vectors $n^{r_h}$ and $n^\theta$ are obtained by 
\begin{equation}
    n^{r_h}=\frac{\phi^{r_h}}{||\phi||},\quad n^{\theta}=\frac{\phi^{\theta}}{||\phi||}.
\end{equation}

\begin{figure}
    \centering
     \includegraphics[scale=0.55]{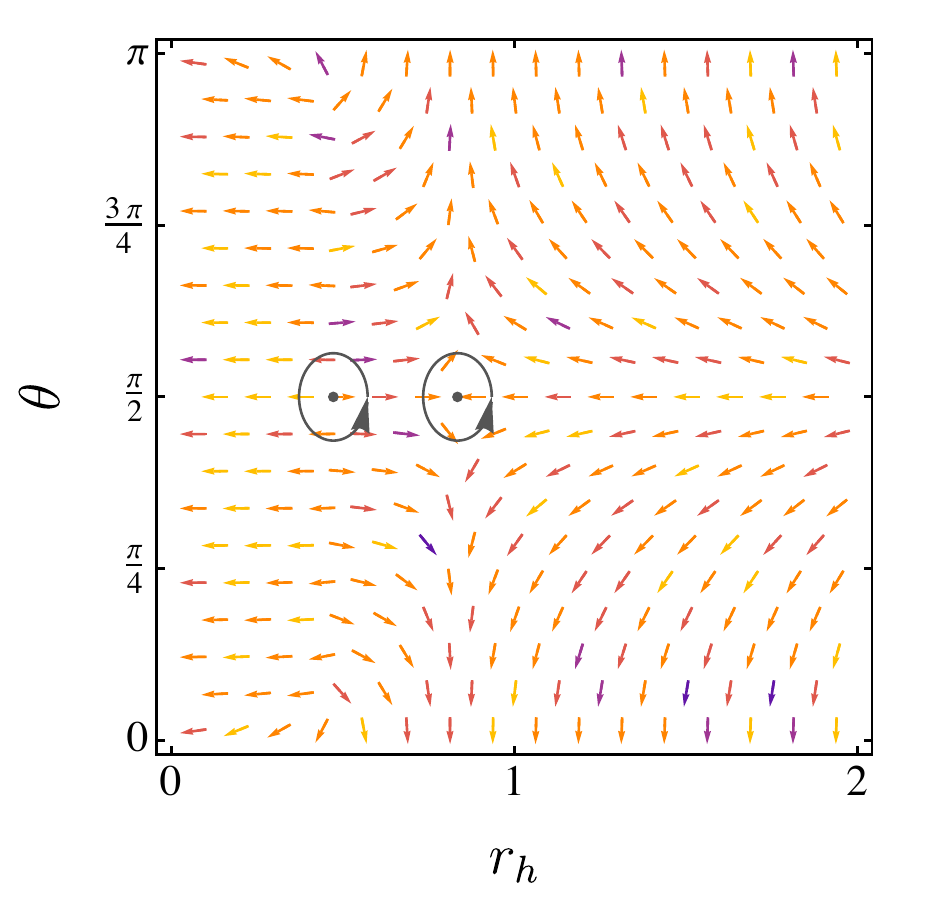}
     \caption{Normalized vector field of temperature potential in the plane of ($r_h, \theta$). The critical points $C_1$ and $C_2$ are located at $(0.472473,\frac{\pi}{2})$ and $(0.834131,\frac{\pi}{2})$, with fixed parameters $M = 1$, $Q = 0.5$, $\alpha = \beta = - 0.1$.  }
    \label{TopologicalT}
\end{figure}

The normalized vector field is illustrated in Fig. \ref{TopologicalT}, where the critical point is precisely identified as the zero of the vector field. In this case, $r_{C_1}=0.472473$ and $r_{C_2}=0.834131$ mark the location of the critical points. Two closed contours enclose these points to examine their topological characteristics, through which the topological charges of $+1$ and $-1$ correspond to critical points $C_1$ and $C_2$, respectively. Based on the classification introduced in Ref. \cite{wei2022black}, this result indicates that the critical points are of the novel and conventional type, respectively.

\subsection{Topological photon sphere}

Recently, the method of examining the stability or instability of the photonic sphere with the topological framework has gained significant attention \cite{Wei2020,Cunha2020,Sadeghi2024,BahrozBrzo2025,alipour2024weak}. Building upon the analysis of thermodynamic topology in the previous section, we extend our investigation to the topological structure of the photon radius. 

The potential for the topological model of the photonic sphere is defined as follows

\begin{align}\label{eq:Hr}
H(r, \theta) &=  \sqrt{\frac{\mathrm{A}(r)}{\mathrm{D}(r)}}\\
&=\frac{\csc \theta  \sqrt{r^5 (r-2 M)+Q^2 r^4+\frac{\alpha  (1-2 \beta )}{10}  Q^4}}{r^4}.
\end{align}

\begin{figure}[ht!]
	\centering
    \includegraphics[width=90mm]{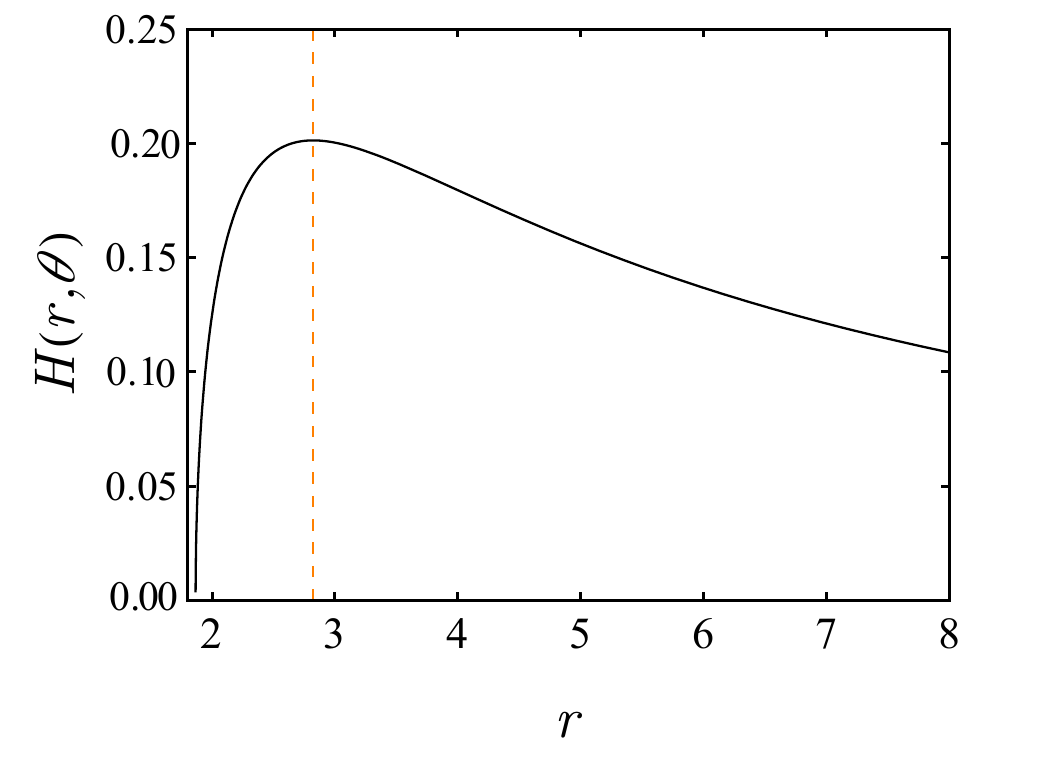}
	\caption{The topological potential $H(r,\theta)$ is plotted concerning the radial coordinate $r$. The dashed line represents the location of the photonic radius.}
	\label{fig:Hr}
\end{figure}

The behavior of potential $H(r,\theta)$ concerning the radius $r$ is shown in Fig. \ref{fig:Hr}. The location of the photon sphere corresponds to critical points that satisfy $\partial_rH = 0$. The maximum of the potential, for $M = 1$, $Q = 0.5$ and $\alpha=\beta=-0.1$ is located at $r_{ph}= 2.82289$ and shows instability of the equilibrium for the photonic radius. As depicted in this figure, minor perturbations can cause the photon to escape outward or fall into the black hole.

The corresponding vector field $(\varphi^r,\varphi^\theta)$ is constructed as 
\begin{align}
\varphi_ r &= \frac{1}{\sqrt{g_{rr}}} \, \partial_r H(r, \theta), \label{eq:phi_r} \\
\varphi_\theta &= \frac{1}{\sqrt{{g_{\theta \theta}}}}\, \partial_\theta H(r, \theta). \label{eq:phi_theta}
\end{align}
The normalized vector field is defined as 
$ n_a = \frac{\varphi_a}{\|{\varphi}\|}$,
where $a$ can take the values $1,2$ which corresponds to $r$ and $\theta$, respectively. 

\begin{figure}[ht!]
	\centering
	\includegraphics[width=90mm]{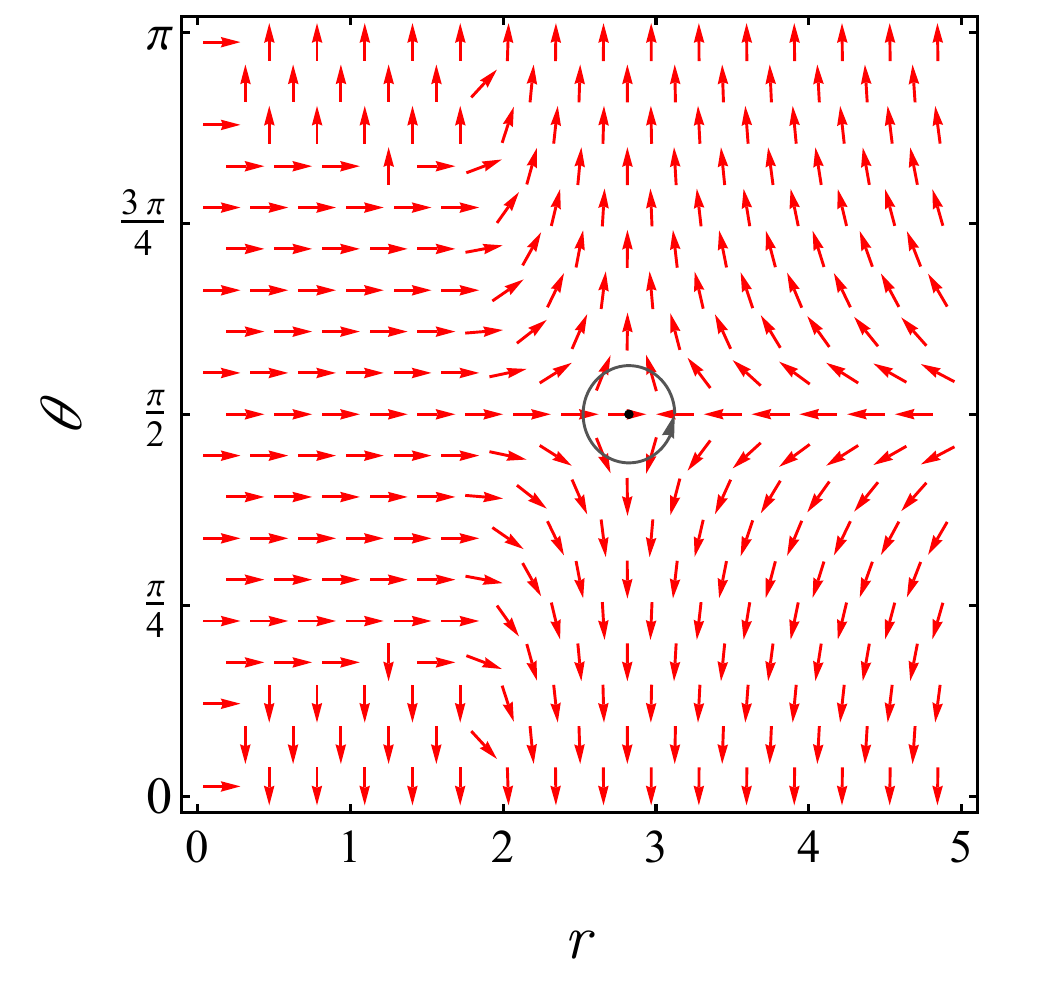}
    \caption{The vector field $n$ on the $(r, \theta)$ plane for $M = 1$, $Q = 0.5$ and $\alpha=\beta=-0.1$. The black loop encircles the photonic sphere point $(2.82286,\frac{\pi}{2})$. }
	\label{fig:Rphvector}
\end{figure}
Photon spheres can be interpreted as topological defects that emerge at points where the vector field vanishes. Following Duane's topological approach \cite{Duane1984}, when a closed contour encircles such a zero point, the resulting net topological charge corresponds to the winding number around that loop. 
The winding number is obtained by considering a closed,
smooth, positively oriented loop $C_i$ enclosing the $i$--th critical point. Then, the winding number is calculated by
\begin{equation}
    \omega_i=\frac{1}{2\pi}\oint \mathrm{d}\Omega,
\end{equation}
where $\Omega=\frac{\varphi_r}{\varphi_\theta}$, and the total topological charge of the black hole system is expressed as
\begin{equation}
    \mathcal{Q}= \sum_{i}^{} \omega_i \, .
\end{equation}
Thus, each photon sphere can be characterized by a distinct topological charge, which takes discrete values of $\mathcal{Q}=\pm1$. According to the classification introduced in~\cite{Wei2020,Sadeghi2024}, a photon sphere with topological charge $\mathcal{Q}=-1$ is associated with unstable, while one with topological charge $\mathcal{Q}=+1$ is associated with stability. It is worth mentioning that when a closed contour does not include any zero points of the vector field, the associated topological charge is zero, signifying that no photon spheres are present within the region. The vector space of photon sphere field is illustrated in Fig. \ref{fig:Rphvector} for fixed values of $M = 1$, $Q = 0.5$ and $\alpha = \beta = -0.1$. Figure~\ref{fig:Rphvector} displays a critical point situated outside the event horizon, indicated by a dot at the photonic radius $r_{ph} = 2.82286$. This point possesses a topological charge of $-1$, which characterizes it as an unstable photonic radius.


\section{Conclusion}

This study focused on exploring the gravitational signatures associated with a specific solution of modified electrodynamics in the framework of $f(R,T)$ gravity, as proposed in Ref. \cite{Rois:2024iiu}. We began by commenting on the derivation of the black hole solution and examined the conditions under which the event horizon $r_h$ remains both real and positive. The analysis revealed two important constraints: $2 \alpha \beta Q^4 - Q^4 > 0$ and $M > Q$.

We then shifted attention to the behavior of null geodesics. By numerically solving the corresponding equations, we analyzed the full trajectory of light rays. Particular emphasis was given to the photon sphere radius $r_{ph}$, whose magnitude was found to decrease as the charge $Q$ increased, for fixed negative values of $\alpha = \beta = -0.01$. Conversely, when $Q$ was held constant at $0.99$, a reduction in $\alpha$ and $\beta$ slightly enlarged $r_{ph}$. The evolution of the shadow radius followed a similar trend, displaying comparable dependencies on the parameters. These results enabled us to derive bounds on $Q$ for selected $\alpha$ and $\beta$ values by confronting our model with observational data from the Event Horizon Telescope.

The thermodynamic properties of the system were subsequently addressed. We computed the Hawking temperature, entropy, and heat capacity, identifying the presence of a remnant mass in the final evaporation stage. By considering the regime where $\alpha$, $\beta$, and $Q$ were small, we derived an analytical approximation for this remnant mass:
$M_{rem} \approx \, \frac{Q}{2} + \frac{\alpha  (1-2 \beta )}{20 Q}.$
Furthermore, the heat capacity analysis indicated the possibility of phase transitions in the system.

To investigate the resonant behavior of the background, we computed quasinormal modes for scalar, vector, tensor, and spinor perturbations. We also carried out a time--domain analysis to understand the temporal evolution of the perturbations, covering all spin types considered.

The gravitational lensing effects were analyzed in both the weak and strong deflection limits. For the former, we employed the Gauss--Bonnet approach formulated by Gibbons and Werner \cite{Gibbons:2008rj}, which enabled us to compute the deflection angle and assess the stability of photon spheres via the Gaussian curvature— found them to be unstable. The weak--field deflection angle $\hat{\alpha}(b, \alpha, \beta, Q)$ increased with rising $Q$ and also responded to decreasing $\alpha$ and $\beta$, for fixed $Q$.

In the strong lensing regime, we adopted the method introduced by Tsukamoto \cite{tsukamoto2017deflection} to evaluate the deflection angle, which turned to be an analytical expression. Interestingly, the strong field deflection $\hat{\alpha}^{\text{s}}(b, \alpha, \beta, Q)$ showed an opposite behavior compared to the weak case: increasing $Q$ led to a decrease in the angle when $\alpha$ and $\beta$ were kept fixed.

The topological features of the black hole configuration were also addressed. This included an analysis of topological thermodynamics as well as the concept of a topological photon sphere. In addition, the photon sphere’s stability was re--evaluated through an alternative method (topological one), serving as a complementary investigation to the Gaussian curvature approach.

As a future direction, we plan to explore the mechanisms of particle creation and the complete evaporation process in this $f(R,T)$ scenario, being analogous to the studies: \cite{araujo2025particle,araujo2025does,araujo2025particle2}.


\section*{Acknowledgments}
\hspace{0.5cm} A. A. Araújo Filho is supported by Conselho Nacional de Desenvolvimento Cient\'{\i}fico e Tecnol\'{o}gico (CNPq) and Fundação de Apoio à Pesquisa do Estado da Paraíba (FAPESQ), project No. 150891/2023-7. Also, V. B. Bezerra is partially supported by the Conselho Nacional de Desenvolvimento Científico e Tecnológico (CNPq) grant number 307211/2020-7. I. P. L. was partially supported by the National Council for Scientific and Technological Development - CNPq, grant 312547/2023-4. I. P. L.  would like to acknowledge networking support by the COST Action BridgeQG (CA23130) and the COST Action RQI (CA23115), supported by COST (European Cooperation in Science and Technology). N. H. would like to acknowledge the contribution of the COST Action CA21106 - COSMIC WISPers in the Dark Universe: Theory, astrophysics and experiments (CosmicWISPers), the COST Action CA21136 - Addressing observational tensions in cosmology with systematics and fundamental physics (CosmoVerse), the COST Action CA23130 - Bridging high and low energies in search of quantum gravity (BridgeQG).

\section{Data Availability Statement}

Data Availability Statement: No Data associated with the manuscript

\bibliographystyle{ieeetr}
\bibliography{main}

\end{document}